\documentclass[smallcondensed]{svjour3}     
\smartqed  
\usepackage[T1]{fontenc}
\usepackage{units}
\usepackage{geometry}
\usepackage{booktabs}
\usepackage{amsmath}
\usepackage{amssymb}
\usepackage{graphicx}
\usepackage{setspace}
\usepackage{esint}
\journalname{}
\usepackage{subfig}
\usepackage{lineno}
\makeatother

\begin{document}

\title{Zonal Jet Creation from Secondary Instability of Drift Waves for Plasma Edge Turbulence}
\subtitle{\normalfont{\emph{Dedicated to Professor Andrew J. Majda on the occasion of his seventieth birthday}}}


\author{Di Qi \and Andrew J. Majda}


\institute{Di Qi \at Department of Mathematics and Center for Atmosphere
and Ocean Science, Courant Institute of Mathematical Sciences, New
York University, New York, NY \email{qidi@cims.nyu.edu}
           \and
              Andrew J. Majda \at Department of Mathematics and Center for Atmosphere
and Ocean Science, Courant Institute of Mathematical Sciences, New
York University, New York, NY \email{jonjon@cims.nyu.edu}
}

\date{Received: date / Accepted: date}

\maketitle

\begin{abstract}
A new strategy is presented to explain the creation and persistence
of zonal flows widely observed in plasma edge turbulence. The core
physics in the edge regime of the magnetic-fusion tokamaks can be
described qualitatively by the one-state modified Hasegawa-Mima (MHM)
model, which creates enhanced zonal flows and more physically relevant
features in comparison with the familiar Charney-Hasegawa-Mima (CHM) model for both plasma and geophysical flows.
The generation mechanism of zonal jets is displayed from the secondary
instability analysis via nonlinear interactions with a background
base state. Strong exponential growth in the zonal modes is induced
due to a non-zonal drift wave base state in the MHM model, while stabilizing
damping effect is shown with a zonal flow base state. Together with the
selective decay effect from the dissipation, the secondary instability
offers a complete characterization of the convergence process to the purely
zonal structure. Direct numerical simulations with and without dissipation
are carried out to confirm the instability theory. It shows clearly
the emergence of a dominant zonal flow from pure non-zonal drift
waves with small perturbation in the initial configuration. In comparison,
the CHM model does not create instability in the zonal modes and usually
converges to homogeneous turbulence.
\end{abstract}

\section{Introduction}

Persistent zonal flows have been widely observed from the nature,
experiments, and numerical simulations of various rotating fluids
\cite{majda2003introduction,pedlosky2013geophysical,rhines1975waves,lin1998,diamond2005zonal,fujizawa2009}.
In fusion plasma, poloidally extended zonal jets in the edge region
of magnetically confined tokamak devices are of particular interest
where the turbulent transport severely limits plasma confinement and
leads to disastrous particle transport towards the boundary regime.
The anomalous particle transport along the radial direction due to
drift wave turbulence is found to be regulated and suppressed by the
generation of poloidal zonal structures \cite{hortonreview1999,diamond2005zonal,majda2018flux,xanthopoulos2011,pushkarev2013}.
It has been suggested from several theoretical and numerical results
\cite{smolyakov2000coherent,manfredi2001zonal} that zonal flows are
generated spontaneously by interacting with the drift waves. The drift
wave in plasma edge turbulence is also analogous to the Rossby wave
in geostrophic fluids where similar zonal jet structures are observed
\cite{majda2003introduction,dewar2007zonal}.

In understanding the drift wave -- zonal flow interacting dynamics,
it is useful to adopt simplified models where the most relevant physical
mechanism is identified. The Hasegawa-Mima (HM) \cite{hasegawa1978pseudo,dewar2007zonal}
and Hasegawa-Wakatani (HW) \cite{hasegawa1983plasma,numata2007bifurcation}
models are two groups of the simplified models which are capable to
qualitatively capture the energy-conserving nonlinear dynamics for
the formation of zonal jets. The HM models contain most essential
physical features in the drift wave -- zonal flow feedback loop mechanism,
while the HW models include a drift wave instability driving the turbulence.
Striking new features are generated in a newly developed \emph{flux-balanced
Hasegawa-Wakatani} (BHW) model \cite{majda2018flux,2018arXiv181200131Q},
where corrected treatment for the electron responses parallel to the
magnetic field lines is introduced as a more physical improvement
from the \emph{modified Hasegawa-Wakatani} (MHW) model \cite{hasegawa1983plasma,numata2007bifurcation}.
One important observation from the BHW model simulations is the enhanced
stronger zonal jets persistent in all the dynamical regimes even with
high particle resistivity \cite{2018arXiv181200131Q,majda2018flux}.
In contrast, the MHW model lacks the skill to maintain such strong
zonal jets and ceases to homogeneous drift wave turbulence at the
low resistivity limit.

In analyzing zonal flows from drift wave turbulence, the BHW model
consists of the interplay of the linear drift wave instability and
the nonlinear coupling between drift waves and zonal states. The modified
Hasegawa-Mima (MHM) model, as the exact adiabatic one-state limit
\cite{majda2018flux} of the BHW model, gives a cleaner setup by filtering
out the linear instability, thus offers a more desirable starting
model for investigating the central mechanism in flow self-organization from drift
waves to coherent zonal states through nonlinear interactions. The MHM model is modified
from the original Charney-Hasegawa-Mima (CHM) model \cite{hasegawa1978pseudo} for plasma and geophysical flows
which is also known as the quasi-geostrophic model \cite{dewar2007zonal,majda2003introduction}.
Modulational instability of drift waves offers a feedback mechanism
for the generation of zonal flows through the nonlinear interactions.
Theories and numerical experiments have been attempted \cite{manfredi2001zonal,smolyakov2000coherent,manz2013}
for describing the emergence of zonal flows by Reynolds stress in
both CHM and MHM models.

In this paper, we provide a precise explanation for the underlying
mechanism in creating the dominant zonal jets observed in the flux-balanced
models using secondary instability analysis about a background base
state of drift wave solutions. To identify the important nonlinear
impact between interactions of the drift wave states and the zonal
modes, we stay in the simple one-state Hasegawa-Mima (HM) models at
the adiabatic limit of the two-state BHW model, where no internal
instability due to the particle resistivity is included to add extra
complexity in the flow turbulence. The generation and persistence
of zonal flows in the HMH model is investigated by demonstrating that:
first a non-zonal drift wave base state induces strong instability
in the zonal modes, implying nonlinear energy transfer to the zonal
states; and then the generated zonal structure as a base state stays
stable to perturbations thus is maintained in time as the system evolves.
The secondary instability results are first illustrated by numerical
computation of the largest growth exponent from the Floquet theory.
Further, we use direct numerical simulations to confirm the jet creation
mechanism. Zonal flows are induced from a pure drift wave state adding
small isotropic fluctuations in the MHM model even without any dissipation
effect. In the case with dissipation, selective decay principle developed
in \cite{qi2018selective} helps to work together with the secondary
instability mechanism to drive the state to a final purely zonal structure.
In contrast in the CHM model, none of these instability and zonal
jets are created due to the improper treatment in the electron flux
response.

In the structure of this paper, we first briefly describe the BHM
and MHM models with a balanced averaged flux creating strong zonal
jets. Section 2 introduces the basic MHM model properties with its
major physical interpretation. The exact single mode drift wave solution
as well as the zonal mean dynamics is derived in Section 3 for the
background base mode in generating the zonal states. The precise energy
transfer mechanism to the zonal modes is explained through the secondary
instability about the background state with numerical computations
of the growth rate in Section 4. Section 5 uses direction numerical
simulations with and without dissipation effects for confirming the
developed theories. The conclusion and further discussion are given
in the final Section 6.

\subsection{The flux balanced models for plasma edge turbulence}

In tokamak devices, the realistic geometry would be a circular domain
with a predominant magnetic field $\mathbf{B}$ along the toroidal
$z$-direction. However, the shape of the plasma edge can be approximated
on a slab geometry under a Cartesian coordinate where the toroidal
magnetic surfaces are embedded. The Hasegawa-Wakatani models  describe
the drift wave -- zonal flow interactions of a two state coupled
system on the 2D slab geometry \cite{majda2018flux,dewar2007zonal},
with $x$-axis corresponding to the radial direction and $y$-axis
representing the poloidal direction. The\emph{ }flux-balanced Hasegawa-Wakatani
(BHW) model is introduced in \cite{majda2018flux} based on the flux-balanced
potential vorticity $q=\nabla^{2}\varphi-\tilde{n}$ and the density
fluctuation $n$ in the following form\addtocounter{equation}{0}\begin{subequations}\label{plasma_balance}
\begin{eqnarray}
\frac{\partial q}{\partial t}+\nabla^{\bot}\varphi\cdot\nabla q-\kappa\frac{\partial\varphi}{\partial y} & = & D\Delta q,\quad q=\nabla^{2}\varphi-\tilde{n},\label{eq:plasma_balance1}\\
\frac{\partial n}{\partial t}+\nabla^{\bot}\varphi\cdot\nabla n+\kappa\frac{\partial\varphi}{\partial y} & = & \alpha\left(\tilde{\varphi}-\tilde{n}\right)+D\Delta n,\label{eq:plasma_balance2}
\end{eqnarray}
\end{subequations}where $\varphi$ is the electrostatic potential,
$n$ is the density fluctuation from background density $n_{0}\left(x\right)$,
and $\mathbf{u}\equiv\nabla^{\bot}\varphi=\left(-\partial_{y}\varphi,\partial_{x}\varphi\right)$
is the velocity field. The parameter $\alpha$ is for adiabatic resistivity
of parallel electrons. It determines the degree to which electrons
can move rapidly along the magnetic field lines. The constant background
density gradient $\kappa=-\nabla\ln n_{0}$ is defined by the exponential
background density profile near the boundary $n_{0}\left(x\right)$
. $D$ acts on the two states with the Laplace operator as a homogeneous
damping \cite{2018arXiv181200131Q,majda2018flux}. The physical quantities
$\varphi$ and $n$ are decomposed into zonal mean stats $\overline{\varphi},\overline{n}$
and their fluctuations about the mean $\tilde{\varphi},\tilde{n}$
so that
\[
\varphi=\overline{\varphi}+\tilde{\varphi},\;n=\overline{n}+\tilde{n},\quad\overline{f}\left(x\right)=L_{y}^{-1}\int f\left(x,y\right)dy.
\]
In the BHW model, the poloidally averaged density $\overline{n}$
along $y$-direction is removed from the potential vorticity $q$.
In contrast, the original Hasegawa-Wakatani model introduced in \cite{hasegawa1983plasma}
as well as the modified version (MHW) \cite{numata2007bifurcation}
uses the `unbalanced' potential density $q=\nabla^{2}\varphi-n$ without
removing the mean state $\overline{n}$ in the potential vorticity,
leading to problems with the convergence at the adiabatic limit $\alpha\rightarrow\infty$.

The BHW model offers a more realistic formulation with several desirable
properties. Most importantly, it is shown from rigorous proof and
numerical confirmation \cite{majda2018flux,2018arXiv181200131Q} that
at the adiabatic limit, $\alpha\rightarrow\infty$, the BHW model
converges to the following equation
\begin{equation}
\frac{\partial q}{\partial t}+\nabla^{\bot}\varphi\cdot\nabla q-\kappa\frac{\partial\varphi}{\partial y}=D\Delta q,\quad q=\nabla^{2}\varphi-\tilde{\varphi},\label{eq:plasma_onelayer}
\end{equation}
which is called the modified Hasegawa-Mima model. Notice
the modification by removing zonal state in $\tilde{\varphi}$ in
the definition of potential vorticity $q$ above. On the other hand,
the MHW model shows performance significantly different from the MHM
model when $\alpha\rightarrow\infty$. The strong zonal jets created
from the BHW and MHM model and the convergence at the adiabatic limit
are discussed with explicit numerical simulations in \cite{majda2018flux}
(see Fig. 4 and 5 there). If we replace the potential vorticity in
(\ref{eq:plasma_onelayer}) by $q=\nabla^{2}\varphi-\varphi$ without
removing the zonal mean state, it recovers the Charney-Hasegawa-Mima model. The CHM model is identical to the quasi-geostrophic model
with $F$-plane effect describing geophysical turbulence with rotation
and stratification \cite{majda2003introduction,pedlosky2013geophysical,qi2016low}.
Then the rigorous theories developed for the geophysical model apply
to the CHM model exactly in the same way. In this paper, we will focus
on the HM models and especially changes in MHM model due to the averaged
flux correction in order to analyze the unstable effect purely from
a background base flow.

\section{The Hasegawa-Mima models and their representing properties}

To offer a better illustration with physical interpretations in
the Hasegawa-Mima models, we start with the original dimensional formulation
with physically related variables and derive the non-dimensionalized
version using the physical scales. The \emph{Charney-Hasegawa-Mima} (CHM)
equation and the \emph{modified Hasegawa-Mima} (MHM) equation can be formulated
under the same framework by defining a switch parameter with $s=0$
for CHM and $s=1$ for MHM as
\begin{equation}
\frac{D}{Dt}\left(\frac{\zeta}{\omega_{\mathrm{ci}}}+\ln\frac{\omega_{\mathrm{ci}}}{n_{0}}-\frac{e}{T_{e}}\left(\tilde{\varphi}+\delta_{s0}\overline{\varphi}\right)\right)=0,\label{eq:plasma_HM_dim}
\end{equation}
where $\varphi$ is the electrostatic potential, $\zeta=\nabla^{2}\varphi/B_{0}$
is the vorticity, $\mathbf{v}_{E}=-\nabla\varphi\times\hat{z}/B_{0}$
is the $\mathbf{E\times B}$ velocity. $D/Dt\equiv\partial_{t}+\mathbf{v}_{E}\cdot\nabla$
represents the material derivative along the velocity. In the parameters, $T_{e}$ is the reference electron temperature, $\omega_{\mathrm{ci}}=eB_{0}/m_{i}$
is the ion cyclotron frequency, and $m_{i}$ is the ion mass \cite{dewar2007zonal}. For model non-dimensionalization,
the new variables are introduced by
\[
e\varphi/T_{e}\rightarrow\varphi,\quad\omega_{\mathrm{ci}}t\rightarrow t,\quad\left(x,y\right)/\rho_{s}\rightarrow\left(x,y\right),
\]
with $\rho_{s}=\omega_{\mathrm{ci}}^{-1}\left(T_{e}/m_{i}\right)^{1/2}=\sqrt{m_{i}T_{e}}/eB_{0}$
the characteristic length scale of drift waves and $\omega_{\mathrm{ci}}^{-1}$
the characteristic time scale from the ion frequency. Accordingly, we
find the non-dimensional velocity and vorticity
\[
\rho_{s}\frac{eB_{0}}{T_{e}}\mathbf{v}_{E}\rightarrow\mathbf{v}_{E}=\nabla^{\bot}\varphi,\quad\rho_{s}^{2}\frac{eB_{0}}{T_{e}}\zeta\rightarrow\zeta.
\]
By substituting the non-dimensionalized quantities back into the dimensional equation
(\ref{eq:plasma_HM_dim}), we can rewrite the original
(with $s=0$) and modified (with $s=1$) Hasegawa-Mima equations in
the non-dimensional form as in (\ref{eq:plasma_onelayer}) so that
\begin{equation}
\left(\frac{\partial}{\partial t}+\nabla^{\bot}\varphi\cdot\nabla\right)q+\left(\partial_{x}\ln n_{0}\right)\frac{\partial}{\partial y}\tilde{\varphi}=0,\quad q=\zeta-\left(\tilde{\varphi}+\delta_{s0}\overline{\varphi}\right).\label{eq:HM_nondim}
\end{equation}
Above we introduce the new variable $q$ as the potential vorticity,
and if we assume a constant exponential decay profile in the background
density $n_{0}\sim\exp\left(-\kappa x\right)$ the coefficient becomes
a constant $\kappa\equiv-\partial_{x}\ln n_{0}$. 

On the magnetic surfaces, the electrons are assumed to respond adiabatically
so that locally thermodynamical equilibrium (with Boltzmann distribution)
is achieved on a given field surface. The electron density fluctuation
does not respond adiabatically on the averaged part of the electrostatic
potential $\overline{\varphi}$, thus only the flux balanced component
$e\tilde{\varphi}/T_{e}$ follows the Boltzmann distribution. This
offers the intuition for removing the zonal mean state $\overline{\varphi}$
in the MHM model. Though simple enough, the modified expansion leads
to much stronger zonal jet structures and more physically consistent
performance \cite{dewar2007zonal,manfredi2001zonal} compared with
the CHM results. 

\subsection{Galilean invariance and model energetics}

We illustrate some representative features especially from the model
flux modification. First, the MHM model enhances the excitation of
zonal flows with more prominent zonal structures. Consider a single
mode plane wave $\varphi=A_{z}\left(x,t\right)\exp\left(i\left(\mathbf{k\cdot x}-\omega t\right)\right)$ decomposed into a slowly varying zonal mean and fast
fluctuation.
The slow mode $A_{z}$ is assumed to be zonal and gives a constant
zonal mean flow $\mathbf{v}_{E}=\bar{v}\hat{y}$. We can find the
linearized dispersion relations for CHM and MHM separately as
\[
\begin{aligned}\mathrm{CHM:\quad} & \omega=\frac{k_{y}\kappa}{1+k^{2}}+\frac{k^{2}}{1+k^{2}}k_{y}\overline{v},\\
\mathrm{MHM:\quad} & \omega=\frac{k_{y}\kappa}{1+k^{2}}+k_{y}\overline{v}.
\end{aligned}
\]
Without the mean flow $\mathbf{v}_{E}$, the HM models generate
no instability with the same dispersion relation in the first term on the
right side. In small scales $k\gg1$, the CHM and MHM models
have similar dispersion relations. In large scales $k\lesssim1$ (that
is, near the scale of $\rho_{s}$), the modified model gets a stronger
feedback from the fluctuation (due to the simple Doppler shift $\mathbf{k\cdot v}_{E}=k_{y}\bar{v}$).
In the unmodified model, the Doppler shift is reduced by a factor
$k^{2}/\left(1+k^{2}\right)$. Detailed discussions about mean flow
interaction in the CHM model can be found in \cite{majda2003introduction}.

Second, the MHM model is Galilean invariant under boosts in the $y$
(poloidal) direction as desired for the symmetry in the poloidal direction
of tokamak devices. If we introduce a poloidal boost $V$ in the flow,
the new states become
\[
y^{\prime}=y-Vt,\quad\varphi^{\prime}=\varphi-VB_{0}x.
\]
Notice that only the fluctuation in the electrostatic potential is
invariant, $\tilde{\varphi}^{\prime}=\tilde{\varphi}$, while the
zonal mean is not invariant, $\overline{\varphi}^{\prime}=\overline{\varphi}-VB_{0}x$,
under the change of coordinate. The CHM model (and also the QG model
in geophysics) does not maintain this invariance due to the last term
$e\overline{\varphi}/T_{e}$ with $s=0$.

At last, we describe the model energetics. In the MHM model, two important
conserved quantities \cite{qi2018selective,numata2007bifurcation}
can be found as the energy $E$ and the enstrophy $W$

\begin{equation}
E=\frac{1}{2}\int\tilde{\varphi}^{2}+\left|\nabla\varphi\right|^{2},\quad W=\frac{1}{2}\int q^{2}=\frac{1}{2}\int\left(\tilde{\varphi}-\nabla^{2}\varphi\right)^{2}.\label{eq:energy_plasma}
\end{equation}
The nonlinear term in (\ref{eq:HM_nondim}) does not alter the value
of both energy and enstrophy. Thus the evolution of energy and enstrophy
can be purely determined by the dissipation effects. Especially with the
homogeneous damping form $D\Delta q$ in (\ref{eq:plasma_onelayer}),
we can derive the dynamical equations
\[
\begin{aligned}\frac{dE}{dt}= & -D\int\left|\nabla\tilde{\varphi}\right|^{2}+\left|\nabla^{2}\varphi\right|^{2},\\
\frac{dW}{dt}= & -D\int\left|\nabla q\right|^{2}=-D\int\left|\nabla\tilde{\varphi}\right|^{2}+2\left|\nabla^{2}\varphi\right|^{2}+\left|\nabla^{3}\varphi\right|^{2}.
\end{aligned}
\]
Similarly, the CHM model also maintains two invariants with $\tilde{\varphi}$
in the definition (\ref{eq:energy_plasma}) and equations replaced
by $\varphi$. The energetic equations play important roles in showing
the stability and decay properties. In particular, a
selective decay to a single dominant mode can be discovered based on the energetics \cite{majda2000selective,qi2018selective}.

\subsection{Selective decay in the flux balanced model}

The persistence of the zonal jets in the MHM model can be first explained
in a rigorous mathematical approach using the selective decay principle
\cite{qi2018selective,majda2000selective}. It states that proper
dissipation operator can dissipate all the non-zero drift wave
states at a much faster rate except a single selected dominant zonal
state in the MHM model. Precisely speaking, we have the convergence
to one of the selective decay zonal states $\overline{\varphi}_{k}$
for the normalized potential function in the $H^{1}$ sense 
\begin{equation}
\lim_{t\rightarrow\infty}\left\Vert \nabla\phi-\nabla\overline{\phi}_{k}\right\Vert _{0}=0,\quad\phi=\frac{\varphi}{\left\Vert \nabla\varphi\right\Vert _{0}}.\label{eq:converge}
\end{equation}
In the CHM model, the selective decay state $\varphi_{k}$ in a single
wavenumber can be also reached under the dissipation operator, while
the final converged state is one drift wave mode without zonal structure.
Proof for the selective decay results using different dissipation
operators including the Landau damping with detailed numerical simulations
are shown in \cite{qi2018selective}. Still, the generation of the
zonal structures from any arbitrary initial states is directly related
with the nonlinear interaction mechanism between different modes before
the selective decay effect takes over.

\section{Exact drift wave solutions and the zonal mean dynamics}

Now we propose the precise model framework for analyzing the instability, creation and stabilization of zonal jets through the nonlinear
interacting mechanism with the background base states. First, we introduce
additional rescaling for the HM models so that the important parameters
that determine the solution structures are identified. Starting with
the previous model formulation (\ref{eq:HM_nondim})
\[
\frac{\partial q}{\partial t}+\nabla^{\bot}\varphi\cdot\nabla q-\kappa\frac{\partial\varphi}{\partial y}=D\Delta q,\quad q=\nabla^{2}\varphi-\left(\tilde{\varphi}+\delta_{s0}\overline{\varphi}\right),
\]
with $s=1$ for the MHM model and $s=0$ for the CHM model, we propose
the rescaled set of variables $\left(q^{\prime},\varphi^{\prime},\mathbf{x}^{\prime},t^{\prime}\right)$
based on the characteristic length scale $L$ and the characteristic
flow velocity scale $U$
\[
\mathbf{x}=L\mathbf{x}^{\prime},\;\mathbf{u}=\nabla^{\bot}\varphi=U\mathbf{u}^{\prime},\quad t=Tt^{\prime},\;\varphi=\Phi\varphi^{\prime},\;q=Qq^{\prime}.
\]
The scales of the other variables can be found based on the values
of $L,U$ as
\[
T=\frac{L}{U},\;\Phi=UL,\;Q=\frac{\Phi}{L^{2}}=\frac{U}{L}.
\]
With the above rescaling, the unit wavenumber mode $\mathbf{p},\left|\mathbf{p}\right|=1$
for the new state represents the inverse length scale $L^{-1}$, and
the flow state with unit amplitude $u=\exp\left(\mathbf{p\cdot x}\right)$
represents the velocity with strength $U$. Accordingly, we derive
the rescaled HM models for the normalized states $\left(q^{\prime},\varphi^{\prime},\mathbf{x}^{\prime},t^{\prime}\right)$
based on the proposed characteristic scales
\begin{equation}
\frac{\partial q^{\prime}}{\partial t^{\prime}}+\nabla_{\mathbf{x}^{\prime}}^{\bot}\varphi^{\prime}\cdot\nabla_{\mathbf{x}^{\prime}}q^{\prime}-\kappa^{\prime}\frac{\partial\varphi^{\prime}}{\partial y^{\prime}}=D^{\prime}\Delta_{\mathbf{x}^{\prime}}q^{\prime},\quad q^{\prime}=\nabla_{\mathbf{x}^{\prime}}^{2}\varphi^{\prime}-L^{2}\left(\tilde{\varphi}^{\prime}+\delta_{s0}\overline{\varphi}^{\prime}\right).\label{eq:HM_rescale}
\end{equation}
The above rescaled equation (\ref{eq:HM_rescale}) is not changed
much just with new non-dimensional parameters $\kappa^{\prime},D^{\prime}$.
Notice that the length scale $L$ now appears explicitly in the potential
vorticity $q^{\prime}$. The flow solution is entirely
determined by the two characteristic coefficients, $\kappa^{\prime}=\frac{\kappa L^{2}}{U}$
and $D^{\prime}=\frac{D}{UL}$. $\kappa^{\prime}$ has the same role
as the Rhines number $\mathrm{Rh}^{-1}$ in geophysical flows, showing
the anisotropic effect in the drift waves; and $D^{\prime}$ as the
Reynolds number $\mathrm{Re}^{-1}$ for the dissipation effect \cite{rhines1975waves,pedlosky2013geophysical}.
We will focus on the MHM model with $s=1$ in (\ref{eq:HM_rescale})
and neglect the primes on the states in the rest part of the
paper (the CHM case can be easily implied and detailed theories
for the CHM model have actually been developed in the geophysical literatures
\cite{majda2003introduction,majda2016introduction,majda2018strategies}).

\subsection{General base flow state from the exact solution}

Exact drift wave solutions of the MHM equation in (\ref{eq:HM_rescale})
can be found by considering a single mode base state. We assume the
base mode in the electrostatic potential and the potential vorticity
for a single wavenumber $\mathbf{k}=\left(k^{x},k^{y}\right)$ as
\begin{equation}
\varphi\left(\mathbf{x},t\right)=\hat{\varphi}\exp\left(i\left(\mathbf{k\cdot x}-\omega\left(\mathbf{k}\right)t\right)\right),\quad q\left(\mathbf{x},t\right)=-\left[k^{2}+L^{2}\left(1-\delta_{k^{y},0}\right)\right]\hat{\varphi}\exp\left(i\left(\mathbf{k\cdot x}-\omega\left(\mathbf{k}\right)t\right)\right),\label{eq:single_mode}
\end{equation}
where $\omega\left(\mathbf{k}\right)$ is the dispersion relation
in the drift waves. The Kronecker delta operator is introduced for
the MHM model modification in the zonal modes $k^{y}=0$. The normalized
wavenumber length $k=\left|\mathbf{k}\right|$ is compared with the
characteristic scale $L$, i.e., wavenumbers $k<1$ characterize the
scales larger than $L$ and wavenumbers $k>1$ for scales smaller
than $L$. Since we only consider a single mode solution in the above
form, the nonlinear term $\nabla^{\bot}\varphi\cdot\nabla q$ vanishes
in the equation through the self interaction $\mathbf{k}^{\bot}\cdot\mathbf{k}=0$.
The dispersion relation can be found as 
\begin{equation}
\omega\left(\mathbf{k}\right)=\kappa^{\prime}\frac{k^{y}}{k^{2}+L^{2}},\quad\kappa^{\prime}=\frac{\kappa L^{2}}{U}.\label{eq:disp_relation}
\end{equation}
Notice that the above dispersion relation $\omega\left(\mathbf{k}\right)$
is valid for both the MHM and CHM models. In fact, the MHM model only
adds modifications for the zonal modes with $k^{y}=0$. In the zonal
modes, the dispersion relation becomes $\omega=\kappa^{\prime}k^{y}/k^{2}=0$.
The formula (\ref{eq:disp_relation}) is still valid for both the
MHM ($s=1$) case and CHM ($s=0$) case. Next, we will consider the
instability of fluctuation perturbations added on top of the representative
exact solutions in the form of (\ref{eq:single_mode}).

\subsection{Dynamical equation of the zonal mean state}

Before proceeding to the detailed discussion about secondary instability
to induce zonal structures, it is useful to check the exact dynamical
equation for the zonal state to achieve a first intuition about the
nonlinear interacting mechanism. Evolutions of the zonal components
$\overline{q}$ can be extracted from the above MHM model by directly
taking the zonal average about the original equation (\ref{eq:HM_rescale})
\begin{equation}
\frac{\partial\overline{q}}{\partial t}+\frac{\partial}{\partial x}\overline{uq}=-D^{\prime}\frac{\partial^{4}\overline{q}}{\partial x^{4}},\quad u=-\partial_{y}\varphi,\label{eq:mean_flow}
\end{equation}
with $u$ the zonal velocity fluctuation. The background density gradient
term $\kappa^{\prime}\partial_{y}\varphi$ vanishes after the average
along $y$-direction. If there is no non-zero zonal mode $k^{y}=0$,
we can check that the advection term vanishes
\[
\overline{uq}=\overline{\left(-ik^{y}\hat{\varphi}e^{i\left(\mathbf{k\cdot x}-\omega t\right)}\right)\left(-\left(k^{2}+L^{2}\right)\hat{\varphi}e^{i\left(\mathbf{k\cdot x}-\omega t\right)}\right)+c.c.}=\overline{ik^{y}\left(k^{2}+L^{2}\right)\hat{\varphi}^{2}e^{2i\left(\mathbf{k\cdot x}-\omega t\right)}+c.c.}=0,
\]
after the integration along the $y$-direction, with $c.c.$ as the complex
conjugate part. Therefore single non-zonal fluctuation mode makes
no contribution to the zonal mean structure in the above form, where
an exact solution (\ref{eq:single_mode}) can be reached. 

On the other hand, zonal wave could be generated through the interactions
between different wavenumbers. If we consider a general solution with
multiple drift wave modes, the mean state equation can be derived
in the following form
\[
\left(\frac{d}{dt}+Dk^{2}\right)\overline{\varphi}_{\mathbf{k}}\left(t\right)=k^{-2}\sum_{\mathbf{m+n=k}}C_{\mathbf{kmn}}\left(t\right)\left(n^{2}-m^{2}\right)\overline{\varphi_{\mathbf{m}}\varphi_{\mathbf{n}}},
\]
with the zonal mode $\mathbf{k}=\left(k^{x},0\right)$ and the fluctuation feedback $\overline{\varphi_{\mathbf{m}}\varphi_{\mathbf{n}}}$,
$m^{y}\ne0,n^{y}\ne0$ to the
zonal state. The coupling parameter $C_{\mathbf{kmn}}\left(t\right)$
is time-dependent on the dispersion relations (\ref{eq:disp_relation})
and models the triad coupling between the interacting drift wave modes
\[
C_{\mathbf{kmn}}\left(t\right)=k^{x}n^{x}e^{i\left(\omega\left(\mathbf{k}\right)-\omega\left(\mathbf{m}\right)-\omega\left(\mathbf{n}\right)\right)t},\quad m^{x}+n^{x}=k^{x},\;m^{y}+n^{y}=0.
\]
The right hand side of the above equation describes the nonlinear
flux to the mean mode for the generation of a zonal
jet. In the next section, we illustrate in a rigorous way how this
nonlinear coupling term transfers the fluctuating energy in the non-zonal
drift wave modes to the zonal directions, and maintains the zonal
structures through the secondary stability mechanism.

\section{Secondary instability from a base flow state}

In this section, we provide a precise description for the energy transfer
mechanism from the drift wave states (with $k^{y}\neq0$) to zonal
flows (with $k^{y}=0$). From the discussion in the last section,
energy in the fluctuation drift wave modes is first transferred
to the zonal directions through the resonant triad interactions; 
next the accumulation of energy in the zonal modes gets saturated
and stabilized with the large-scale stability of a zonal base state.
The drift wave -- zonal flow interactions are characterized by the
secondary instability analysis based on a base flow state. A brief
summary for the main result achieved for the MHM model is: a fluctuation
drift wave base state will induce strong instability along a wide
range of zonal modes, implying strong transport of energy from non-zero
drift waves to the zonal directions; in contrast a purely zonal flow
base state will add no instability to zonal modes or the drift waves,
showing stability in the generated zonal mean structure.

\subsection{Formulation of the secondary instability from a base state}

First notice that drift wave instability is filtered out in the one-state
Hasegawa-Mima models (\ref{eq:HM_rescale}), which enables us to focus
on the nonlinear interaction mechanism from a background state. Below
we derive the secondary instability based on the MHM model (the CHM
case can be derived in a similar fashion). The development is motivated
by the secondary instability analysis carried out in \cite{lee2003stability}
for geophysical turbulence on beta-plane and in \cite{meshalkin1962investigation}
for 2D Navier-Stokes equations with a Kolmogorov base flow using \emph{Floquet
theory}. However, the main focus here is the changes introduced through the
flux modification in the MHM model potential vorticity $q=\nabla^{2}\varphi-\tilde{\varphi}$.

For simplicity in the MHM model, we consider a single mode base state
with wavevector $\mathbf{p}=\left(\cos\theta_{p},\sin\theta_{p}\right)$
of unit length (then $\theta_{p}$ defines the characteristic direction
of the base flow with $\theta_{p}=0$ for the zonal flow state and
$\theta_{p}=\frac{\pi}{2}$ for the pure drift wave state). The single
mode base flow potential $\varPhi_{p}$ and vorticity $Q_{p}=\nabla^{2}\varPhi_{p}-L^{2}\tilde{\varPhi}_{p}$
with the unit length wavenumber $\mathbf{p}$ can be defined from
the exact solution formula (\ref{eq:single_mode}) as 
\begin{equation}
\varPhi_{p}=-\frac{1}{2}e^{i\left(\mathbf{p\cdot x}-\omega\left(\mathbf{p}\right)t\right)}-\frac{1}{2}e^{-i\left(\mathbf{p\cdot x}-\omega\left(\mathbf{p}\right)t\right)},\quad Q_{p}=-\left[1+L^{2}\left(1-\delta_{p^{y},0}\right)\right]\varPhi_{p},\label{eq:base_state}
\end{equation}
with the dispersion relation $\omega\left(\mathbf{p}\right)=\kappa^{\prime}\frac{p^{y}}{1+L^{2}}$
defined in (\ref{eq:disp_relation}). From the rescaled equation (\ref{eq:HM_rescale})
using the characteristic scales $\left(L,U\right)$, the base solution
$U_{p}=\nabla^{\bot}\varPhi_{p}$ with unit wavenumber $p=1$ represents
the characteristic length scale $L$ and the characteristic flow velocity
$U$. For the corresponding CHM model solution, we just need to remove
the delta functions in the above formula. With no additional internal
instability in the HM models, the above solution can be simply generated
by a combined forcing and damping effect.

The Floquet theory considers a fluctuation solution with a characteristic
multiplier $e^{\mu t}$. To study the secondary instability at each
wavenumber $\mathbf{k}$ based on the single-mode base flow (\ref{eq:base_state}),
we introduce fluctuations, $\varphi_{p}=\varphi-\varPhi_{p}$, $q_{p}=q-Q_{p}$,
on top of the base state in the form
\begin{equation}
\begin{aligned}\varphi_{p}= & e^{\mu t}e^{i\left(\mathbf{k\cdot x}-\omega\left(\mathbf{k}\right)t\right)}\sum_{l=-N}^{N}\hat{\varphi}_{l}e^{il\left(\mathbf{p}\cdot\mathbf{x}-\omega\left(\mathbf{p}\right)t\right)},\\
q_{p}= & e^{\mu t}e^{i\left(\mathbf{k\cdot x}-\omega\left(\mathbf{k}\right)t\right)}\sum_{l=-N}^{N}-\left[q^{2}\left(l\right)+L^{2}\left(1-\delta_{q^{y},0}\right)\right]\hat{\varphi}_{l}e^{il\left(\mathbf{p}\cdot\mathbf{x}-\omega\left(\mathbf{p}\right)t\right)}.
\end{aligned}
\label{eq:fluc_state}
\end{equation}
The perturbation is added on the wavenumber $\mathbf{k}$ with dispersion
relation $\omega\left(\mathbf{k}\right)$. $\mu$ is the \emph{Floquet
exponent} characterizing the growth or decay rate in this perturbed
mode. $\mathbf{q}\left(l\right)=\mathbf{k}+l\mathbf{p}$ with $q^{2}\left(l\right)=\left|\mathbf{q}\left(l\right)\right|^{2}$
is the full wavenumber with the perturbation. The perturbation mode
$\hat{\varphi}_{l}$ is added on each of the directions as multiples
of the characteristic wavenumber $l\mathbf{p}$ . It refers to a multiplicative
perturbation along the directions of the base state flow. Here we
truncate the perturbed states within the leading modes up to $N$
multiples of the base state. Thus the perturbation coefficients $\left\{ \hat{\varphi}_{l}\right\} $
form a finite dimensional system of $2N+1$ real states. The problem
can be further extended to an infinite dimensional system as the truncation size
$N\rightarrow\infty$. Still, since we are mostly interested in the
instability among the largest scales in the zonal states, a finite
size truncation $N$ will be sufficient for characterizing this instability
(see the instability results illustrated in Section \ref{subsec:Secondary-instability-num}).

By subtracting the base flow solution $\left(\varPhi_{p},Q_{p}\right)$
from the MHM equation (\ref{eq:HM_rescale}), the fluctuation equation
for the perturbed component $q_{p}$ of potential vorticity can be
derived in the following form
\[
\frac{\partial q_{p}}{\partial t}+\nabla^{\bot}\varPhi_{p}\cdot\nabla\left[q_{p}+\varphi_{p}+L^{2}\left(1-\delta_{p^{y},0}\right)\varphi_{p}\right]+\nabla^{\bot}\varphi_{p}\cdot\nabla q_{p}-\kappa^{\prime}\frac{\partial\varphi_{p}}{\partial y}=D^{\prime}\Delta q_{p},\;q_{p}=\nabla^{2}\varphi_{p}-\tilde{\varphi}_{p},
\]
where the relation for the single mode base state vorticity and potential,
$Q_{p}=-\left[1+L^{2}\left(1-\delta_{p^{y},0}\right)\right]\varPhi_{p}$,
is used. Still in the secondary stability analysis, we focus on the
secondary instability induced by the interactions between the background
base state $\left(\varPhi_{p},Q_{p}\right)$ and the fluctuation modes
$\left(\varphi_{p},q_{p}\right)$ with small perturbations. The higher-order
nonlinear term between fluctuation modes, $\nabla^{\bot}\varphi_{p}\cdot\nabla q_{p}$,
is assumed to stay small in the starting transient state and is neglected
in this instability analysis. The unstable growth in perturbations
due to the background base state is represented by the growth parameter
$\mu$ in the fluctuation states (\ref{eq:fluc_state}). With a positive
value in the real part of $\mu$, it infers exponential growth of the perturbation
mode in the transient state on top of the base mode due to the interactions.
The equation for calculating the secondary growth $\mu$ can be achieved
by substituting the fluctuation modes (\ref{eq:fluc_state}) into
the above fluctuation equation for $q_{p}$. It becomes an eigenvalue
problem for each perturbation direction $\mathbf{k}$ individually
with interactions between the triad neighboring modes $\left(\hat{\varphi}_{l-1},\hat{\varphi}_{l},\hat{\varphi}_{l+1}\right)$
\begin{equation}
\begin{aligned}\left[\mu\left(\mathbf{k}\right)-i\omega\left(\mathbf{k}\right)-il\omega\left(\mathbf{p}\right)+i\omega\left(\mathbf{q}\left(l\right)\right)+Dq^{2}\left(l\right)\right]\hat{\varphi}_{l}\\
+\frac{1}{2q^{2}\left(l\right)}\left(\mathbf{p\times k}\right)\cdot\hat{z}\left[q^{2}\left(l-1\right)-1-L^{2}\left(1-\delta_{p^{y},0}\right)\right]\hat{\varphi}_{l-1}\\
-\frac{1}{2q^{2}\left(l\right)}\left(\mathbf{p\times k}\right)\cdot\hat{z}\left[q^{2}\left(l+1\right)-1-L^{2}\left(1-\delta_{p^{y},0}\right)\right]\hat{\varphi}_{l+1}
\end{aligned}
=0,\;l=-N,\cdots,N.\label{eq:sec_instability}
\end{equation}
with the combined wavenumber $\mathbf{q}\left(l\right)=\mathbf{k}+l\mathbf{p}$,
$q^{2}\left(l\right)=\left|\mathbf{q}\left(l\right)\right|^{2}$,
and the coupling coefficient $\left(\mathbf{p\times k}\right)\cdot\hat{z}=p^{x}k^{y}-p^{y}k^{x}$.
The first row above includes the effects from the dispersion relation
and damping, and the second and third rows are due to the interactions
with the neighboring modes through the background state. The equations
in (\ref{eq:sec_instability}) form a $\left(2N+1\right)\times\left(2N+1\right)$
tri-diagonal system (with $N$ the number of base mode perturbations
added) based on the perturbed modes $\hat{\varphi}_{l}$ for each
wavenumber $\mathbf{k}$. The solution of $\mu$ reflecting instability
can be achieved by computing the eigenvalues of the corresponding
tri-diagonal matrix. The maximum positive eigenvalue in real part of $\mu^{\mathrm{max}}\left(\mathbf{k}\right)$
refers to the most unstable growth rate in the mode $\mathbf{k}$
according to the base flow in direction $\mathbf{p}$.

\

As a comment for the general case, the base flow can be generalized
to a combination of the single drift wave solutions (\ref{eq:single_mode})
with a group of characteristic wavevectors $\left\{ \mathbf{p}_{j}\right\} _{j=1}^{J}$.
The base states for the electrostatic potential $\varPhi$ and the
potential vorticity $Q=\nabla^{2}\varPhi-L^{2}\tilde{\varPhi}$ then
can be defined as the combination of all the modes
\[
\varPhi\left(\mathbf{x},t;\left\{ \mathbf{p}_{j}\right\} _{j=1}^{J}\right)=\sum_{j=1}^{J}A_{j}e^{i\left(\mathbf{p}_{j}\cdot\mathbf{x}-\omega\left(\mathbf{p}_{j}\right)\right)},\;Q\left(\mathbf{x},t;\left\{ \mathbf{p}_{j}\right\} _{j=1}^{J}\right)=-\sum_{j=1}^{J}A_{j}\left[p_{j}^{2}+L^{2}\left(1-\delta_{p_{j,0}^{y}}\right)\right]e^{i\left(\mathbf{p}_{j}\cdot\mathbf{x}-\omega\left(\mathbf{p}_{j}\right)\right)}.
\]
Above $J$ is the total number of characteristic modes that are combined
in the background base state $\varPhi$. Accordingly for the fluctuation
state about the combined base solution, we need to combine perturbations
on mode $\mathbf{k}$ for each component of the base state in the
form
\[
\varphi\left(\left\{ \mathbf{p}_{j}\right\} \right)=e^{\mu t}e^{i\left(\mathbf{k\cdot x}-\omega\left(\mathbf{k}\right)t\right)}\cdot\sum_{\mathbf{l}}\hat{\varphi}_{\mathbf{l}}\exp\left[\sum_{j=1}^{J}il_{j}\left(\mathbf{p}_{j}\cdot\mathbf{x}-\omega\left(\mathbf{p}_{j}\right)t\right)\right],
\]
where the index $\mathbf{l}=\left(l_{1},l_{2},\cdots,l_{J}\right)\in\mathbb{Z}^{J}$
goes through all the $J$-multiples in the summation. Similar result
can be derived just with much more complicated formulas. We will leave
the multiple base mode case in future investigation and focus on the
central issue about the generation of zonal jets through a single
drift wave mode.

\subsection{Secondary instability about a single mode base state\label{subsec:Secondary-instability-num}}

Now we check the secondary growth rate $\mu$ according to different
types of the base flows through simple numerical tests. Especially,
we consider the single-mode drift wave base state $\mathbf{p}_{1}=\left(0,1\right)$
and the zonal flow base state $\mathbf{p}_{2}=\left(1,0\right)$.
The number of perturbed modes in the fluctuation state (\ref{eq:fluc_state})
is fixed at $N=20$ (thus it forms a $41\times41$ matrix). Larger
truncation sizes of $N$ have been checked and show no significant
difference for the instability in the zonal modes. 

From the scale analysis, we find that the flow solutions can be determined
by the two non-dimensional parameters, $\kappa^{\prime}=\frac{\kappa L^{2}}{U}$
and $D^{\prime}=\frac{D}{UL}$. Especially, $L$ represents the characteristic
scale in the base mode $\mathbf{p}$ and $U$ defines the strength
in the base flow. In the numerical tests, the strategy is that we
fix the model parameters as $\kappa=0.5,D=5\times10^{-4}$, and change
the scale parameters $L$ and $U$. In general, we choose a computational
domain size $L_{D}=40$ used in the direct numerical simulations in
Section \ref{sec:Direct-numerical-simulations}. The characteristic
length scale $L$ represents a drift wave state with wavenumber $s=L_{D}/L$.
We tested two different length scales $L=10,20$ and two velocity
scales $U=1,0.1$. Correspondingly, it gives the non-dimensional parameters
$\kappa^{\prime}=50,200,500$ and $D^{\prime}=5\times10^{-5},2.5\times10^{-5},5\times10^{-4}$
in the three test cases shown in Figure \ref{fig:sec_instability_drift}.

\subsubsection{Instability about drift wave mode $\mathbf{p}=\left(0,1\right)$\label{subsec:Instability-about-drift}}

In the first test case, we consider the secondary instability due
to a single-mode drift wave state $\mathbf{p}=\left(0,1\right)$ varying
only along the $k^{y}$-direction. In this case, it illustrates the
transfer of energy from the purely drift wave modes to the zonal jet
states through the nonlinear interactions between the background state
and the fluctuations. Especially, from the drift wave linear instability
in the two-state HW model, the most linearly unstable modes are always
along the $k^{y}$ axis (see Fig. 2 in \cite{2018arXiv181200131Q}
for the linear instability result). Therefore, starting from the HM
model framework, the pure drift wave background state represents
the first excited states due to the (unresolved) linear instability
effect. We investigate the secondary instability induced due to this
background state from linear drift wave instability.

\begin{figure}
\subfloat{\includegraphics[scale=0.32]{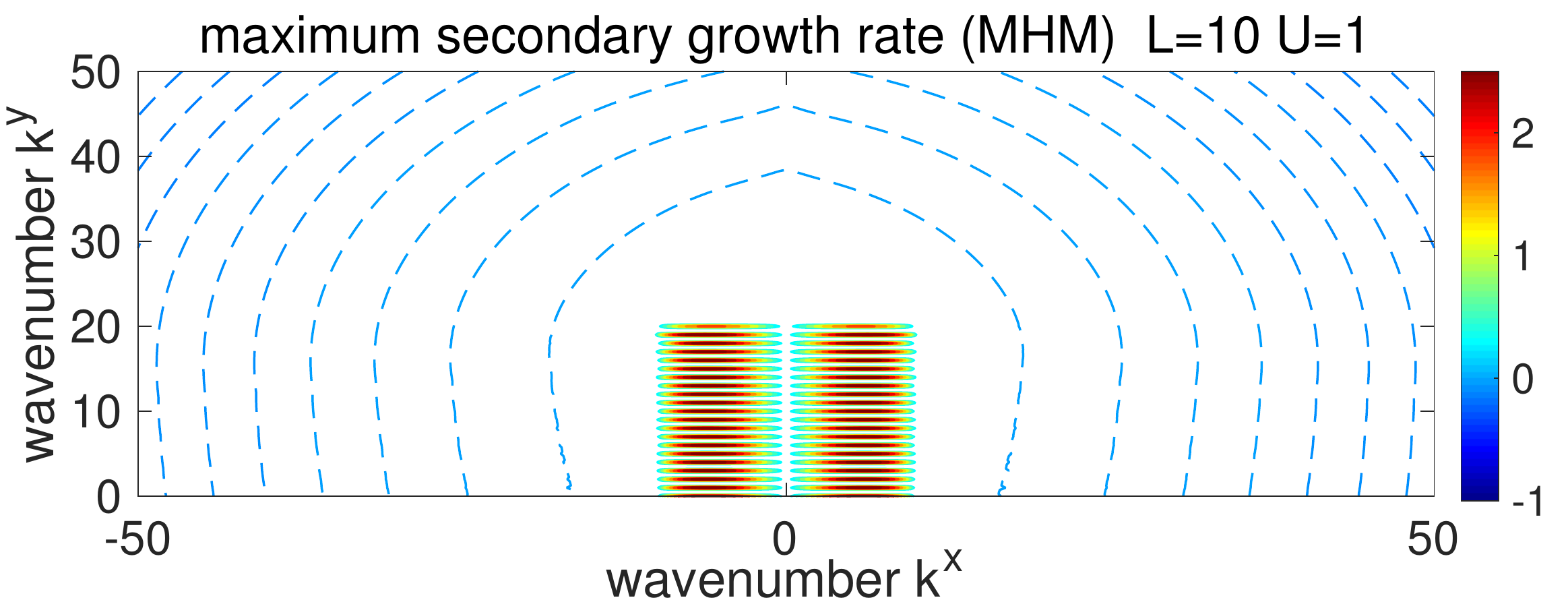}\includegraphics[scale=0.32]{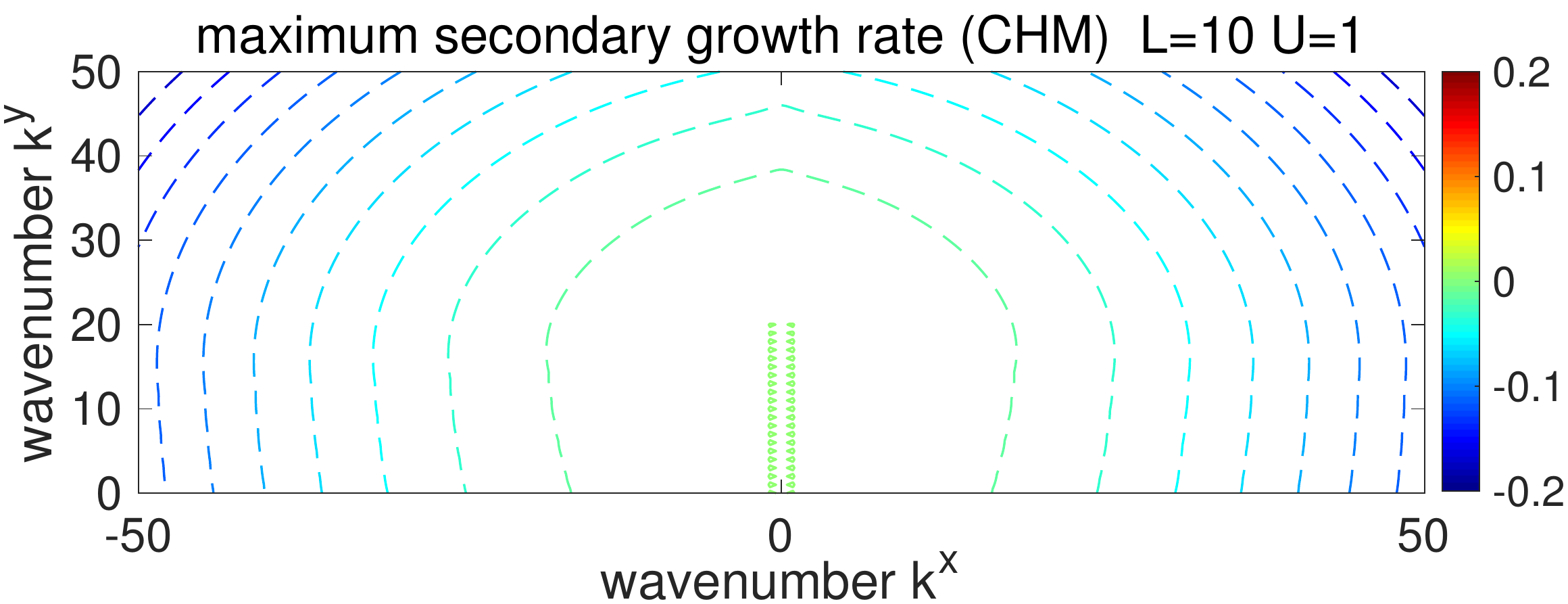}}

\vspace{-1em}

\subfloat{\includegraphics[scale=0.32]{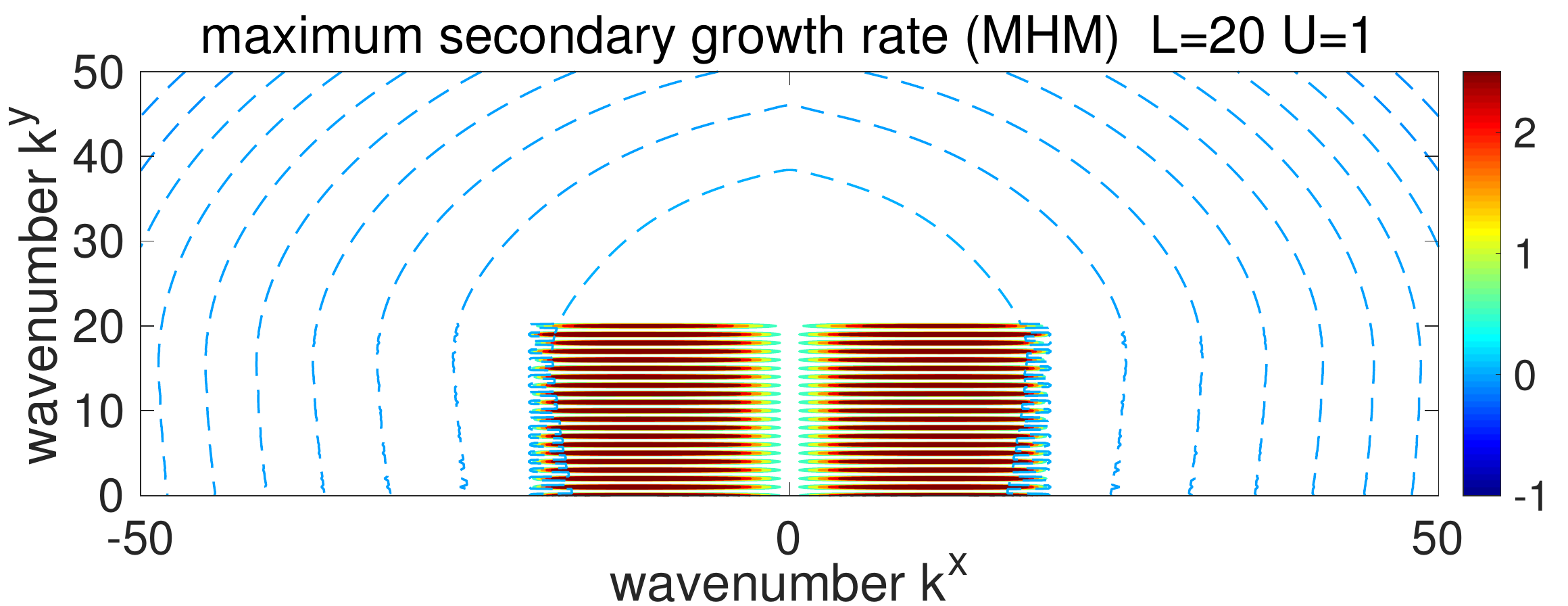}\includegraphics[scale=0.32]{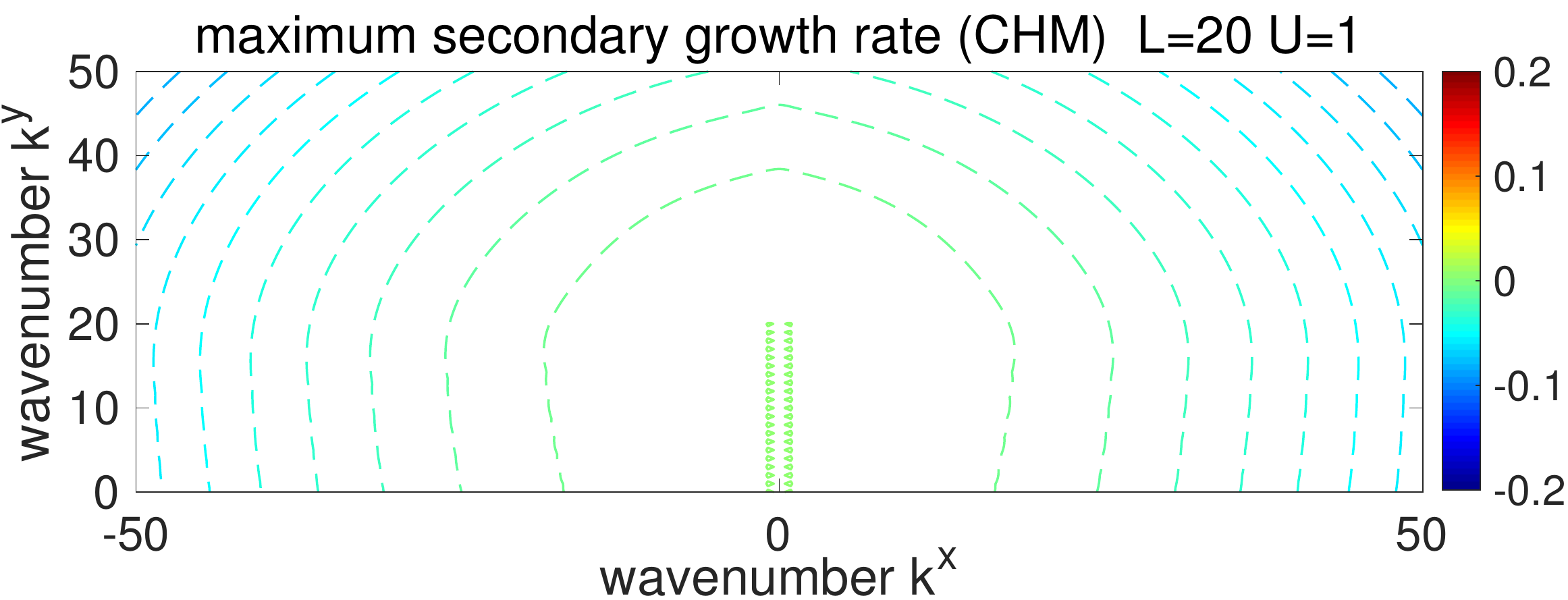}}

\vspace{-1em}

\subfloat{\includegraphics[scale=0.32]{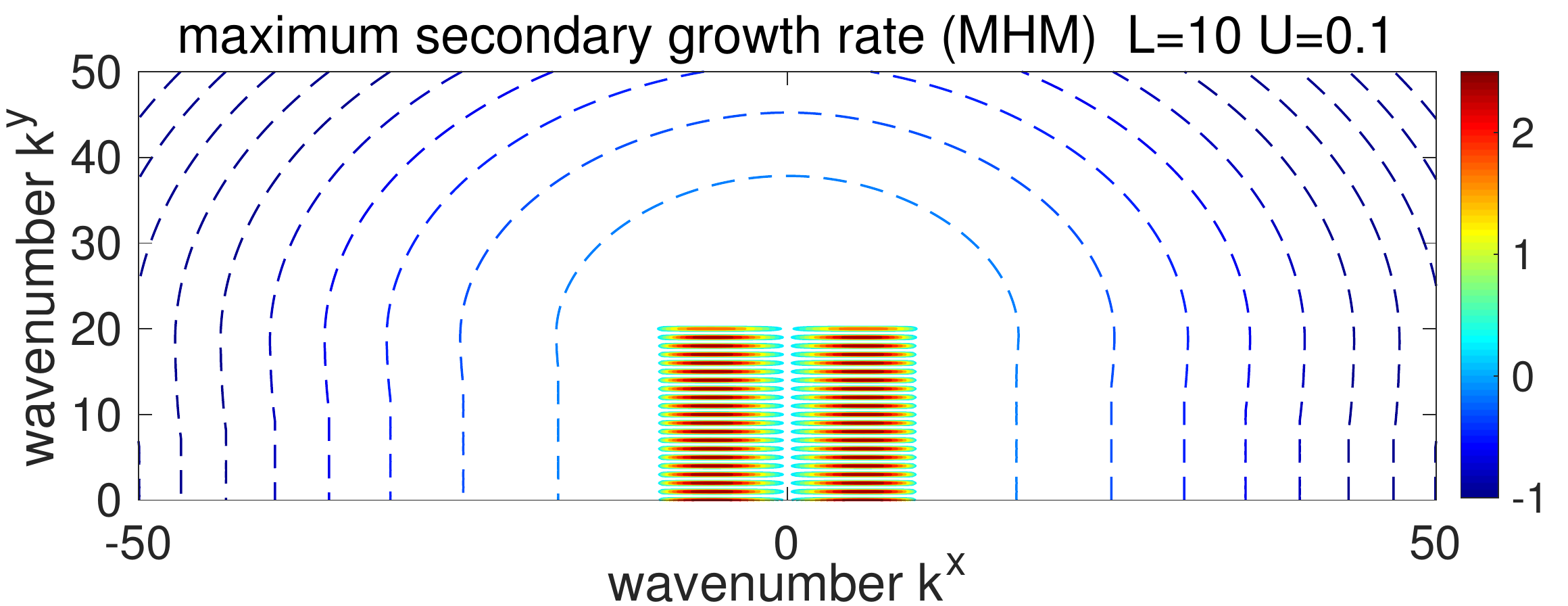}\includegraphics[scale=0.32]{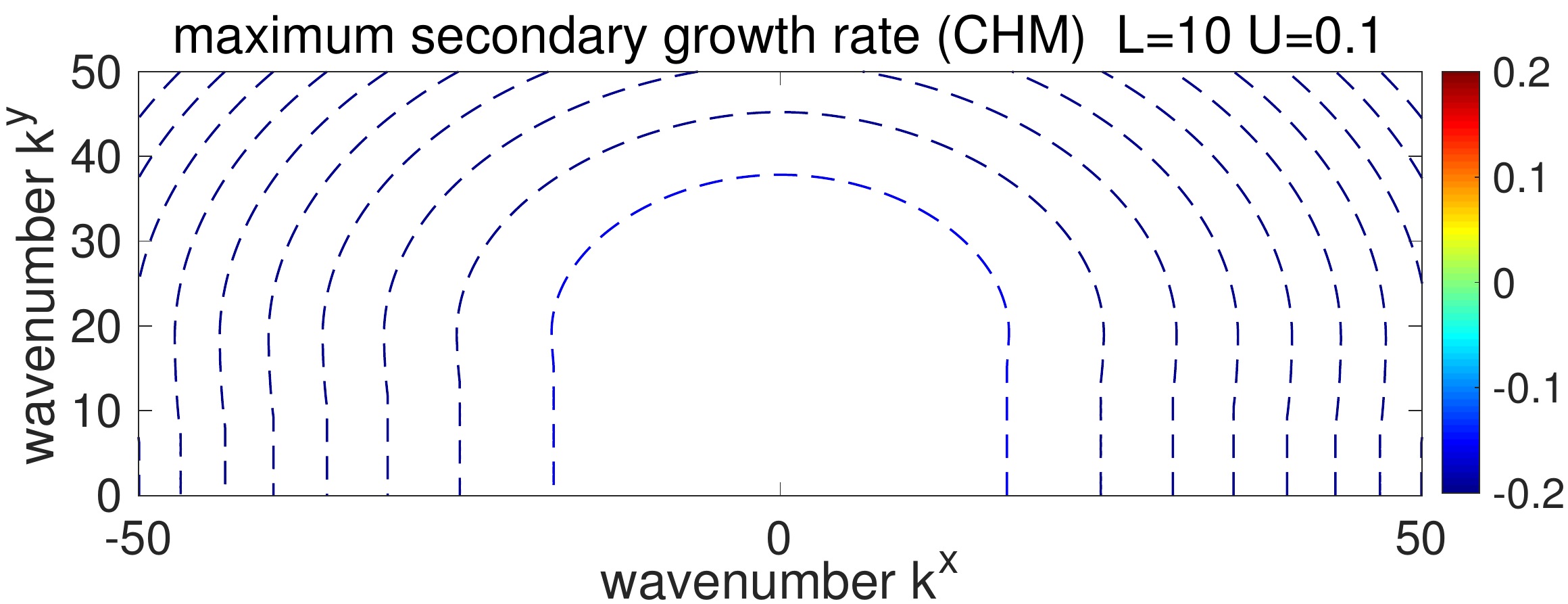}}

\caption{Maximum growth rate at each spectral mode from the secondary stability
analysis according to the drift wave base flow $\mathbf{p}=\left(0,1\right)$.
Solid lines are for positive growth rates and dashed lines are for
negative damping rates. Results for the MHM model (left) and for the
CHM model (right) in same parameter domains are compared. Different
characteristic scales for $\left(L,U\right)$ are compared. The other
parameters used are $\kappa=0.5,D=5\times10^{-4}$. Notice the large
amplitudes in the MHM model and small values in the CHM model from
the colorbars.\label{fig:sec_instability_drift}}
\end{figure}

In Figure \ref{fig:sec_instability_drift}, we plot the contours for
the maximum growth rate in the real part of the exponent $\mu$ with
different wavenumbers $\mathbf{k}$ in the spectral domain. Results
with balanced vorticity $q=\nabla^{2}\varphi-L^{2}\tilde{\varphi}$
in the MHM model and original vorticity $q=\nabla^{2}\varphi-L^{2}\varphi$
in the CHM model are compared. With drift wave mode $\left(0,1\right)$
as the basic flow, strong positive growth rate is generated in the
large-scale zonal modes with $k^{y}=0$ in the MHM model uniformly
for all the tested model scales $L,U$. All the largest positive growth
rates are located near the zonal direction. This corresponds to the
rapid energy transfer from the drift waves to form up zonal structures.
In small scales the real parts of the eigenvalues become negative
due to the much stronger damping in the smaller scales. In contrast,
the CHM model without the flux modification gets no instability but
only has negative decaying effect in the real part of $\mu$. This
shows the inability of creating zonal structures of the CHM model.
Direct numerical simulations starting from drift waves will be shown
in Section \ref{sec:Direct-numerical-simulations} for an explicit
illustration of the difference between the two models.

\begin{figure}
\subfloat{\includegraphics[scale=0.32]{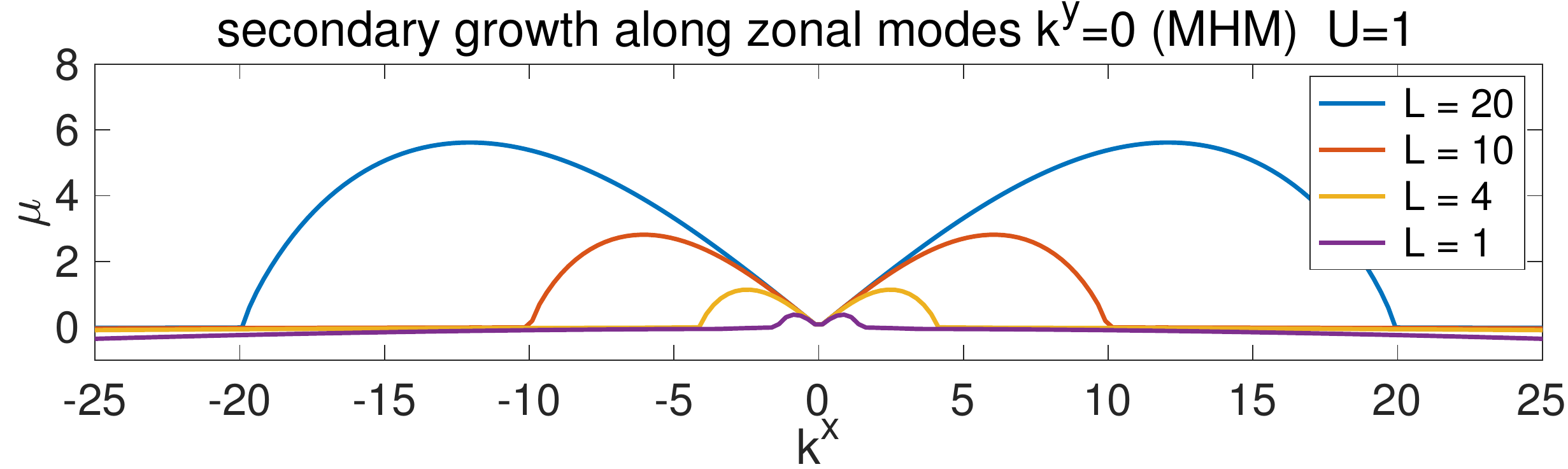}\includegraphics[scale=0.32]{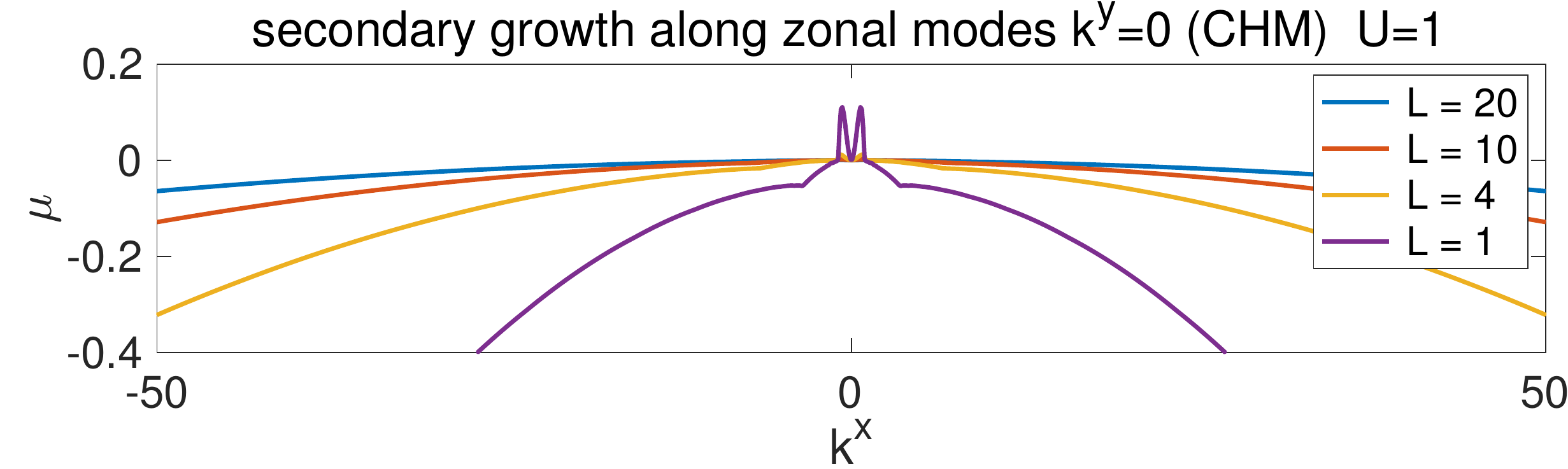}}

\vspace{-1em}

\subfloat{\includegraphics[scale=0.32]{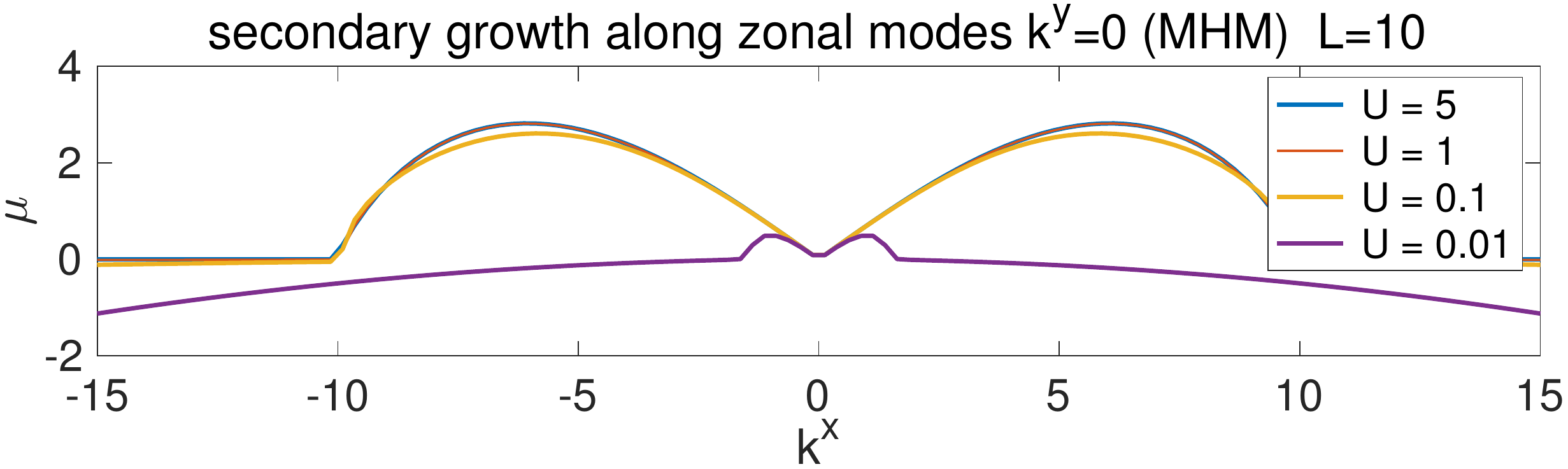}\includegraphics[scale=0.32]{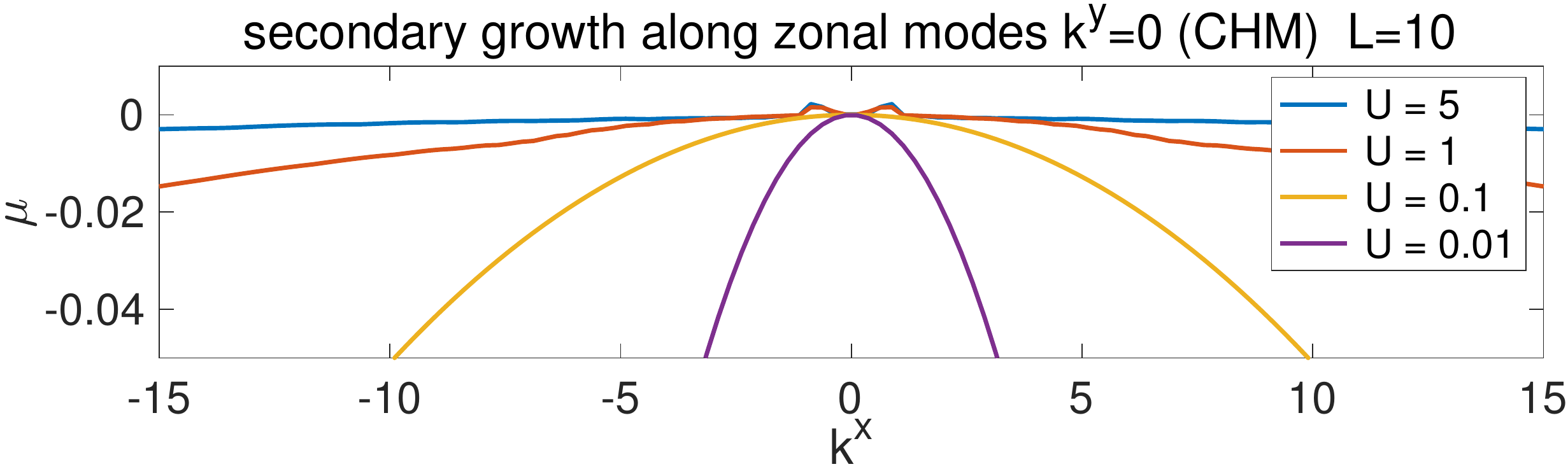}}

\caption{Maximum growth rate from secondary stability analysis along the zonal
mode direction with $k^{y}=0$ for the MHM and CHM models according
to the drift wave base flow $\mathbf{p}=\left(0,1\right)$. Results
with different characteristic length scale $L$ and background flow
strength $U$ are compared.\label{fig:growth_zonal}}

\end{figure}

For a more detailed comparison about the model instability changing
with characteristic scales, the maximum growth with different parameter
values $L$ and $U$ are compared. As shown in the three rows of Figure
\ref{fig:sec_instability_drift}, larger characteristic scale $L=20$
in the drift waves induces larger number of unstable zonal modes in
higher wavenumbers and stronger growth rate. In comparison, the CHM
model results have little change with only negative eigenvalues for
stability. Figure \ref{fig:growth_zonal} shows the maximum growth
rate $\mu$ depending on different model scales $U$ and $L$ from secondary stability analysis along the zonal modes $k^{y}=0$. More clearly, the MHM model gets unstable zonal modes from the drift
wave state, while the CHM model has no instability at all along the
zonal direction.  For the
MHM model, the largest growth gets saturated at large values of flow
amplitude $U$; and with decreasing amplitudes of $U$, the growth
rate drops slowly and will finally vanish at the extremely small value
$U<0.01$ where the effective dissipation $D^{\prime}=\frac{D}{UL}$
becomes strong. On the other hand for the CHM model, at most weak
instability is induced for large value of $U$ around the largest scales.
The positive growth rate quickly vanishes as the flow strength $U$
decreases in value. This is also reflected in the contour plots in
Figure \ref{fig:sec_instability_drift}.

\begin{figure}
\includegraphics[scale=0.32]{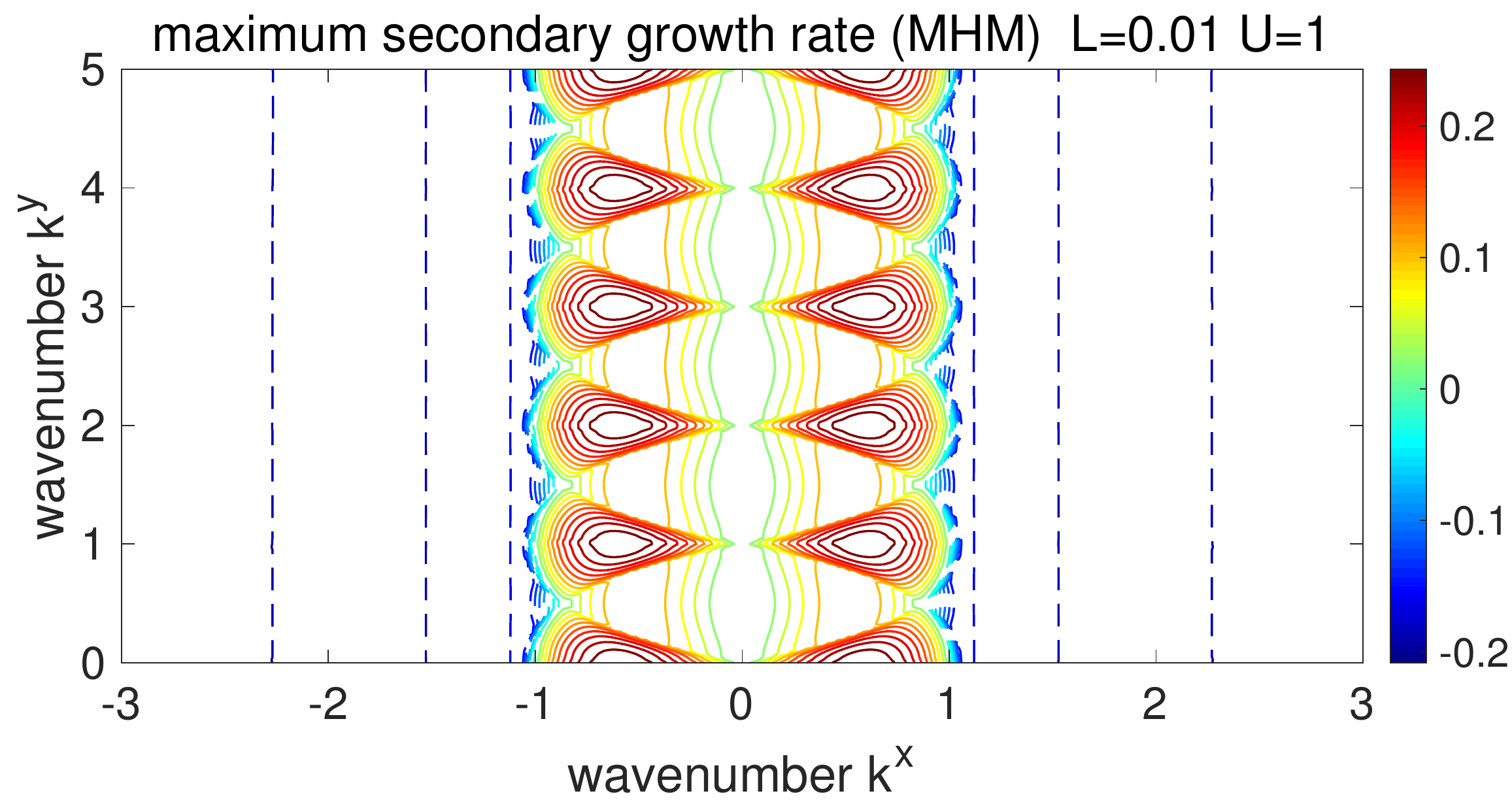}\includegraphics[scale=0.32]{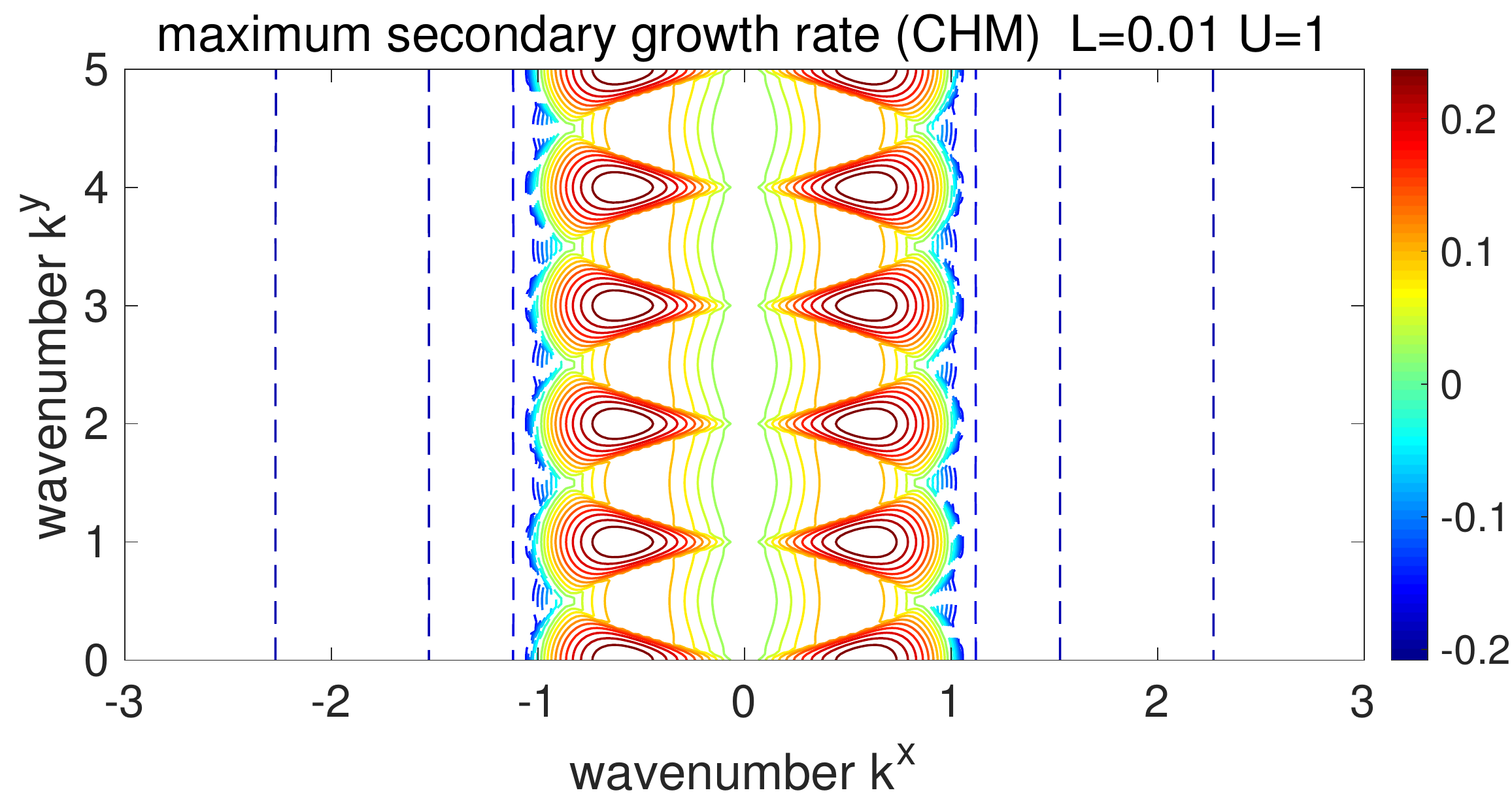}

\caption{Real part of the secondary growth rate $\mu$ according to background
base drift wave flow $\mathbf{p}=\left(0,1\right)$ with small characteristic
length $L=0.01$ and $U=1$. Solid lines are for positive growth rates
and dashed lines are for negative damping rates. Results for the MHM
model (left) and for the CHM model (right) are compared. The other
parameters used are $\kappa=0.5,D=5\times10^{-4}$.\label{fig:sec_instability-baro}}
\end{figure}

Finally, we test the small characteristic length $L=0.01$,
that is, the background drift wave state is in a very small scale.
At this small length scale limit, both MHM and CHM models converge
to the barotropic model with infinite (or large) deformation frequency,
$q\rightarrow\nabla^{2}\varphi$ \cite{majda2003introduction,majda2016introduction}.
Figure \ref{fig:sec_instability-baro} shows the maximum secondary
growth rate from both the MHM and CHM model results. As expected,
the growth rates from the two models perform similarly at this limit
where the balanced flux correction for $\tilde{\varphi}$ becomes
negligible. Also the maximum growth of the perturbations takes place
at the largest scales $k<1$ compared with the small characteristic
scale $L$ in the background drift wave state. Besides, the maximum
secondary growth stays in small amplitude with just weak instability
among all the wavenumbers. This agrees with the results in \cite{lee2003stability}
for barotropic turbulence.

\subsubsection{Stability about zonal flow mode $\mathbf{p}=\left(1,0\right)$}

In this second test case, we consider the secondary instability due
to a background zonal jet state with wave direction $\mathbf{p}=\left(1,0\right)$.
In this case, a positive growth rate along the zonal modes implies
further growth of the fluctuations in zonal states until higher order
nonlinear interactions between modes take over. Usually, if the zonal
jet amplitude keeps growing, it will finally break down due to the
nonlinear interactions and cascade to the smaller scales to get dissipated.
On the other hand, no instability in the zonal directions implies
that the zonal jet structure is maintained stable since perturbations
in zonal modes will not grow and jet structure will persist in time.
This case with a zonal flow base state then characterizes the stability
of the zonal structures, which can be created by the instability
from drift waves through the instability analysis shown in Section \ref{subsec:Instability-about-drift}.

\begin{figure}
\subfloat[maximum eigenvalue]{\includegraphics[scale=0.35]{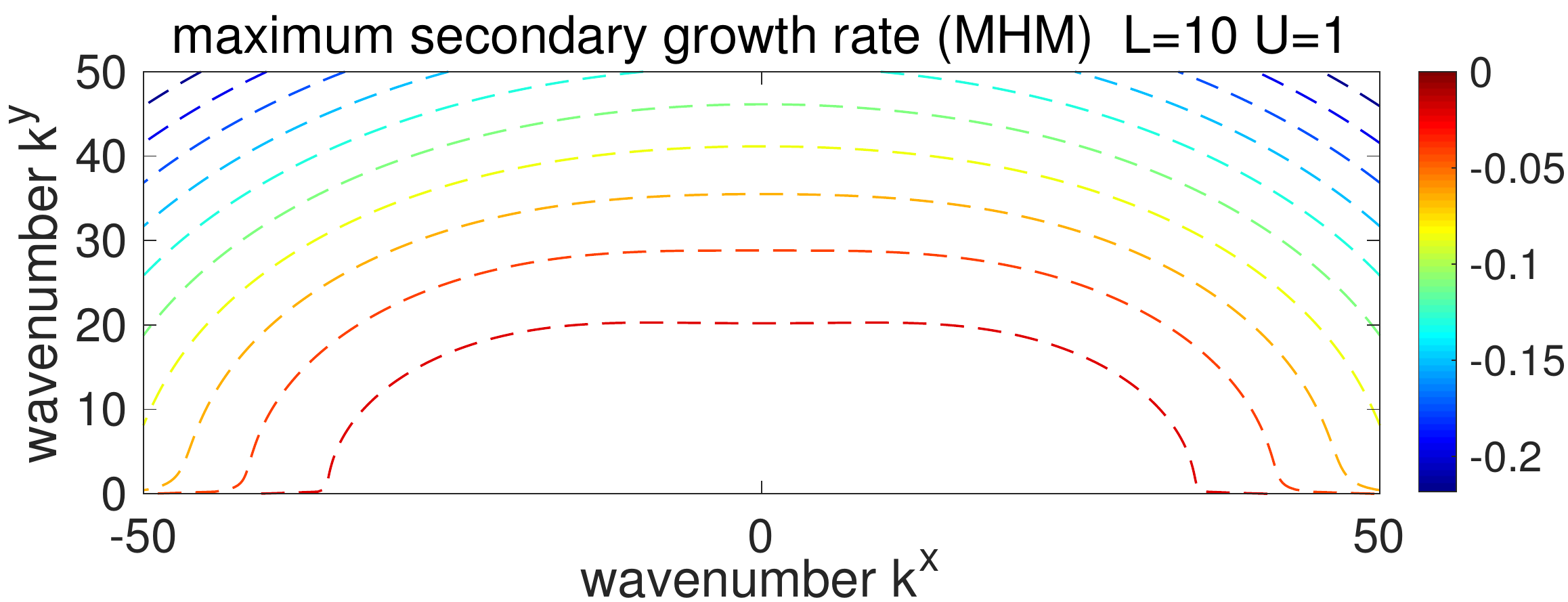}\includegraphics[scale=0.35]{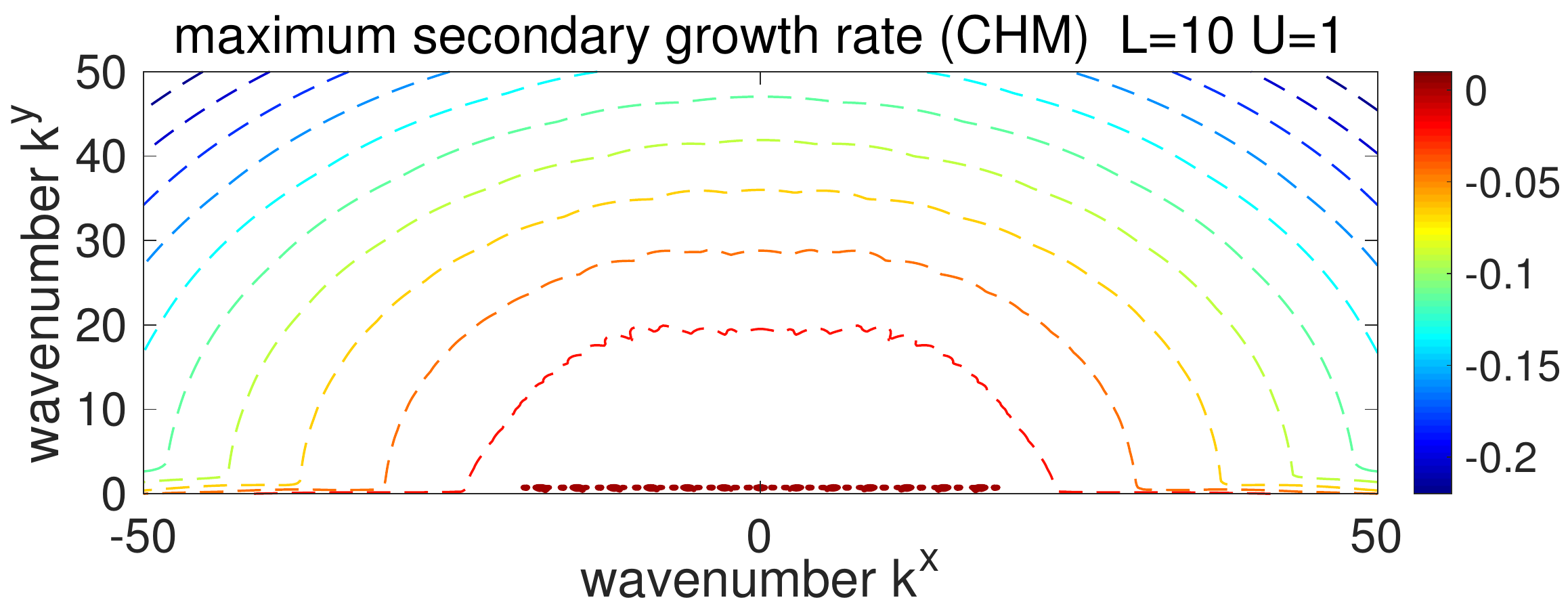}

}

\subfloat[minimum eigenvalue]{\includegraphics[scale=0.35]{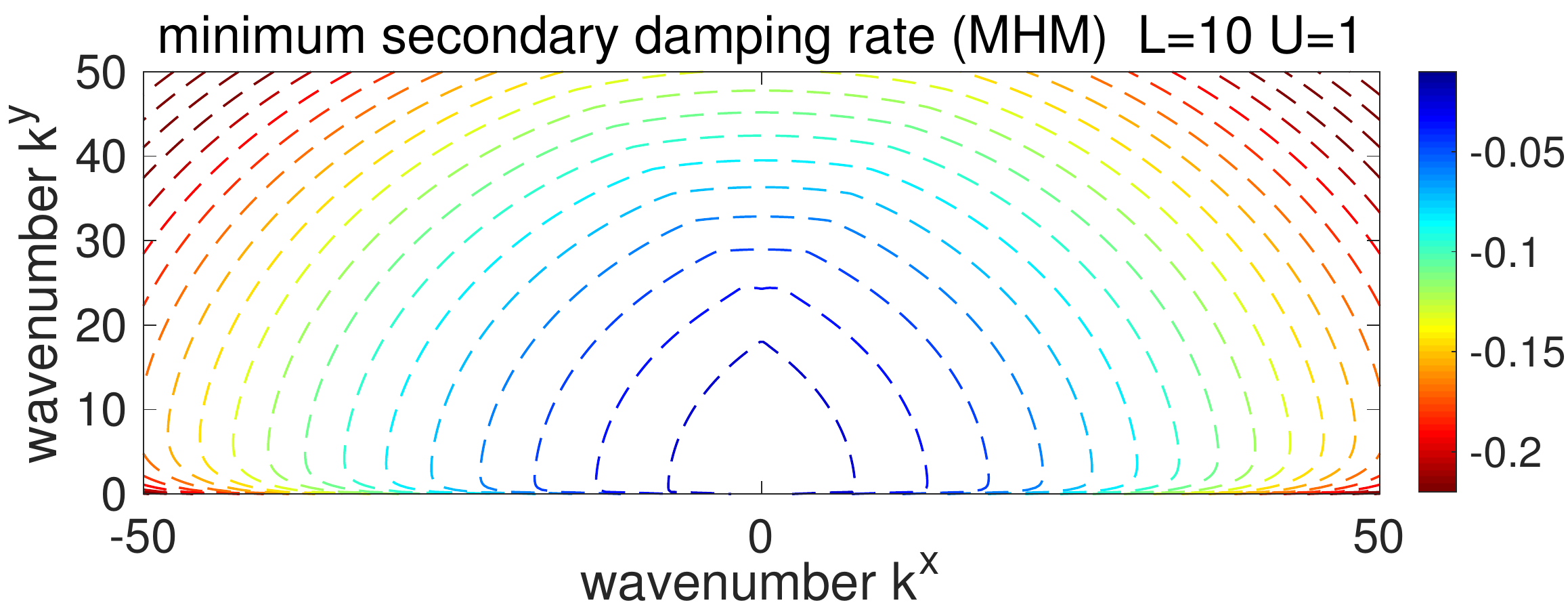}\includegraphics[scale=0.35]{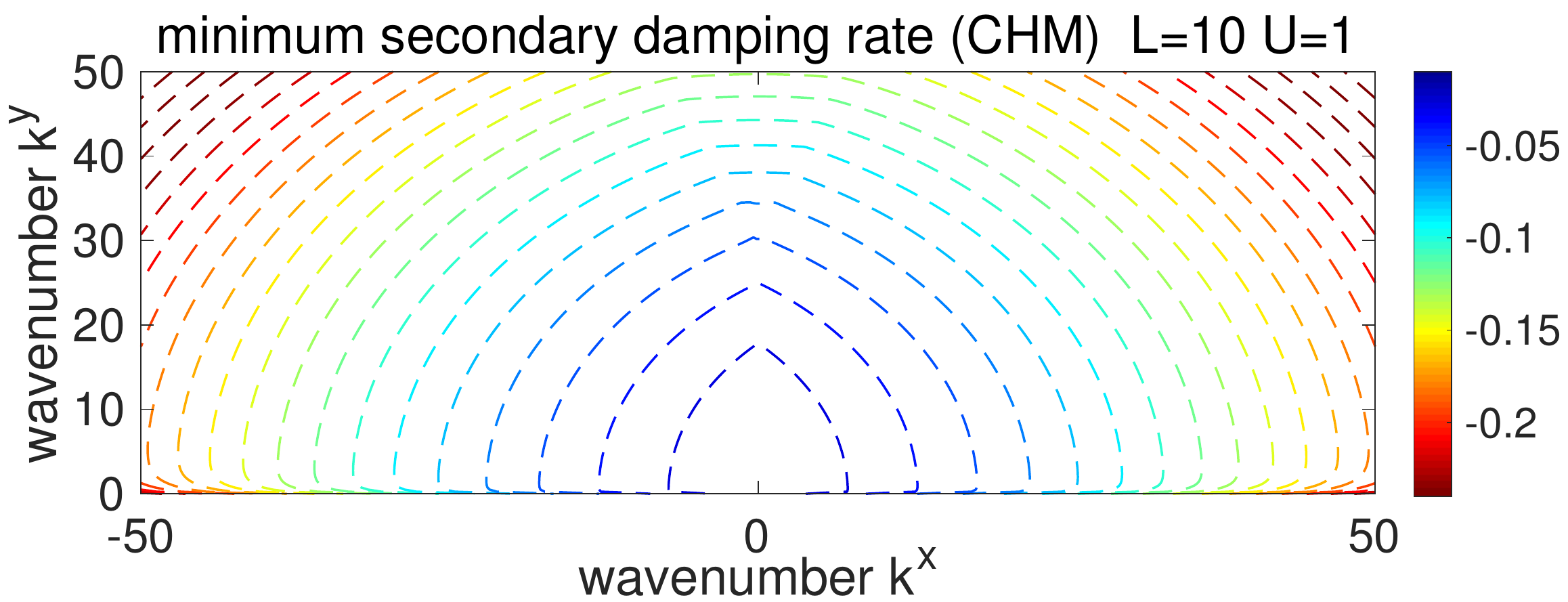}

}

\caption{Maximum growth rates and minimum damping rates for each spectral mode
from largest and smallest eigenvalues in the secondary stability analysis
according to zonal jet base flow $\mathbf{p}=\left(0,1\right)$. All
the eigenvalues are negative in the MHM model shown in dashed lines,
while the CHM model has positive growth rates near the zonal axis
$k^{y}=0$. Results for the MHM model (left) and for the CHM model
(right) are compared. The parameters used are $L=10,U=1$, and $\kappa=0.5,D=5\times10^{-4}$.\label{fig:sec_instability_jet}}
\end{figure}

In Figure \ref{fig:sec_instability_jet}, the maximum and minimum
eigenvalues from the secondary stability analysis according to zonal
jet base flow are plotted. It can be observed that there exists no
positive growth rate in $\mu$ for instability at all throughout the spectral regime due to zonal flows in the MHM model. This supports
the intuition described above that the induced zonal jets can be maintained stable 
in time in response to additional wave perturbations. In the CHM model
case, the growth rate is in a similar shape but gets small positive
growth near the zonal direction $k^{y}=0$. This instability in the
zonal modes may imply the less stable zonal jets and the possible
break down of the zonal structures due to perturbations in the CHW
model. In additional, we also compare the minimum eigenvalue for the
strongest damping rate in each mode from the stability analysis. The
zonal modes get stronger damping at smaller scales. This again confirms
the stability of the zonal jets, so that small perturbations in the
zonal direction will be quickly damped down from the stabilizing effect
in the background base mode.

\

Combining the conclusions of instability in the drift wave state and
stability in the zonal modes, we can draw a complete picture about
the energy mechanism due to the nonlinear transfer of energy in the
transient state. By adopting the Hasegawa-Mima model (\ref{eq:plasma_onelayer})
without drift wave linear instability, we start with a drift wave
state that could be generated from the drift wave turbulence in the higher level 
Hasegawa-Wakatani model (\ref{plasma_balance}). The secondary instability
in the drift wave base mode induces strong growth particularly along
the zonal mode direction, which infers the strong transport of energy
from the drift wave modes to the zonal states. After the formation
of the zonal structure, the strong negative damping with no instability
about the zonal jet background state shows the persistence of the
zonal structure to perturbations. The zonal jets will emergence even
without the help of the selective decay in dissipating small scale fluctuations
described in \cite{qi2018selective}.

\section{Direct numerical simulations to confirm the generation of zonal jets\label{sec:Direct-numerical-simulations}}

In this final part, we use direct numerical simulations of the MHM and
CHM models (\ref{eq:HM_nondim}) to confirm the theory from the
secondary instability induced by the background base mode discussed in
the previous section. A pseudo-spectral code with a 3/2-rule for de-aliasing
the nonlinear term \cite{qi2018selective,qi2016low} is applied on the square domain with side length
$L_{D}=40$ and resolution $N=256$. For the
time integration, a 4th-order Runge-Kutta scheme is adopted. Small
time integration step $\Delta t=1\times10^{-4}$ is taken in all the
simulations to ensure the conservation properties especially for the
non-dissipative case. The same model parameter values are taken as
in Section \ref{subsec:Secondary-instability-num} for instability
analysis.

To check the energy transfer mechanism from drift waves, the initial
state of the simulations is set as a pure drift wave adding homogenous
perturbations
\begin{equation}
\varphi_{0}=\frac{L_{D}}{s}\cos\left(\frac{2\pi s}{L_{D}}y\right)+\epsilon\sum_{\left|\mathbf{k}\right|\leq\Lambda}k^{-2}\hat{\xi}_{\mathbf{k}}e^{i\mathbf{k\cdot x}},\label{eq:init}
\end{equation}
with $\hat{\xi}_{\mathbf{k}}\sim\mathcal{N}\left(0,1\right)$ sampled
independently from the standard normal distribution. In practice,
we add the perturbations up to wavenumber $\Lambda=5$ and the perturbation
amplitude is set in a small value $\epsilon=0.01$. The parameter $s$ determines
the scale of the background drift wave. We test two values with
$s=2$ representing drift waves of two wavelengths and $s=10$ representing
drift waves of wavenumber 10 (see the electrostatic potential $\varphi_{0}$
shown in Figure \ref{fig:Snapshots} for these two initial states).
Besides, we consider two different situations without and with the
dissipation operator $D\Delta q$ in the model.

\subsection{Time evolution of energy and enstrophy in full and zonal modes}

In the first numerical simulations, we introduce no dissipation effect
$D=0$ in the model. Thus the conservation
of total kinetic energy $E$ and potential enstrophy $W$ defined
in (\ref{eq:energy_plasma}) should be guaranteed. We run the model
in this way so that the selective decay effect \cite{qi2018selective}
will be excluded. Therefore if zonal structures are generated in the final steady state from
the simple drift wave initial state (\ref{eq:init}), the mechanism
can only be the secondary interactions between the initial background
state $\varphi_{0}$ and the perturbed small modes. 

\begin{figure}
\subfloat{\includegraphics[scale=0.35]{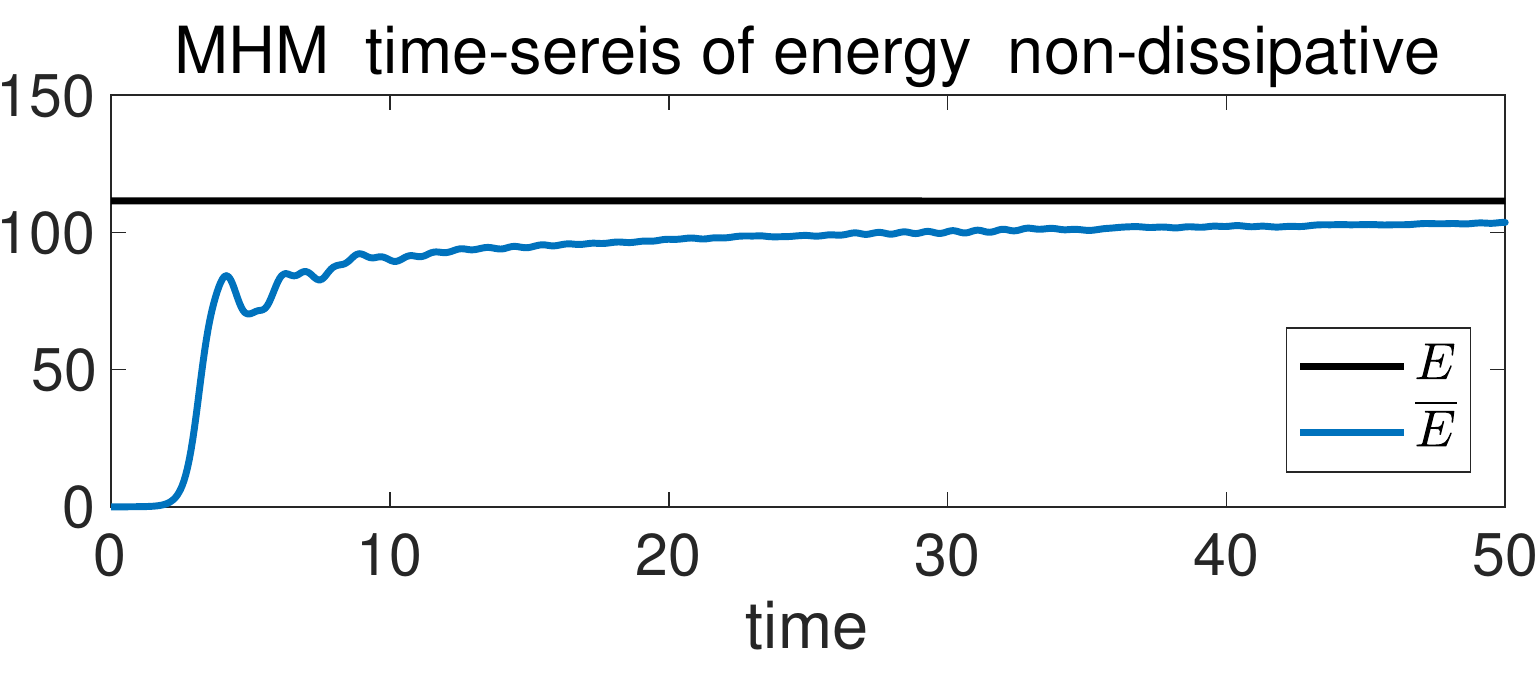}\includegraphics[scale=0.35]{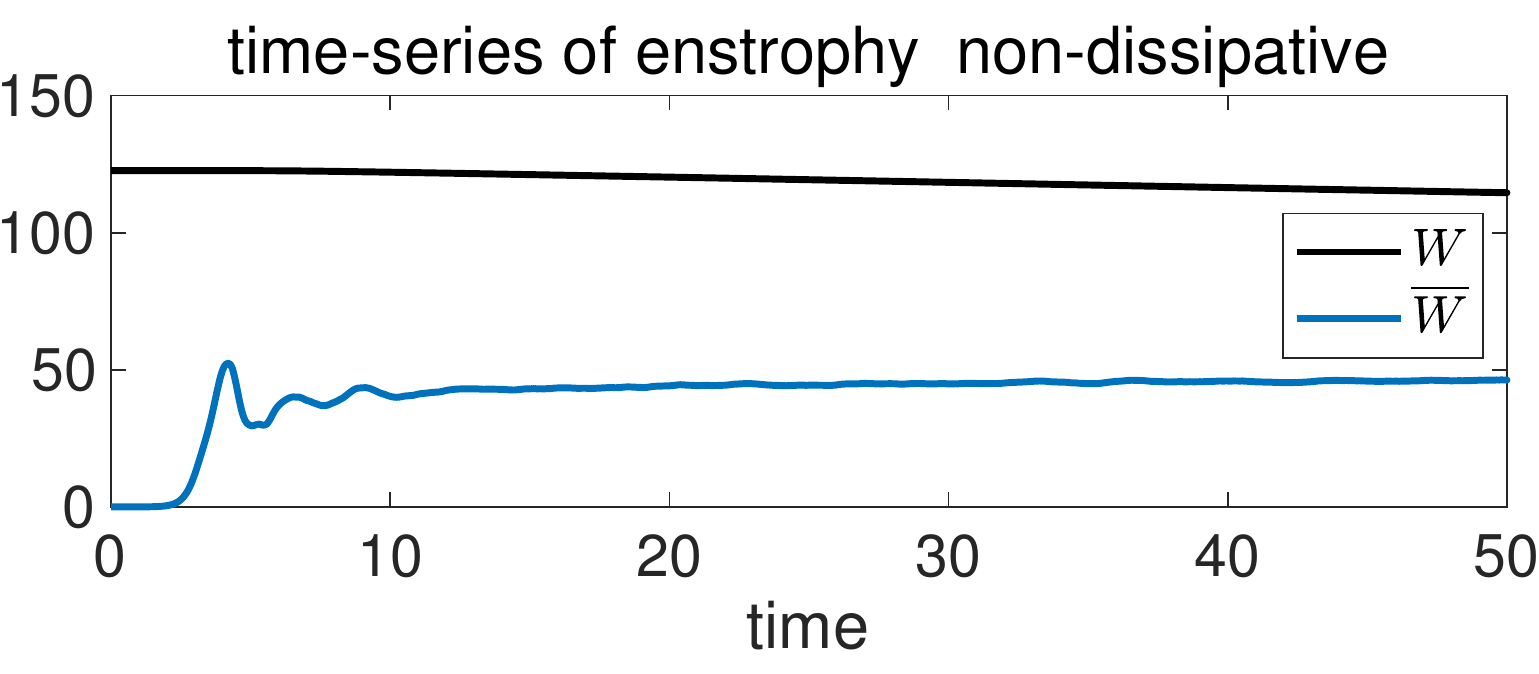}\includegraphics[scale=0.35]{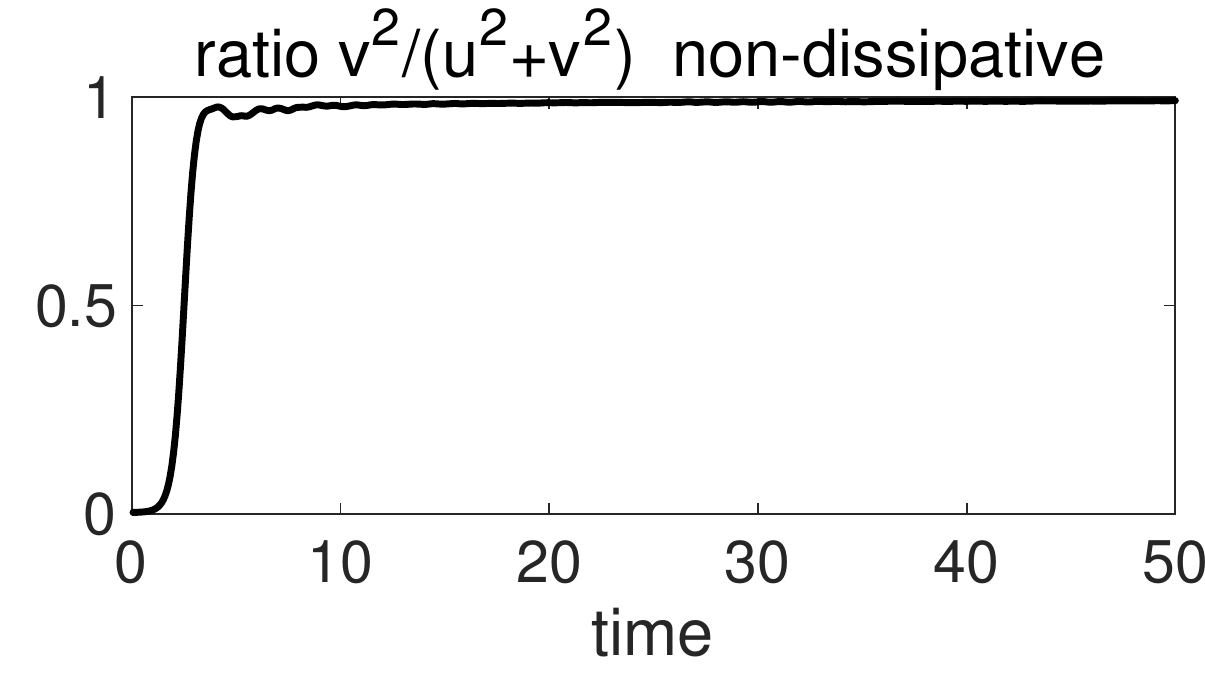}}

\vspace{-1.8em}

\subfloat{\includegraphics[scale=0.35]{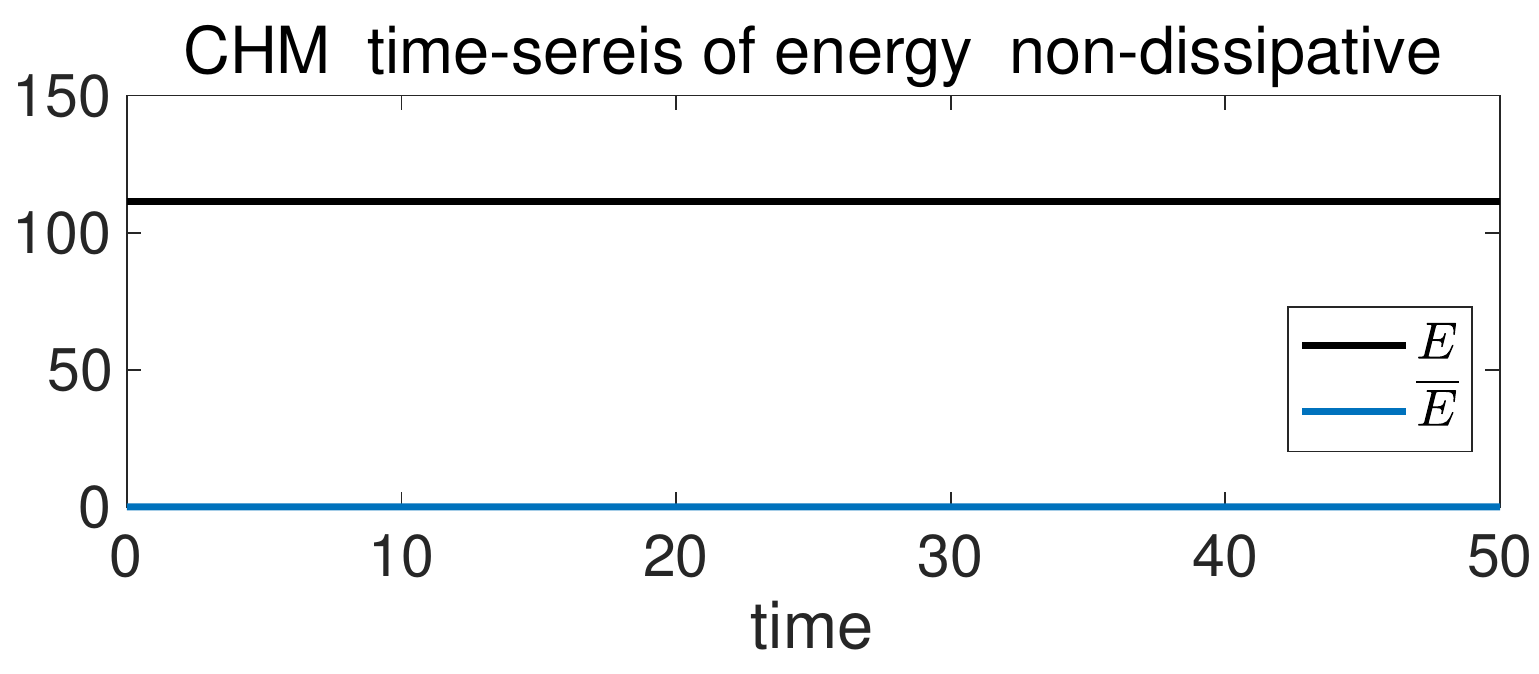}\includegraphics[scale=0.35]{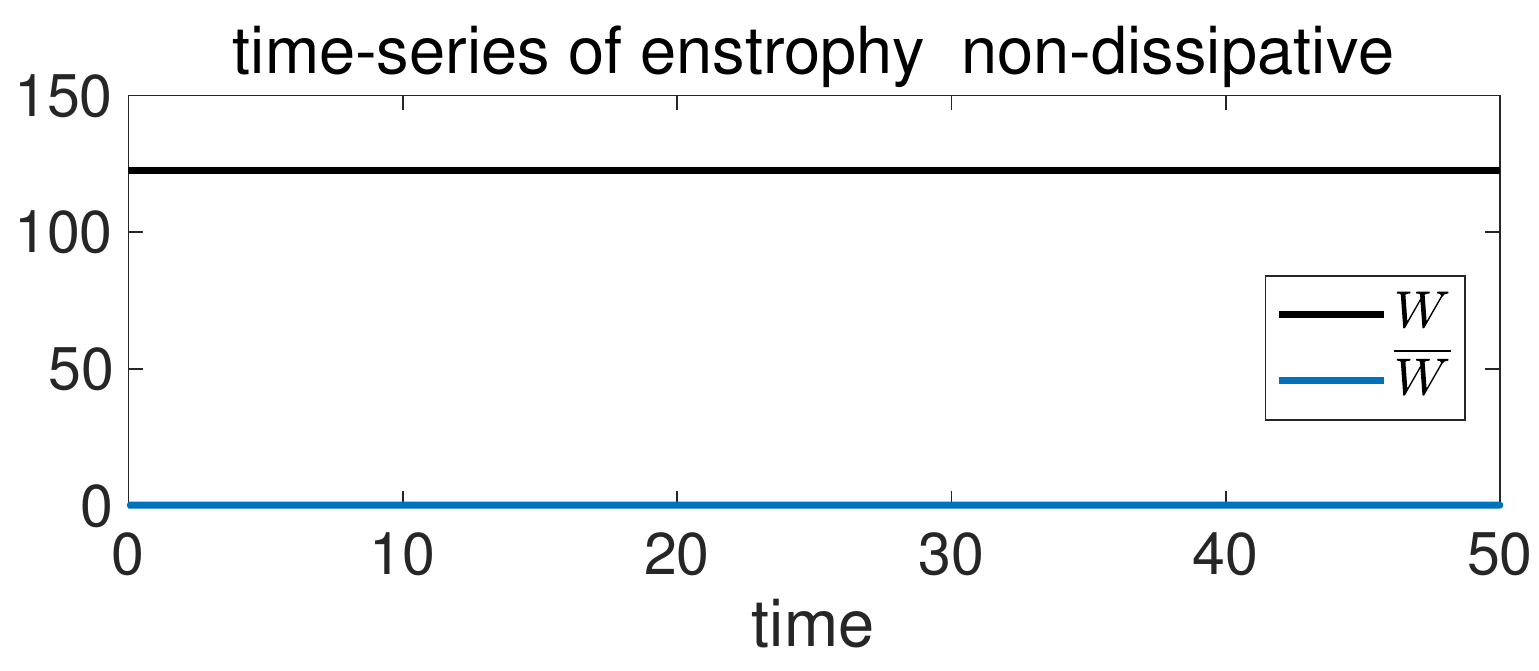}\includegraphics[scale=0.35]{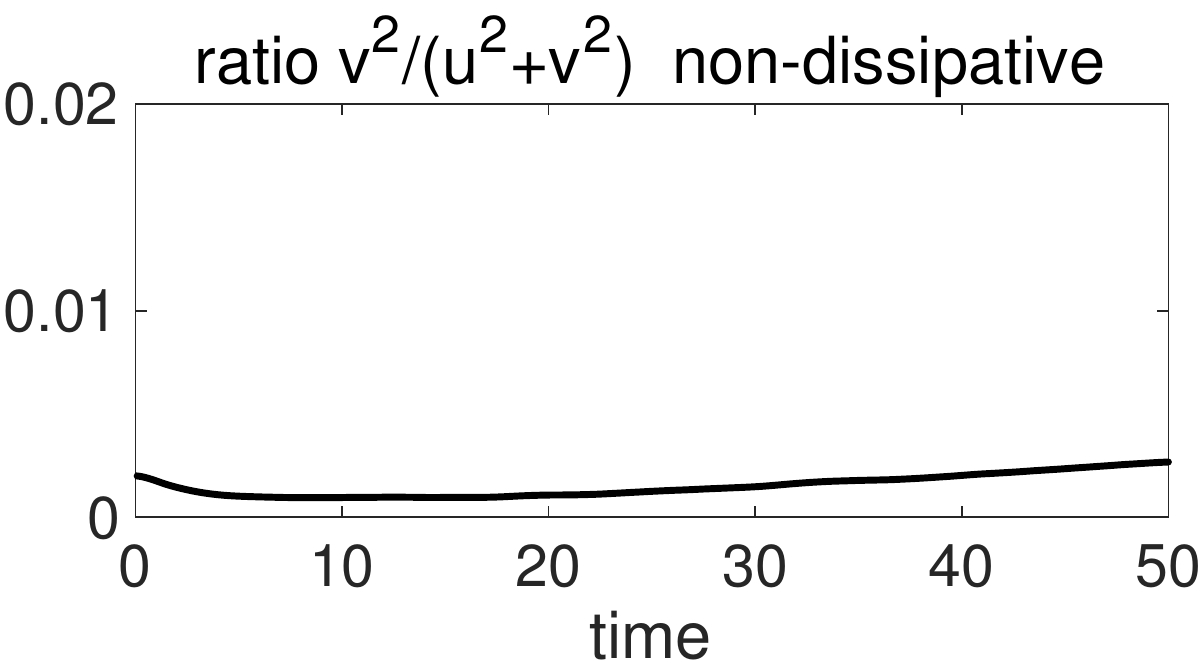}}

\vspace{-1.8em}

\subfloat{\includegraphics[scale=0.35]{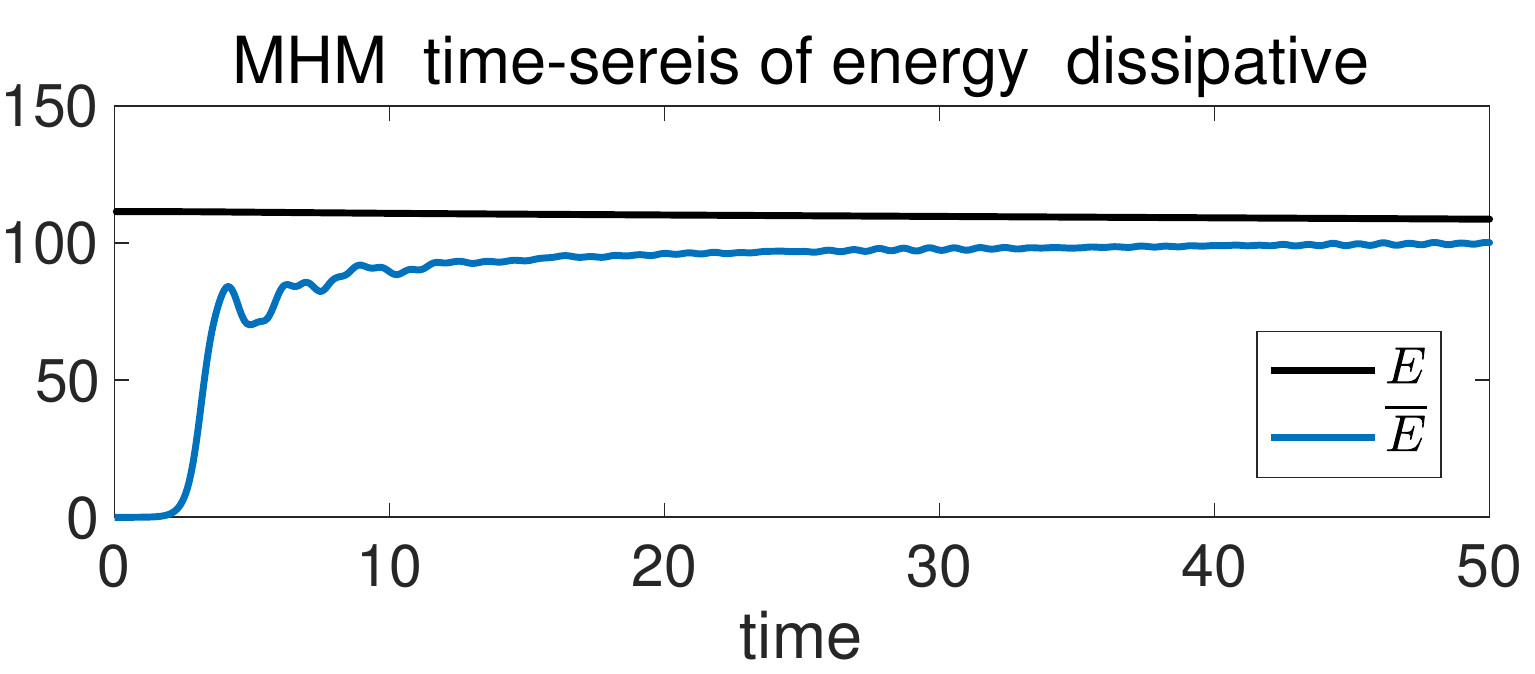}\includegraphics[scale=0.35]{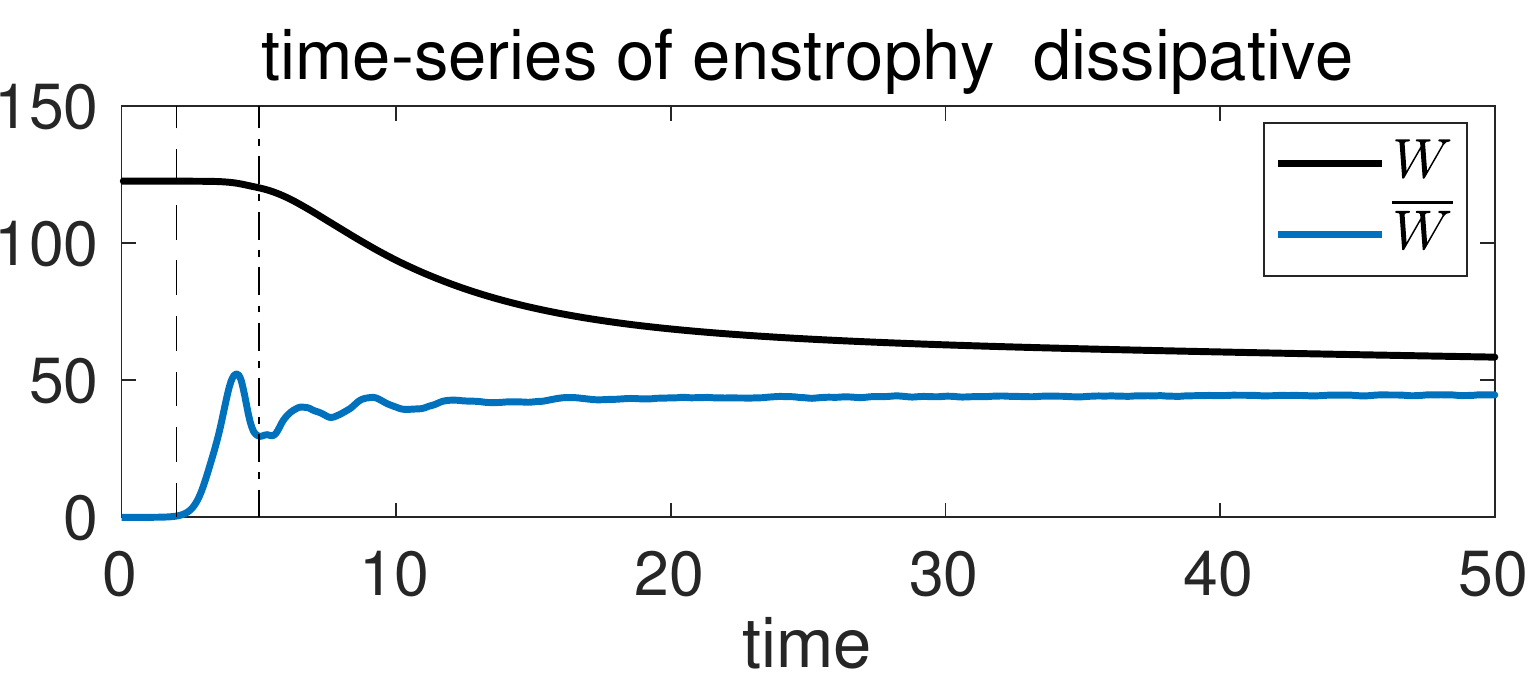}\includegraphics[scale=0.35]{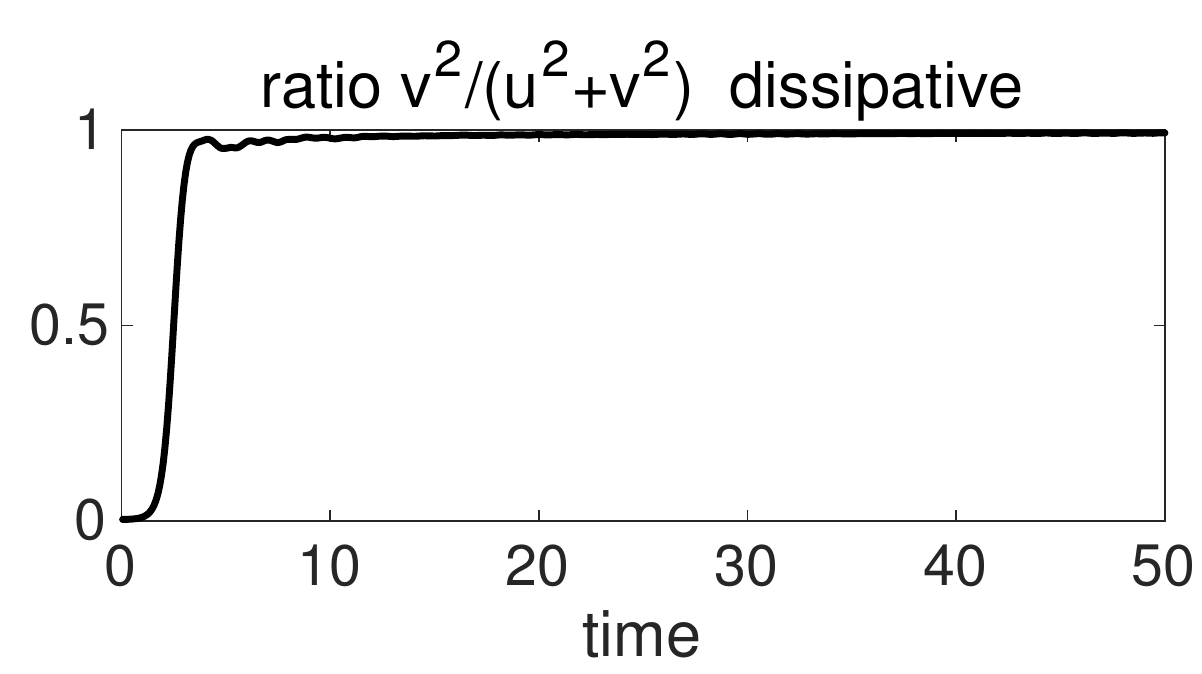}}

\vspace{-1.8em}

\subfloat{\includegraphics[scale=0.35]{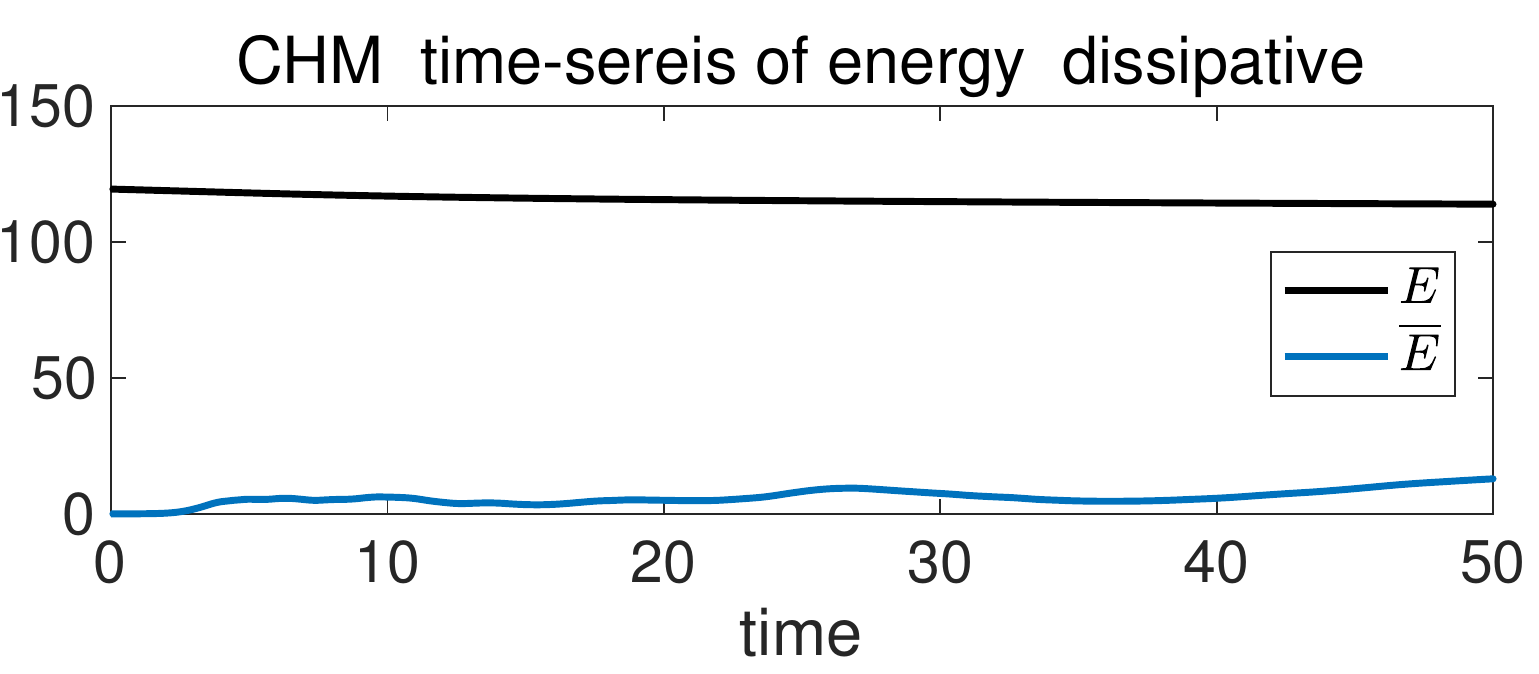}\includegraphics[scale=0.35]{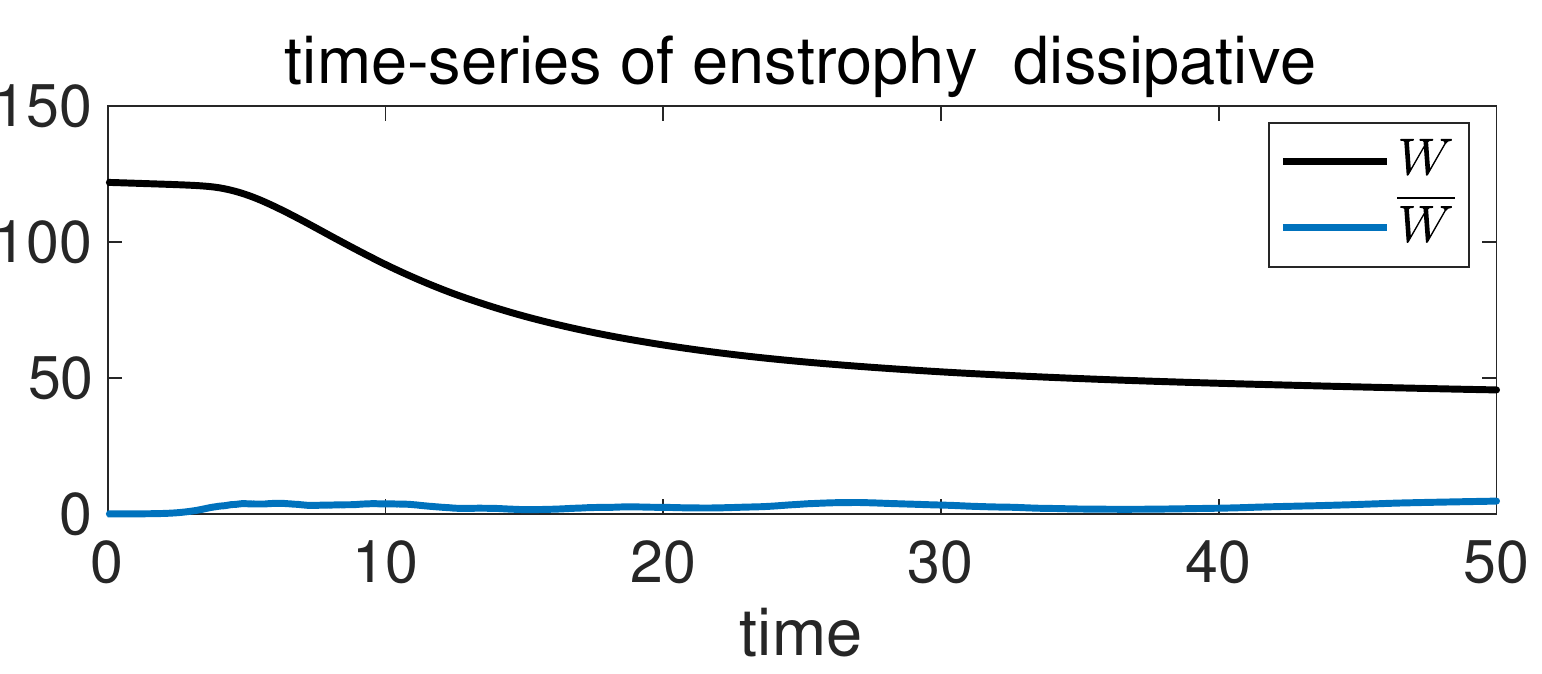}\includegraphics[scale=0.35]{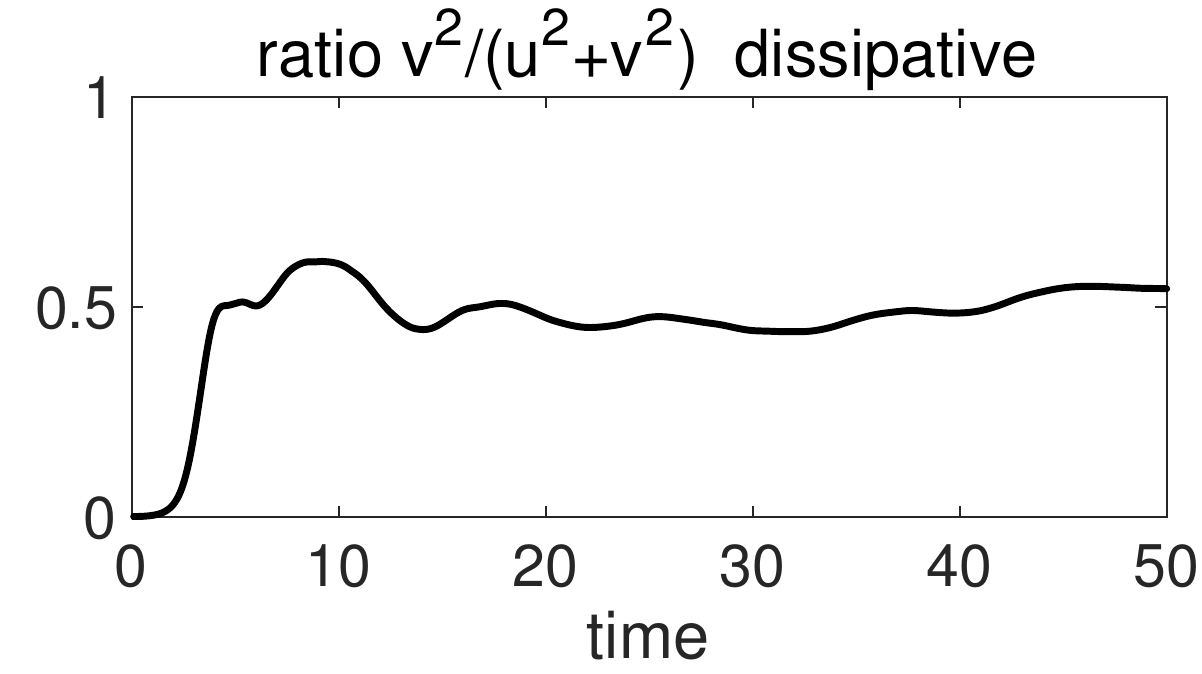}}

\caption{Time-series of the total energy $E$ and total enstrophy $W$ as well
as the energy and enstrophy contained in the zonal state $\left(\overline{\varphi},\overline{q}\right)$
from the MHM and CHM model simulations. The first two rows show the
results for MHM and CHM models without dissipation effect $D=0$,
and the last two rows are the results for both models including weak
dissipation $D=5\times10^{-4}$. The last column plots the ratio of
energy in the zonal state $v^{2}/\left(u^{2}+v^{2}\right)$.\label{fig:Time-series}}

\end{figure}

We need to confirm in the first place that the numerical dissipations
have little effect in changing the model energy and enstrophy and
offer no contribution in the final state of the model. For checking
the conservation in the model simulations,
the first two rows of Figure \ref{fig:Time-series} plot the time-series
of the total energy $E$ and enstrophy $W$ from the direct model
simulations. In both the MHM and CHM model results, the total energy
and enstrophy are conserved in time with at most small decrease in
the enstrophy due to the numerical dissipation strongest at the smallest
scales. Further we plot the energy and enstrophy only contained in
the zonal state $\overline{\varphi}$ and $\overline{q}$. The ratio
of energy in zonal velocity $v^{2}/\left(u^{2}+v^{2}\right)$ is used
to characterize the flow structure, which reaches 1 when the purely
zonal flow is reached. In the MHM model, the zonal energy and enstrophy
start near zero in the initial time due to the initial setup, then
the secondary instability takes over and the zonal energy and enstrophy
jump to a large non-zero value through the nonlinear interaction.
This infers the strong instability from drift wave modes and stability
in the zonal jet states. In contrast, the zonal energy and enstrophy
in the CHM stay in small values near zero throughout the time evolution.
Then no zonal state is excited in the CHM model from the nonlinear
effect.

In the last two rows of Figure \ref{fig:Time-series}, the time-series
of energy and enstrophy as well as the zonal energy ratio with a small
dissipation $D=5\times10^{-4}$ are plotted. In comparison with the
non-dissipative case before, the energy and enstrophy are no longer
conserved. Especially, the enstrophy decays in a faster rate than
the energy, implying that the dissipation is stronger on damping the smaller
scale modes. Still, the MHM case induces strong zonal structures as
the system approaches the final state. The CHM model still lacks the skill
in generating zonal jets, while a pure single drift wave mode is converged
consistent with the selective decay principle \cite{qi2018selective,majda2000selective}. 

Most importantly, observe that in the time-series of enstrophy for
the MHM model, the zonal enstrophy starts to rise at $t=2$ (marked
by dashed line in the figure) while the total enstrophy begins to
drop at a later time at $t=5$ (marked by dotted-dashed line). This illustrates
the competition between the secondary instability and selective decay:
i) during the starting time $t<2$, the initial state maintains with no linear
instability; ii) between the time $2<t<5$, the secondary instability
comes into effect to generate a strong zonal structure while the dissipation
has no obvious effect on the smaller scale modes; iii) finally after time
$t>5$, the selective decay becomes dominant and strongly dissipates
the smaller scale fluctuations while maintains the created zonal jet.

\subsection{Convergence to the final steady state without dissipation}

\begin{figure}
\subfloat[initial drift wave state with wavenumber $s=2$]{\includegraphics[scale=0.26]{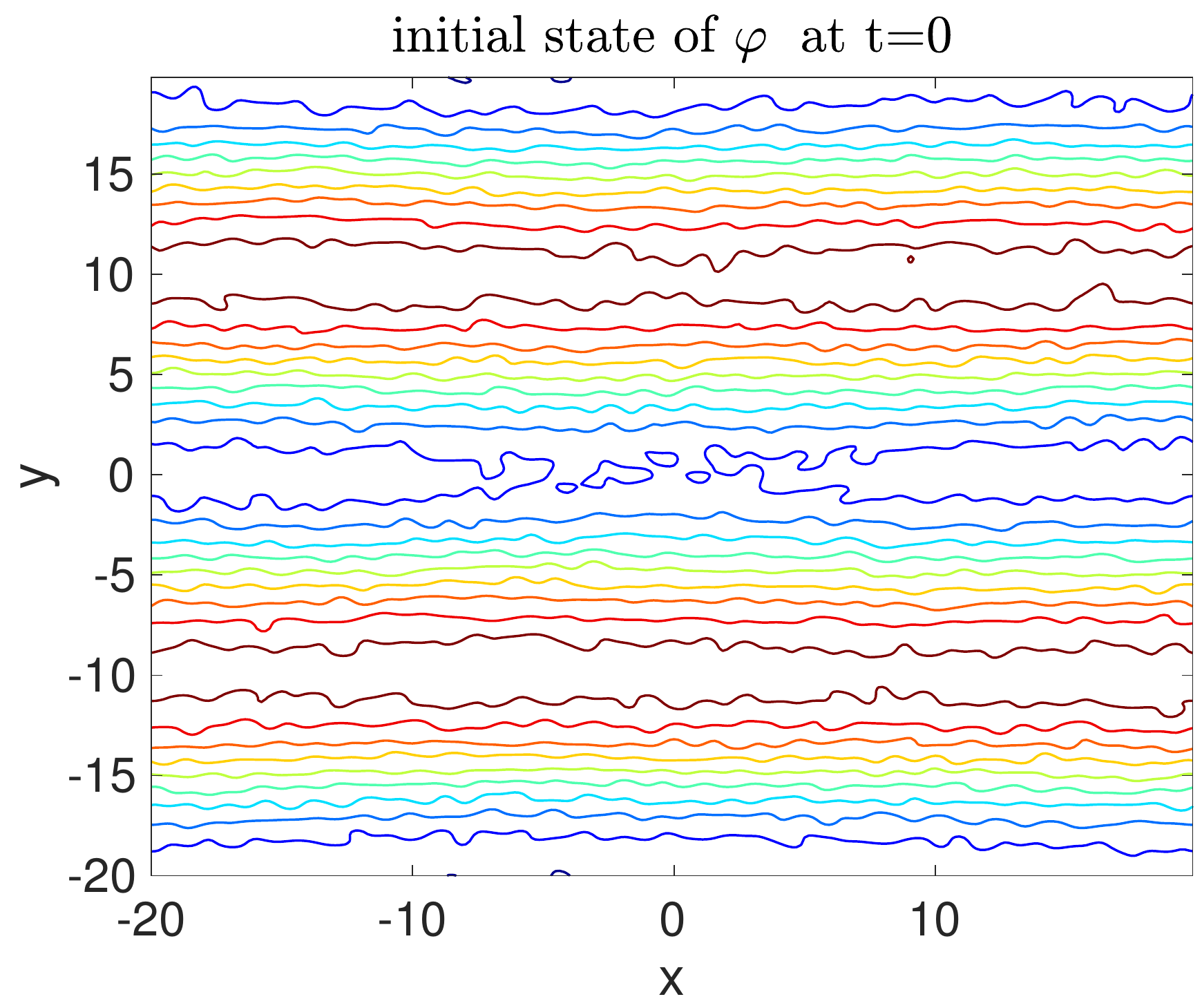}\includegraphics[scale=0.26]{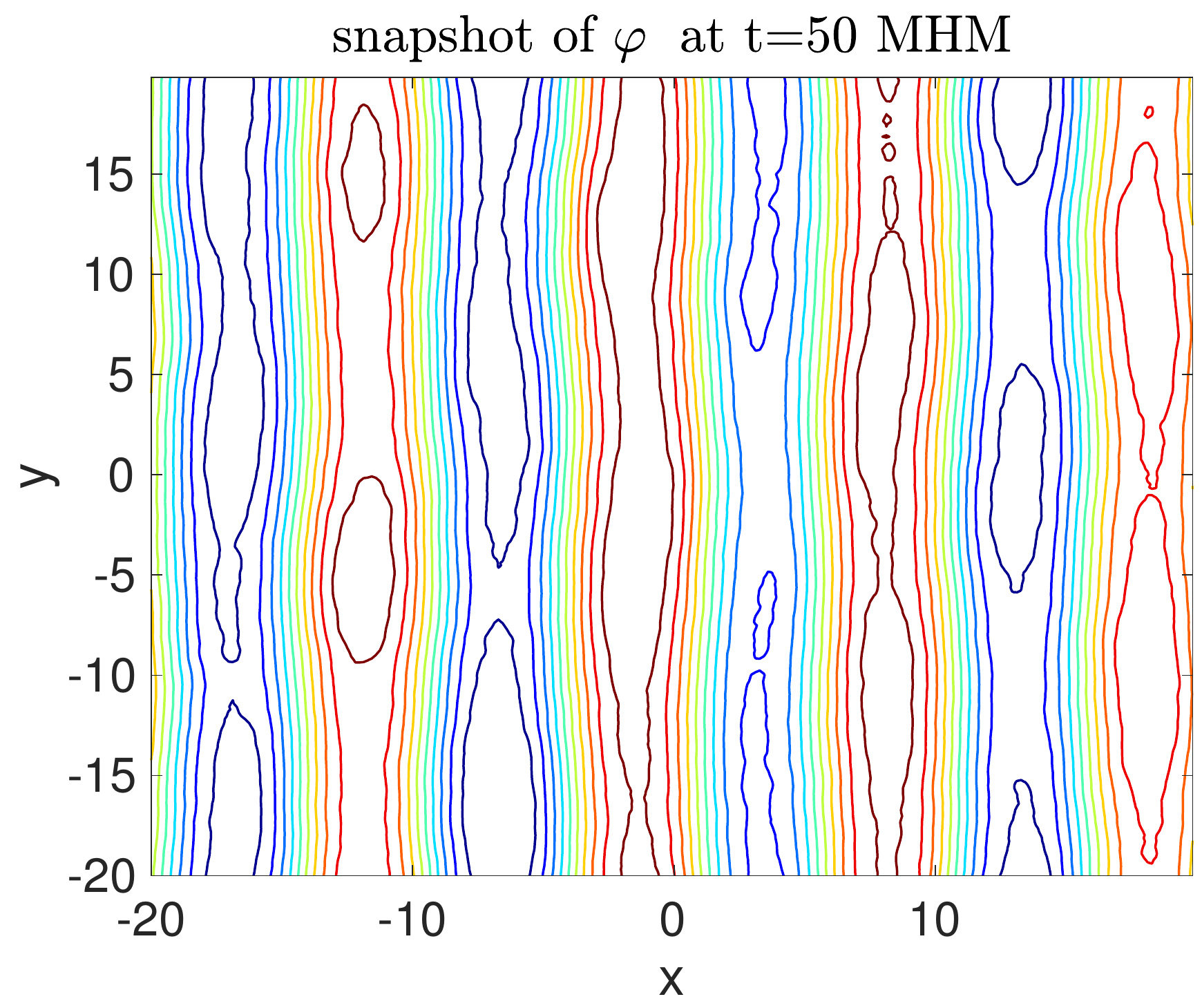}\includegraphics[scale=0.26]{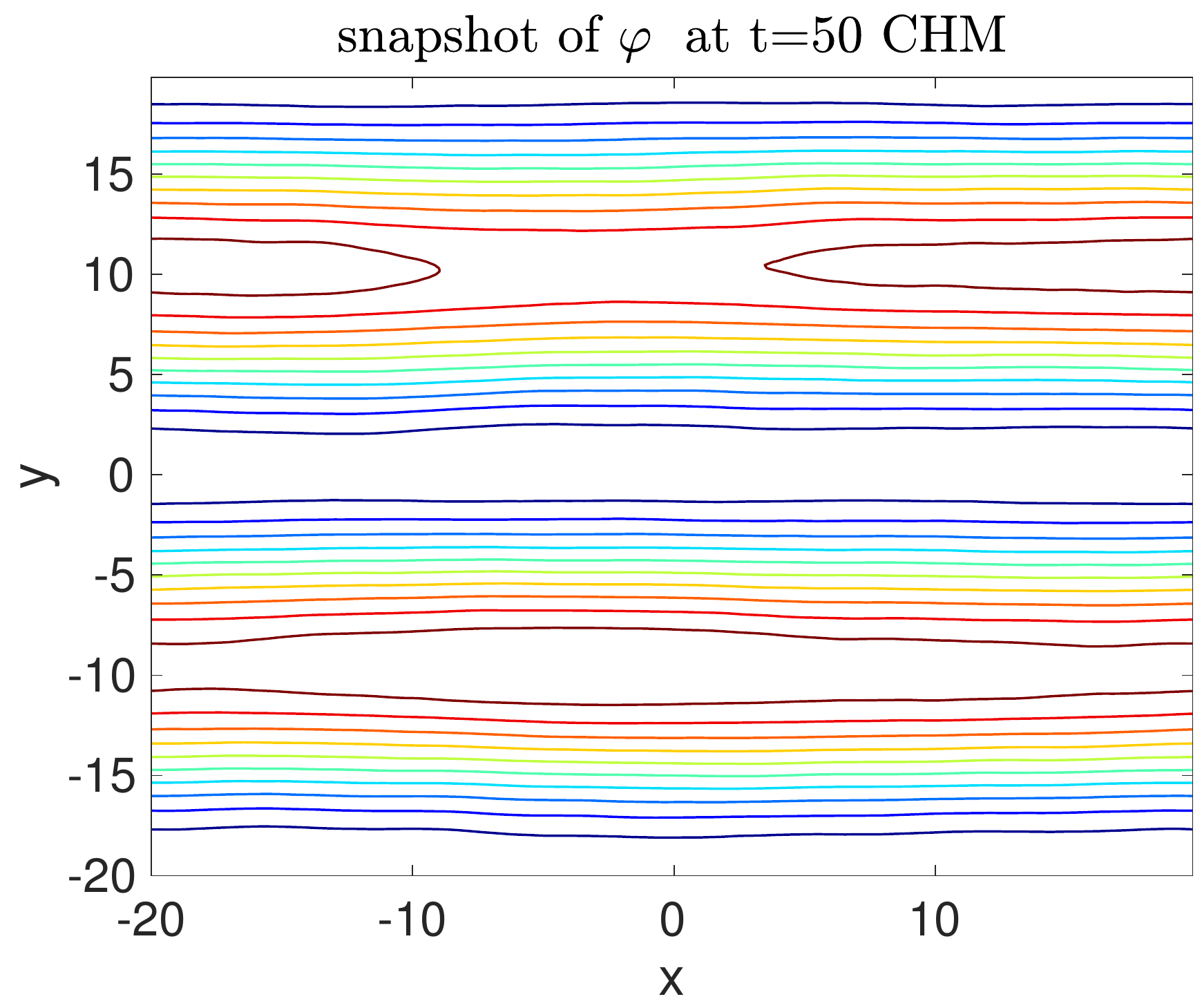}

}

\subfloat[initial drift wave state with wavenumber $s=10$]{\includegraphics[scale=0.26]{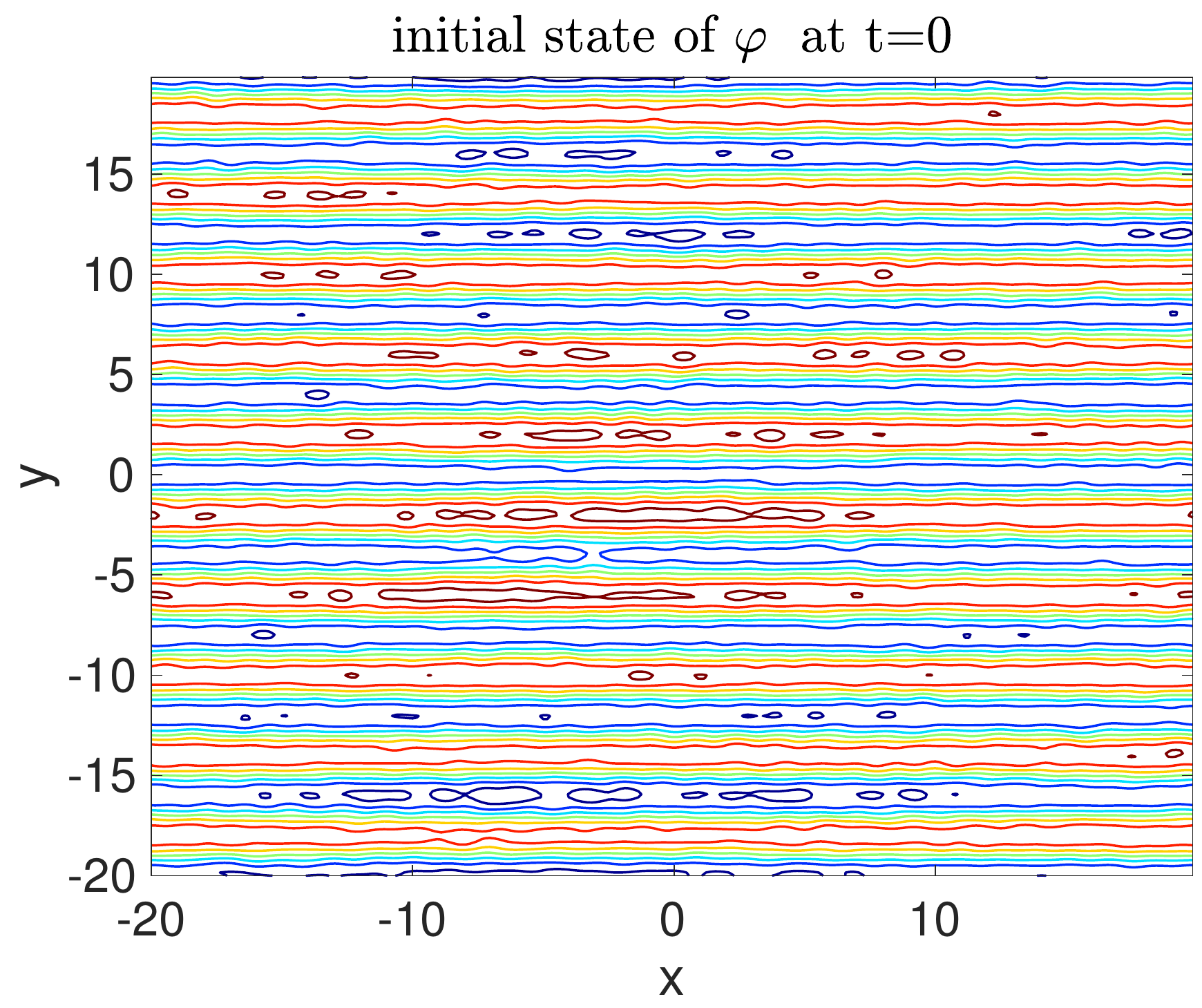}\includegraphics[scale=0.26]{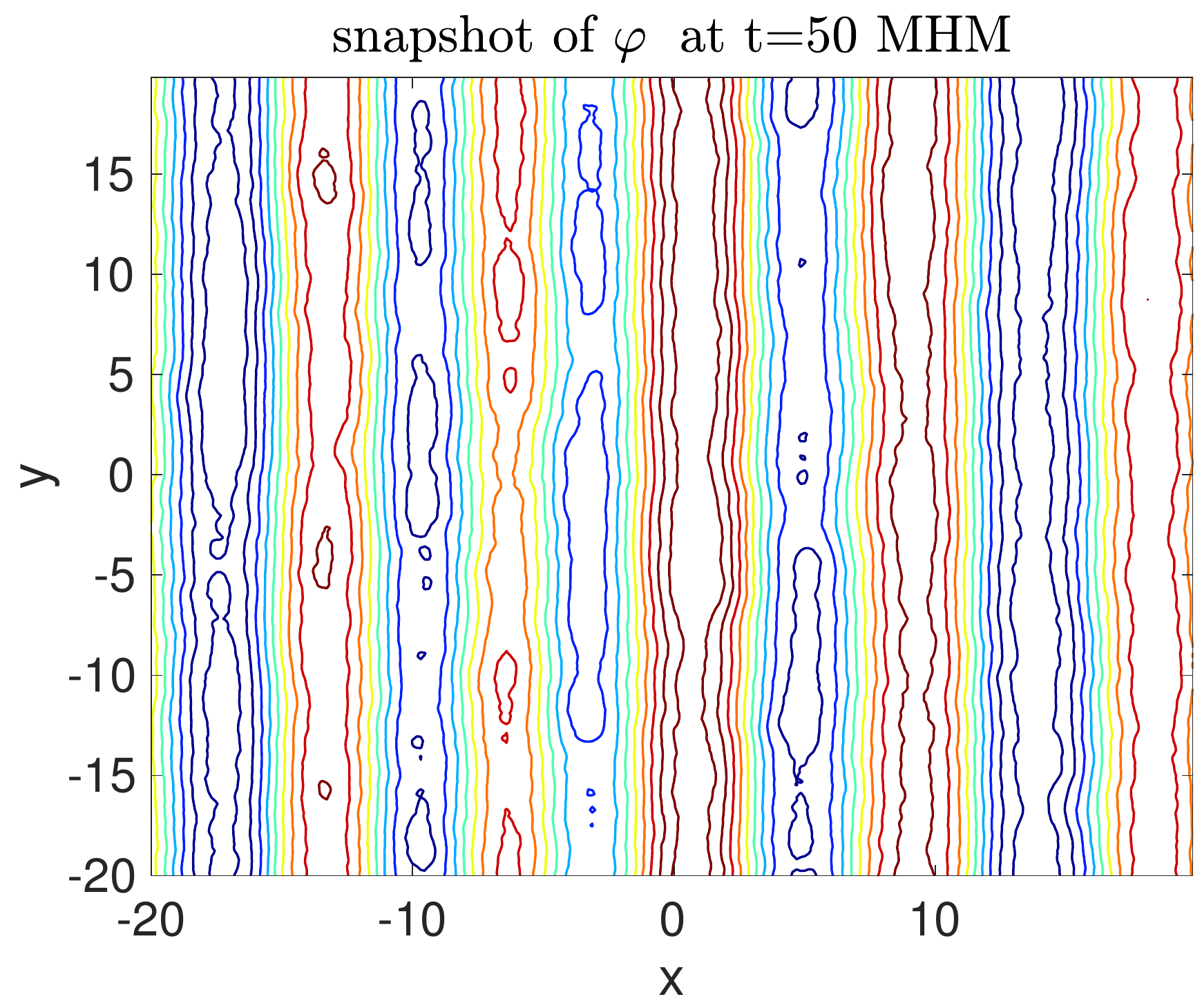}\includegraphics[scale=0.26]{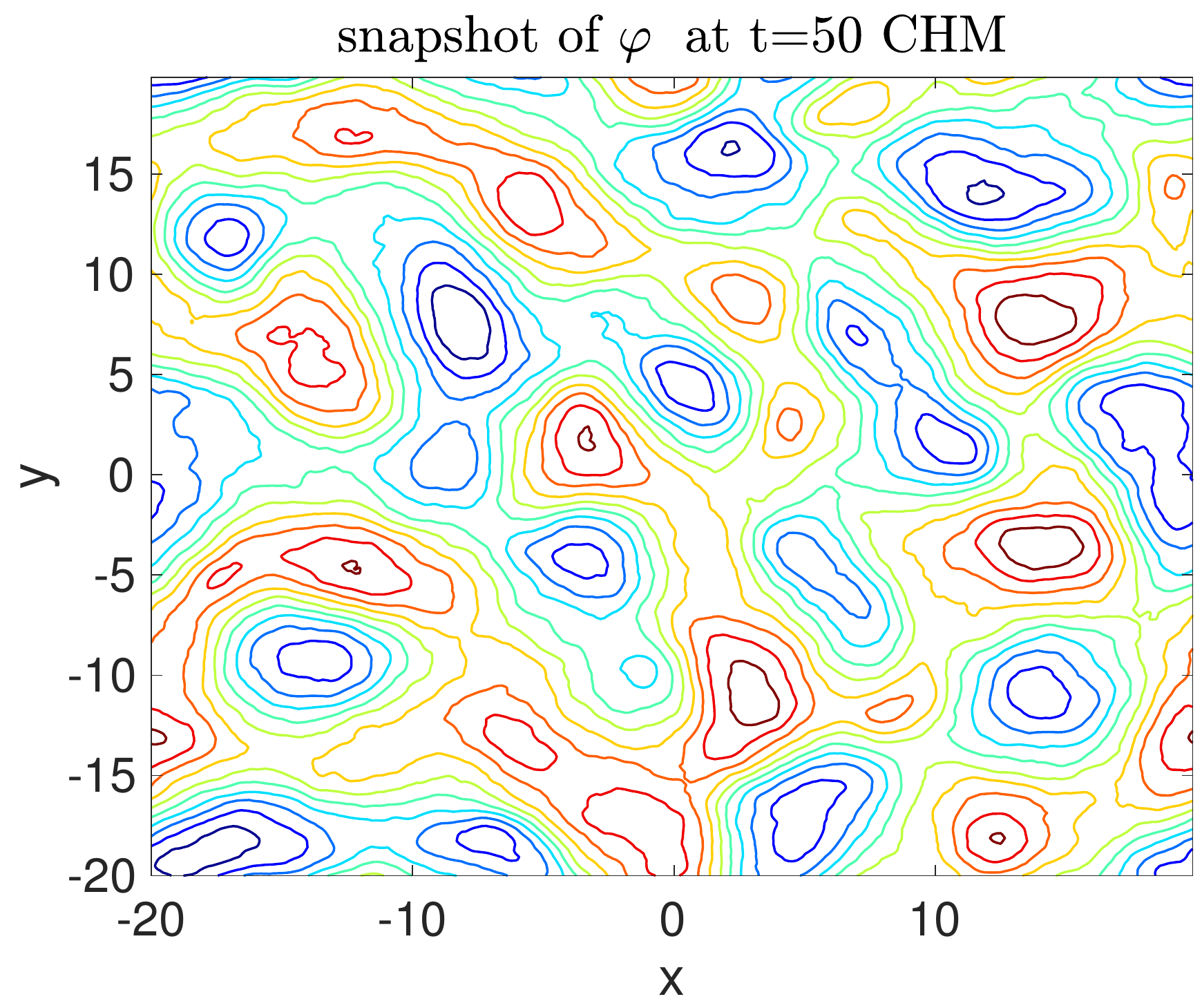}

}

\caption{Snapshots of the electrostatic potential function $\varphi$ in the
initial state (left) and the final numerical states from MHM (middle)
and CHM (right) model simulations without dissipation $D=0$. Two
different initial drift wave states with $s=2$ wavelengths (upper)
and with $s=10$ wavelengths (lower) are compared.\label{fig:Snapshots}}
\end{figure}

We check the explicit flow structures from the direct numerical simulations
of the MHM and CHM models. Figure \ref{fig:Snapshots} plots the snapshots
of the electrostatic potential functions $\varphi$ from both the MHM
and CHM model simulations starting from the same initial drift wave
structure (\ref{eq:init}). First in the MHM model in the second column,
starting from the pure drift wave state with only small isotropic
perturbations (first column of Figure \ref{fig:Snapshots}), the solutions
always generate strong zonal jets in the end. This confirms the transfer
of energy through the secondary instability shown in Figure \ref{fig:sec_instability_drift}
since no dissipation and other effect are included to generate the zonal
structures. Also since there is no dissipation, the small scale non-zonal
fluctuations always exist in the system on top of the zonal jets.

In comparison, the CHM model shown in the third column of Figure \ref{fig:Snapshots}
has difficulty in generating the zonal structures. In the first case
with a large scale wavenumber of two, the drift wave structure is maintained
in time as the system evolves. This is consistent with the secondary
stability result in the drift wave case where no positive growth rate is observed in the CHM
model with a large scale base flow (see the right column of Figure
\ref{fig:sec_instability_drift}). In the second test case with smaller
scale drift wave state of wavenumber ten initially, finally the pure
drift wave structure is destroyed due to the relatively stronger instability
with smaller scale drift waves. Still the flow breaks into homogeneous
drift wave turbulence without any zonal jet structure. This confirms
the little instability in the zonal modes (only in the largest scales)
in the CHM model case.

\subsection{Combined effects with secondary instability and selective decay}

In the previous test cases, we run the models without dissipation
effect. As described in \cite{qi2018selective}, the damping operator
usually adds stronger selective effect on the non-zonal modes and
drives the flow solution to a single wavenumber state at the long
time limit. Thus in this final test case, we consider the combined
contributions from both the secondary instability and the selective
damping. For simplicity, we introduce the simple dissipation operator,
$D\Delta q$, as in (\ref{eq:plasma_onelayer}) for both MHM and CHM models.
The damping rate is kept in small value $D=5\times10^{-4}$. In the
last two rows of Figure \ref{fig:Time-series}, we already show the
time-series of total energy and enstrophy in this case with dissipation.
Both energy and enstrophy are no longer conserved in time. Still the
energy/enstrophy in the zonal mean goes to the same level as the total energy/enstrophy at the long time limit 
in the MHM case, while the CHM only gets little energy in the zonal
state.

\begin{figure}
\subfloat[initial drift wave state with wavenumber $s=2$]{\includegraphics[scale=0.24]{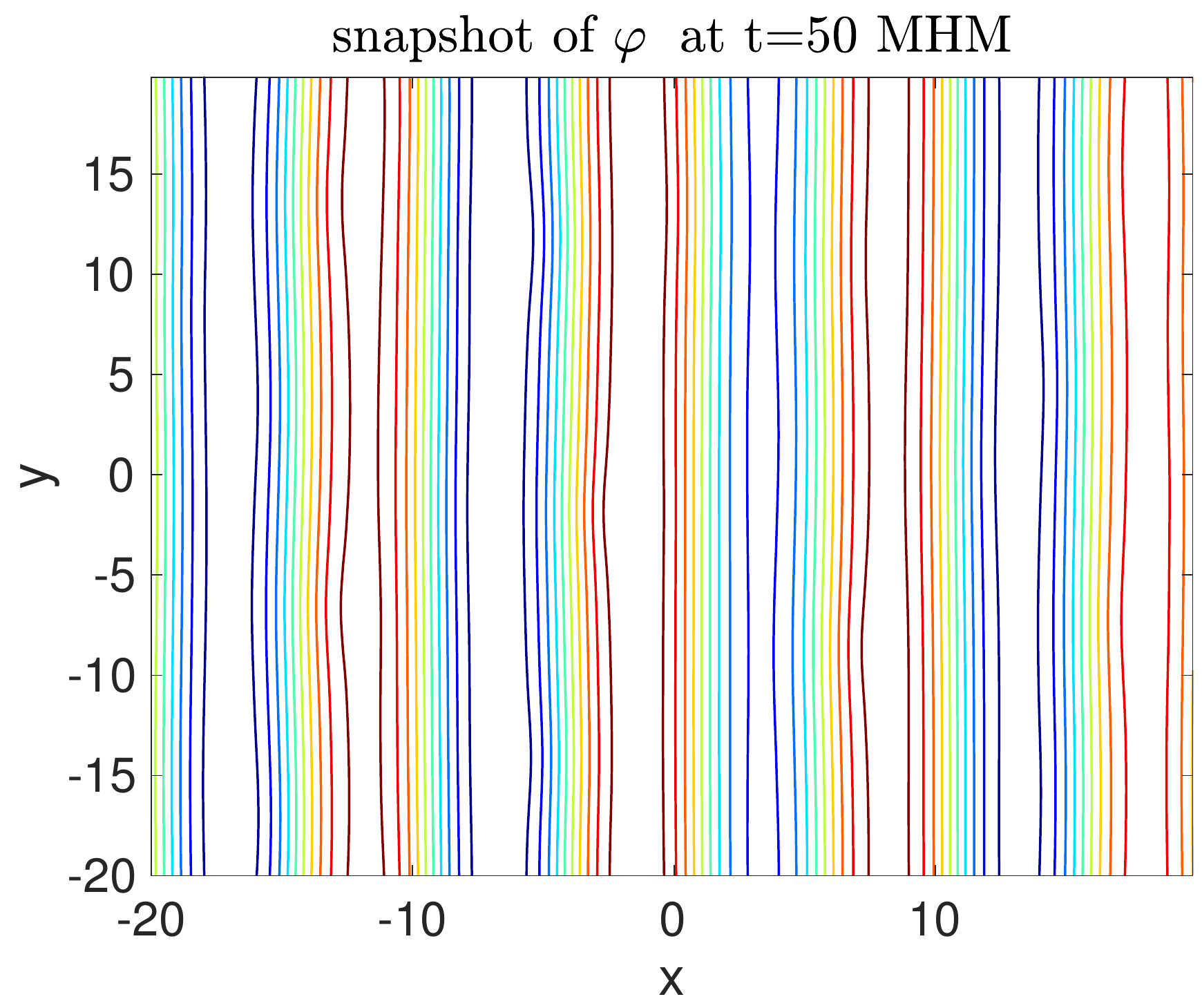}\includegraphics[scale=0.24]{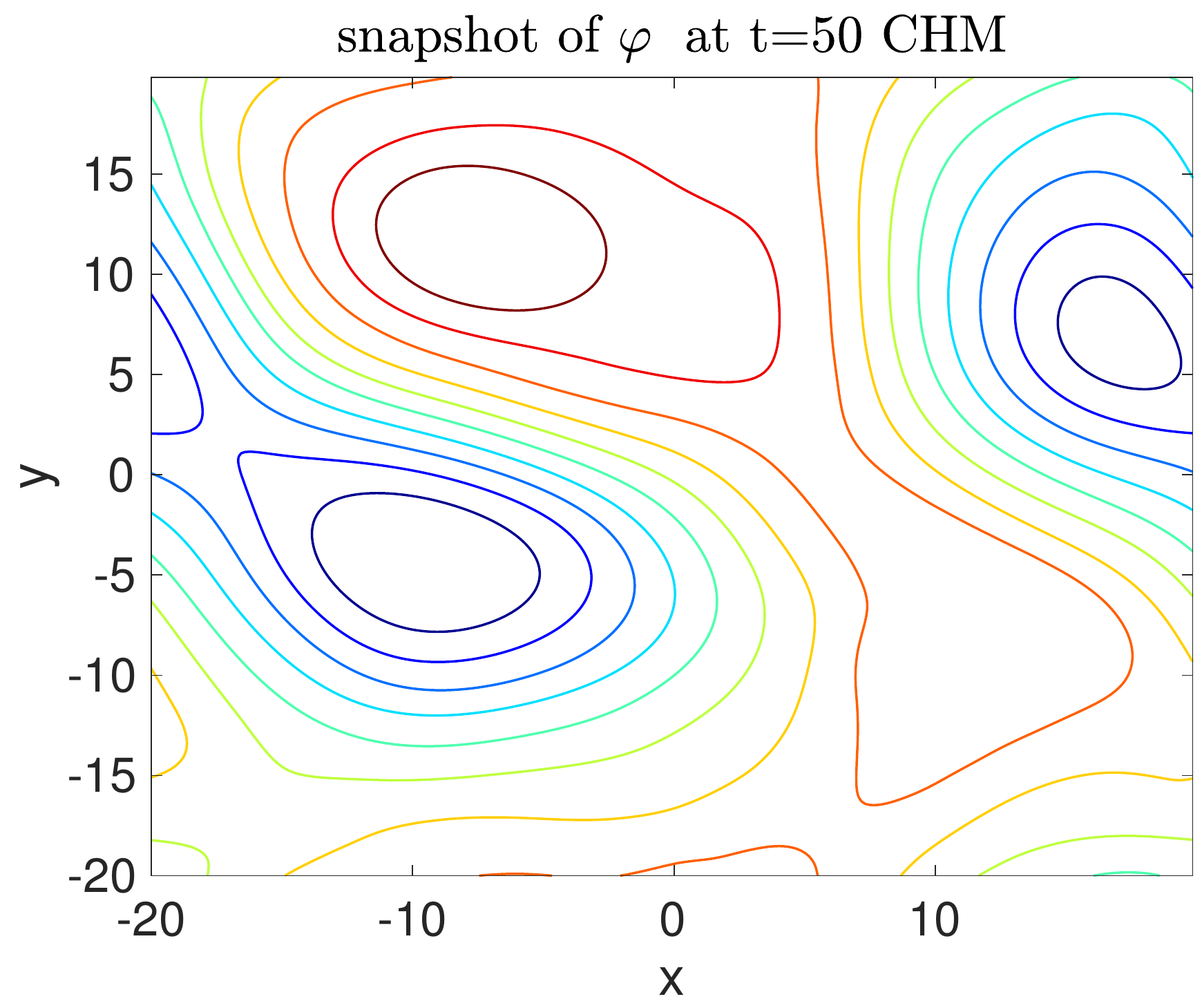}

}\:\subfloat[initial drift wave state with wavenumber $s=10$]{\includegraphics[scale=0.24]{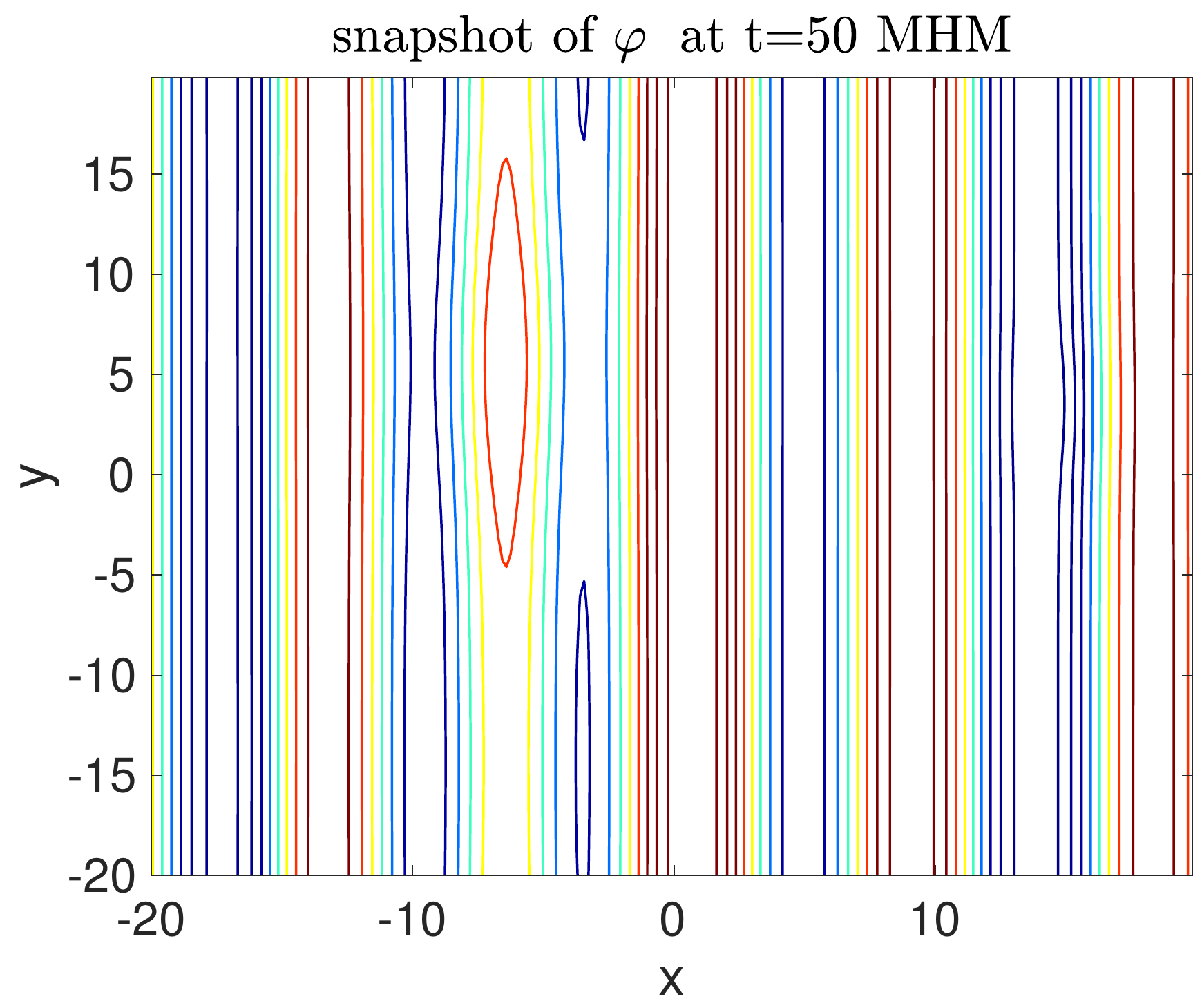}\includegraphics[scale=0.24]{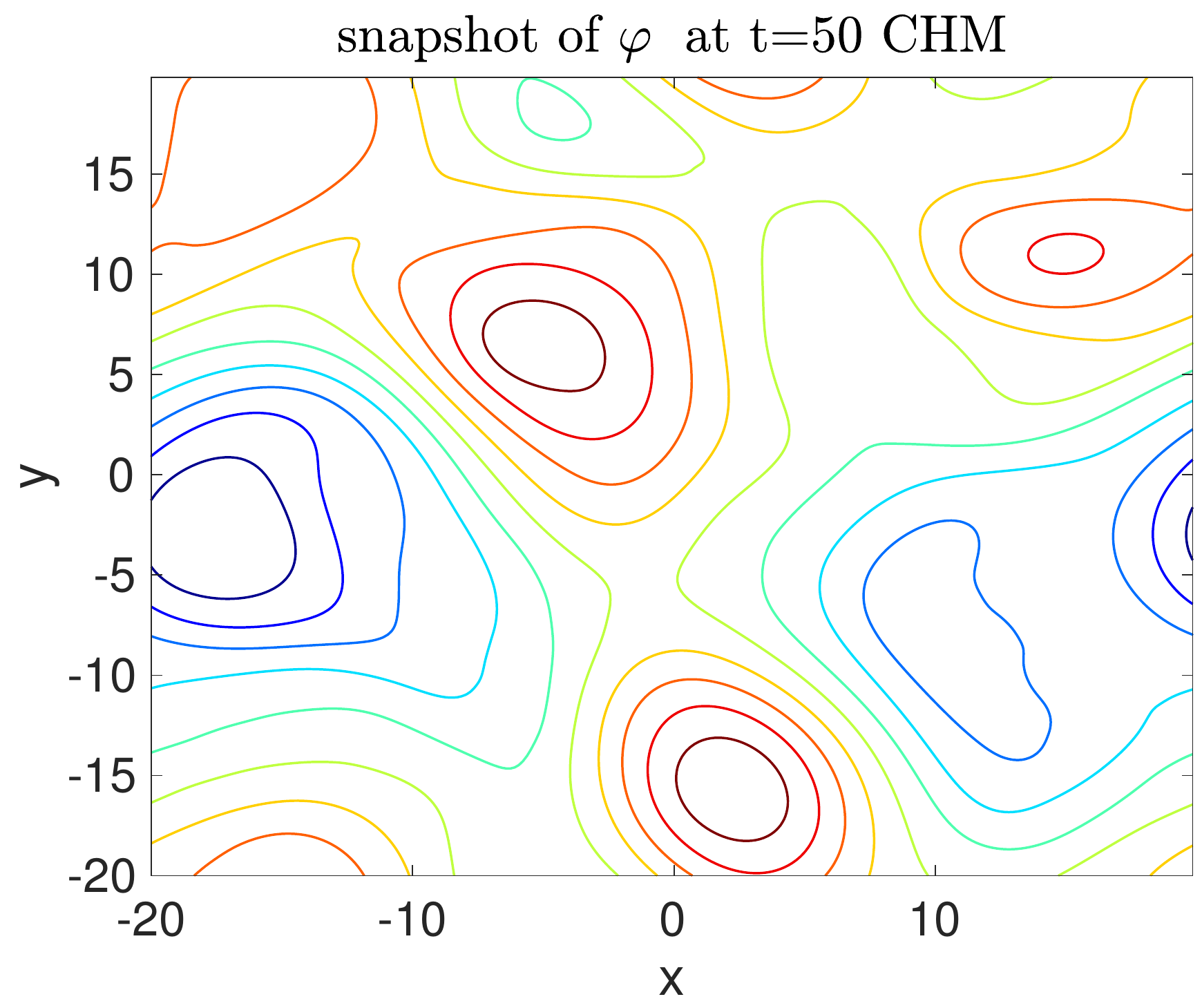}

}

\caption{Snapshots of the electrostatic potential function $\varphi$ in the
final states from MHM and CHM model simulations with dissipation $D=5\times10^{-4}$.
The same two initial drift wave states with $s=2$ and $s=10$ are
compared.\label{fig:Snapshots-dissip}}
\end{figure}

Again, we check the final dissipated solutions from the direct numerical
simulations. We plot in Figure \ref{fig:Snapshots-dissip} the snapshots
of the electrostatic potential $\varphi$ at the final simulation
time starting from the same two initial states with different drift
wave scales with the inclusion of dissipation effect. Comparing with
the the non-dissipative results in Figure \ref{fig:Snapshots}, the
fluctuating small-scale structures are damped down in this case while
the dominant zonal structure is still maintained in the MHM model.
This is the typical selective decay solution shown in Fig. 3 of \cite{qi2018selective},
while here the detailed energy exchanging mechanism is discovered
by the secondary instability. Observe that same number of zonal jets
in the two cases is reached as in the non-dissipative results. This
shows (together with the time-series of enstrophy in Figure \ref{fig:Time-series})
that the instability determines the final zonal structure in the first
place, then the selective decay takes over to dissipate all the other
non-zonal fluctuation modes to reach a clean single mode zonal state.
In comparison, the CHM model converges to a drift wave selective decay
state without zonal flows. With long enough time, the CHM flows will
finally converge to a single drift wave mode. More detailed results
about selective decay in the CHM model can be found in Fig. 1 of
\cite{qi2018selective} and \cite{majda2000selective}.

\subsubsection{Energy transfer mechanism in the decaying process}

\begin{figure}
\subfloat[energy spectra in radial averaged modes]{\includegraphics[scale=0.35]{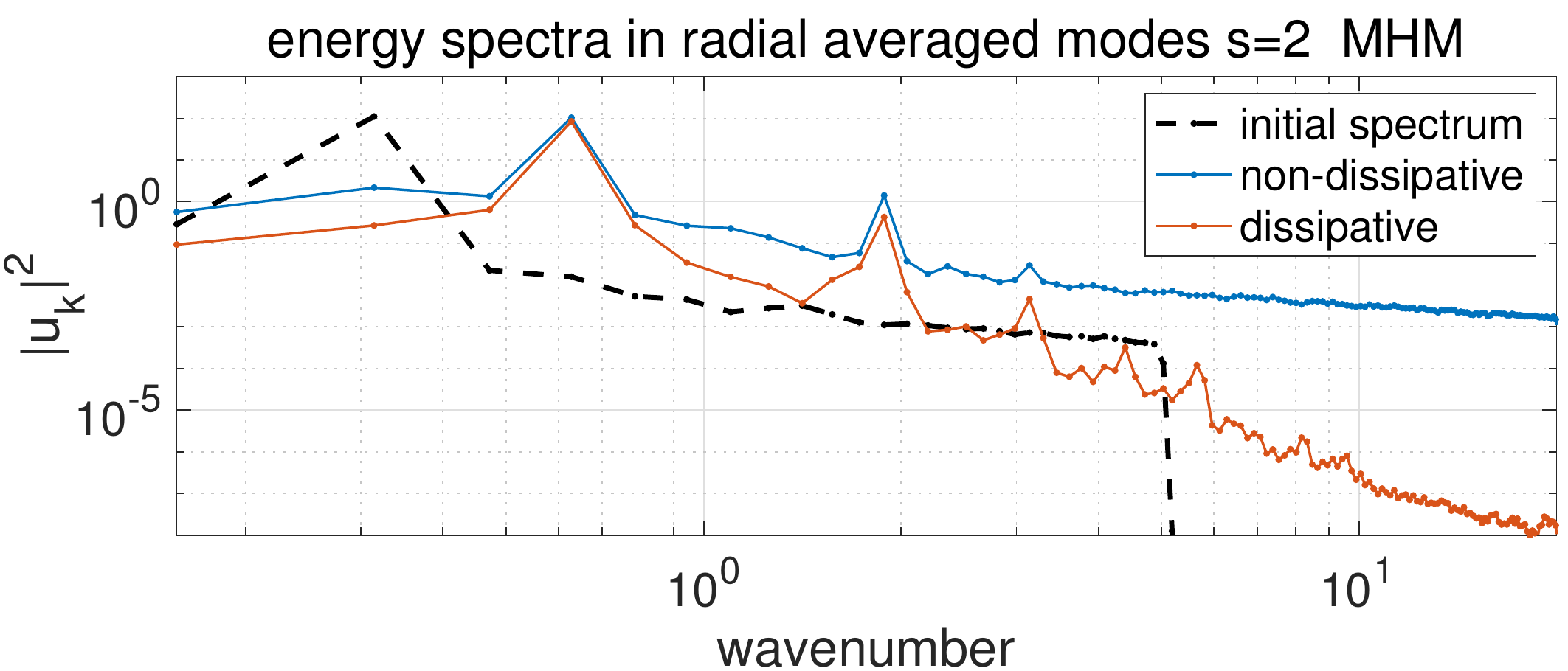}\includegraphics[scale=0.35]{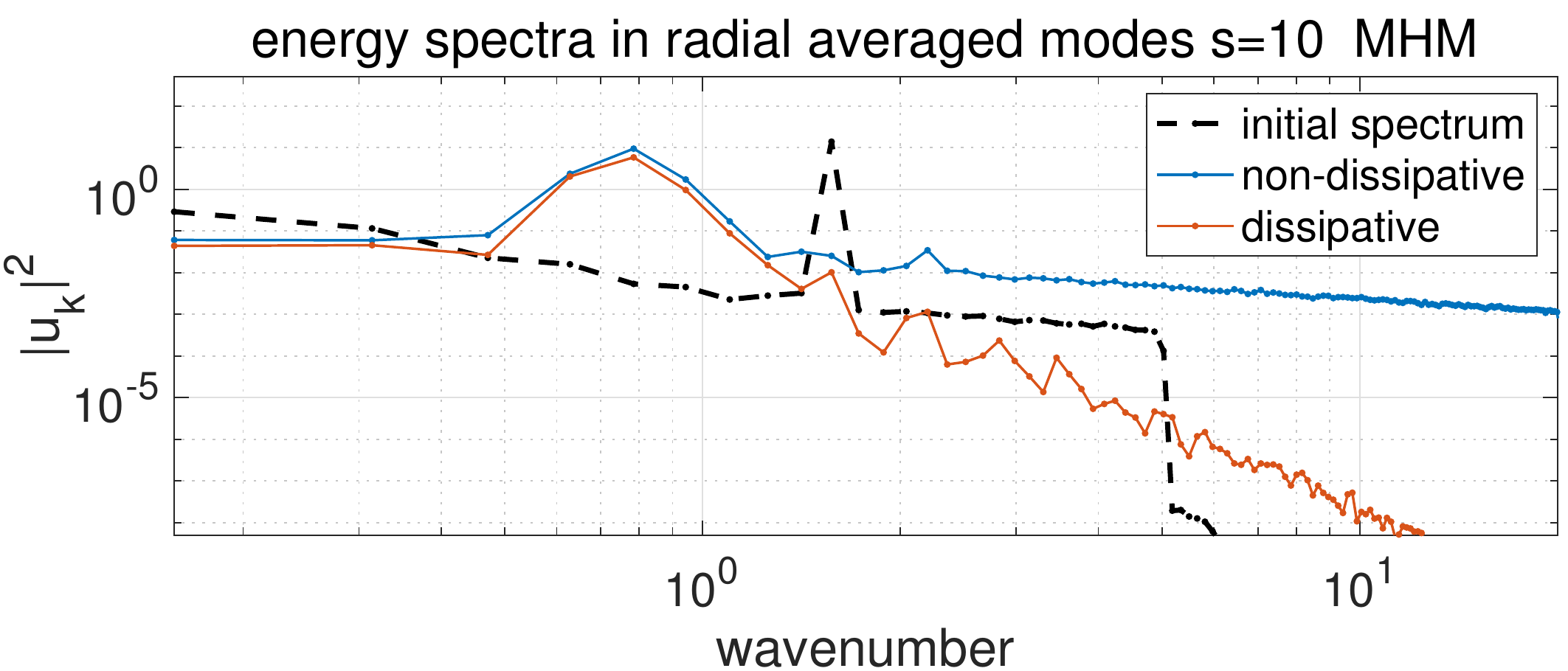}}

\subfloat[energy spectra in zonal modes]{\includegraphics[scale=0.35]{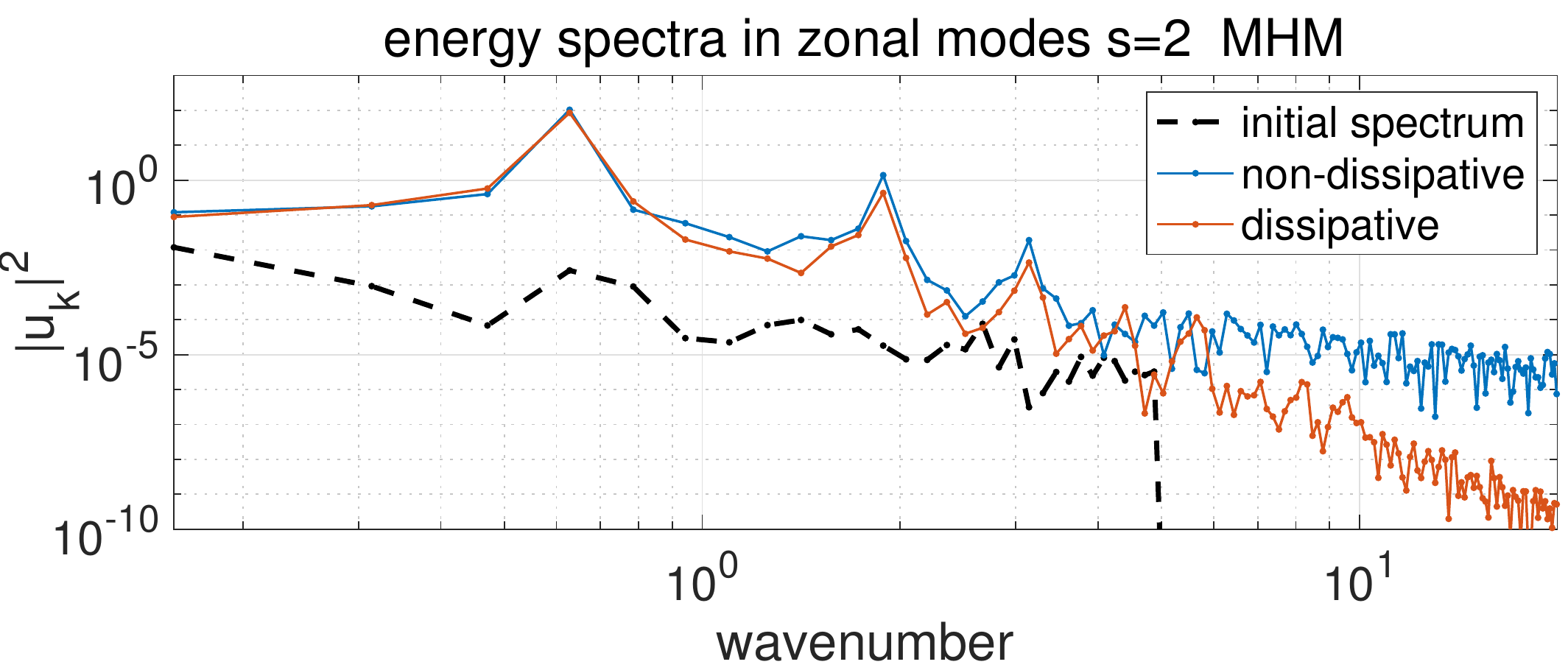}\includegraphics[scale=0.35]{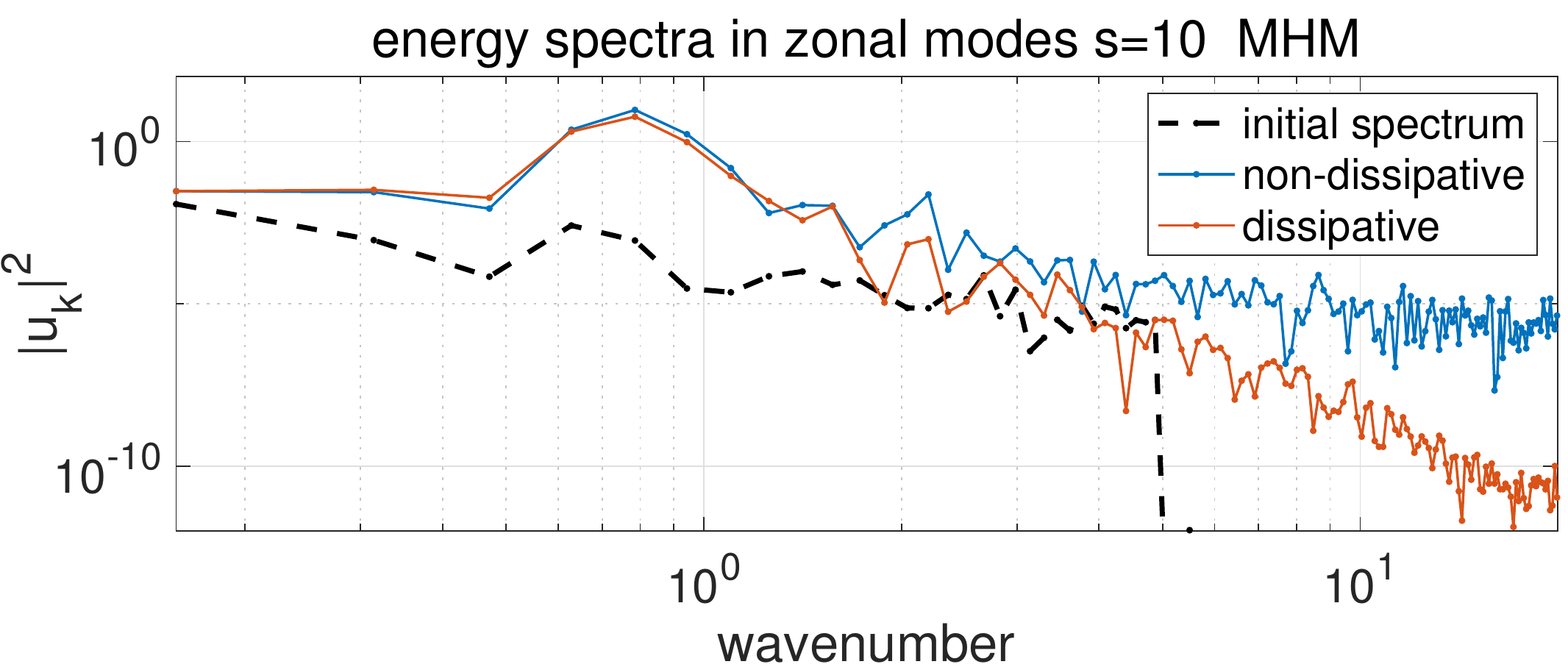}}

\caption{Energy spectra in radial averaged modes (upper) and purely zonal modes
(lower) from the MHM model simulations. The initial energy spectra
in the two test cases are shown together with the final spectra achieved
with and without dissipation effect.\label{fig:Energy-spectra}}
\end{figure}

We offer a more detailed illustration about the energy transfer mechanism
in time between different modes in the
MHM model through comparing the energy spectra. In Figure \ref{fig:Energy-spectra}, we plot the energy
spectra in the two initial cases with and without dissipation from
the MHM model simulation results. To display the transfer of energy
from the non-zonal drift wave modes to the zonal modes, we compare
the spectra in radially averaged modes in the first row and in zonal
modes $k^{y}=0$ only in the second row. The initial spectra get a
dominant second or tenth wavenumber from the initial setup (\ref{eq:init})
with small fluctuations and a high wavenumber truncation. The energy
will gradually cascade to smaller scales in the transient state. A
dominant zonal mode with largest energy emerges finally. With dissipation,
the selective damping effect only strongly dissipates the smaller
scale modes. The dominant zonal mode gets maintained at exactly the
same wavenumber as the non-dissipative case and stays with large energy
for all the time.

\begin{figure}
\subfloat{\includegraphics[scale=0.37]{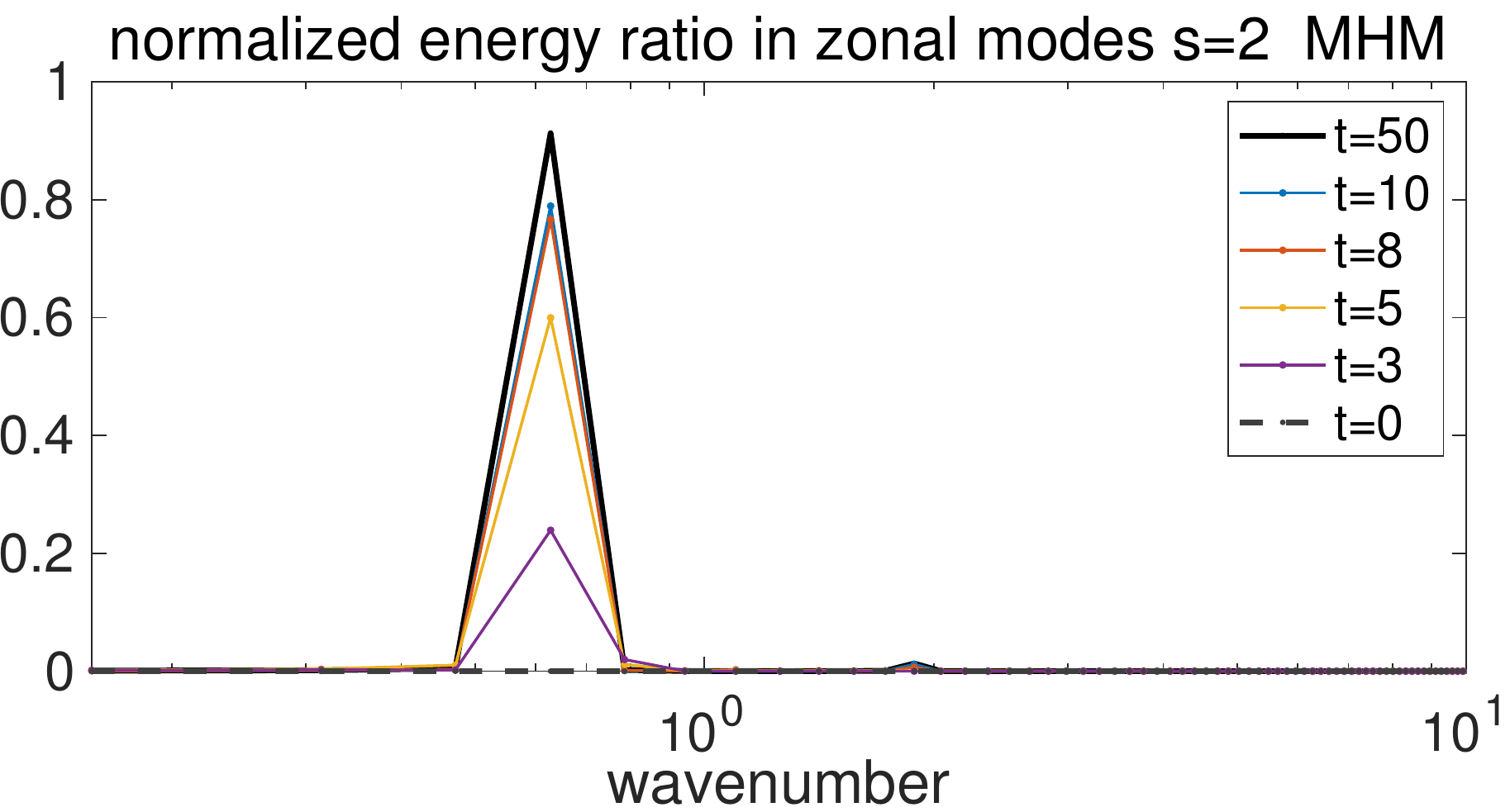}\includegraphics[scale=0.37]{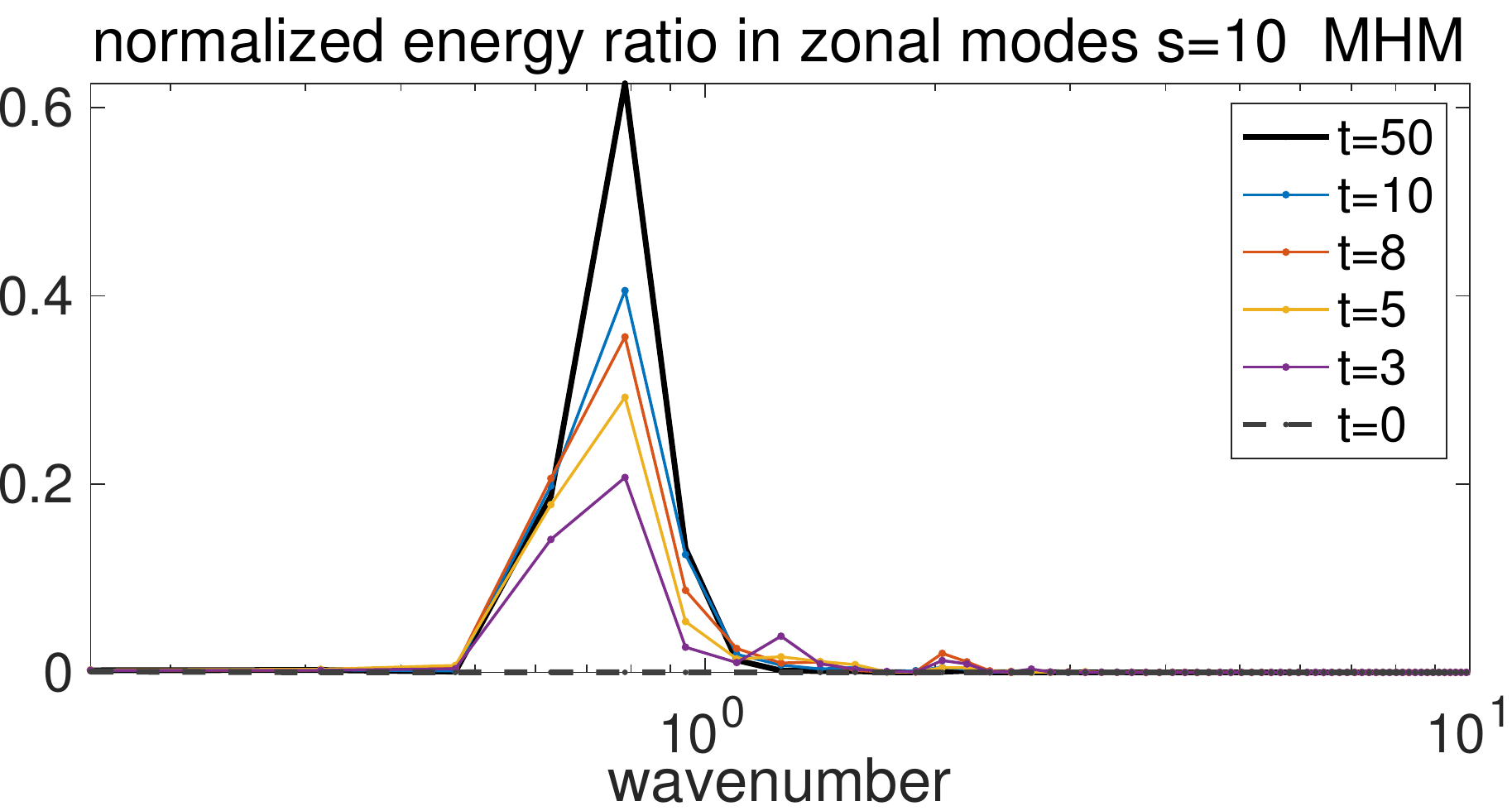}}

\vspace{-1em}

\subfloat{\includegraphics[scale=0.37]{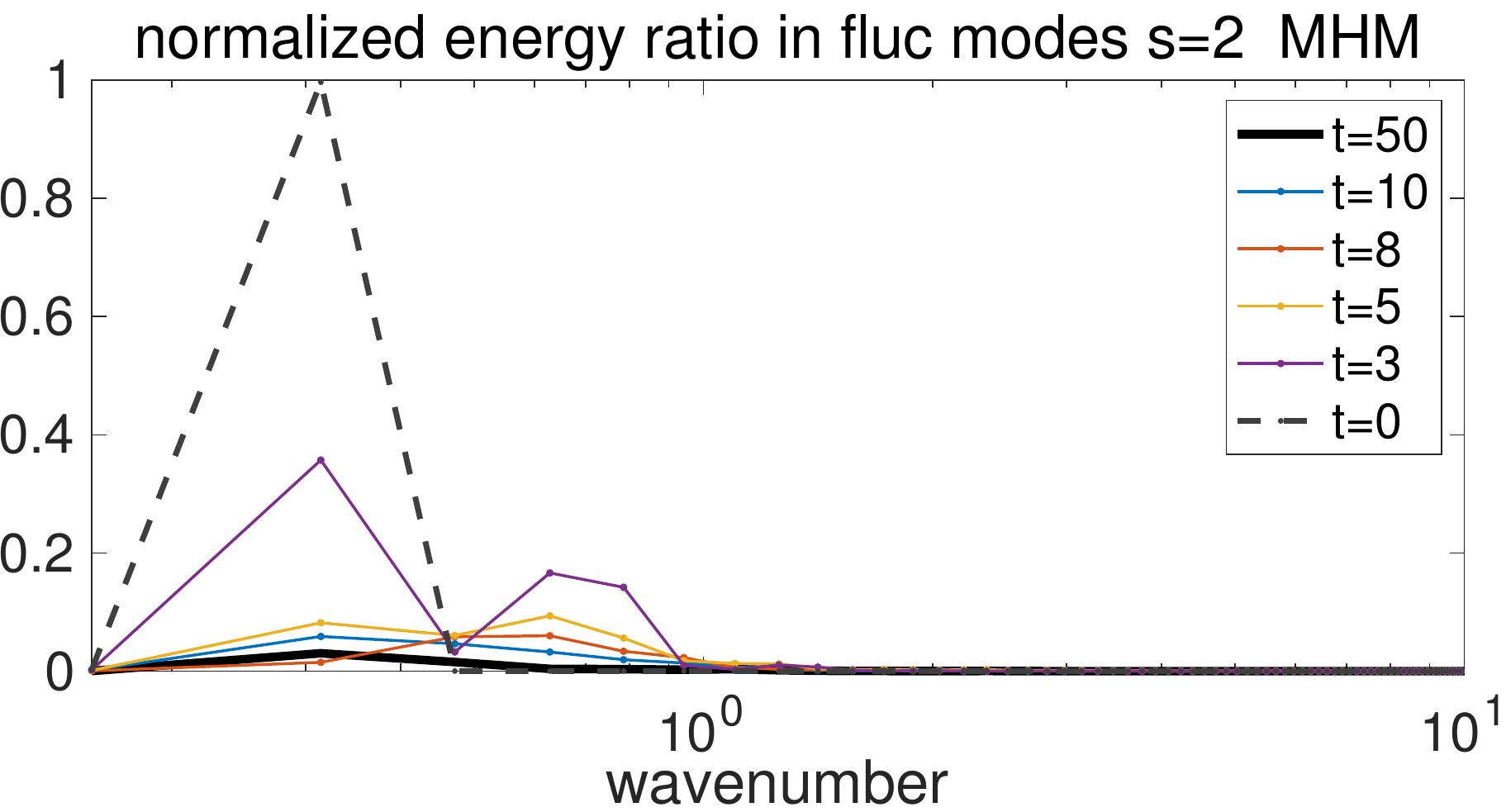}\includegraphics[scale=0.37]{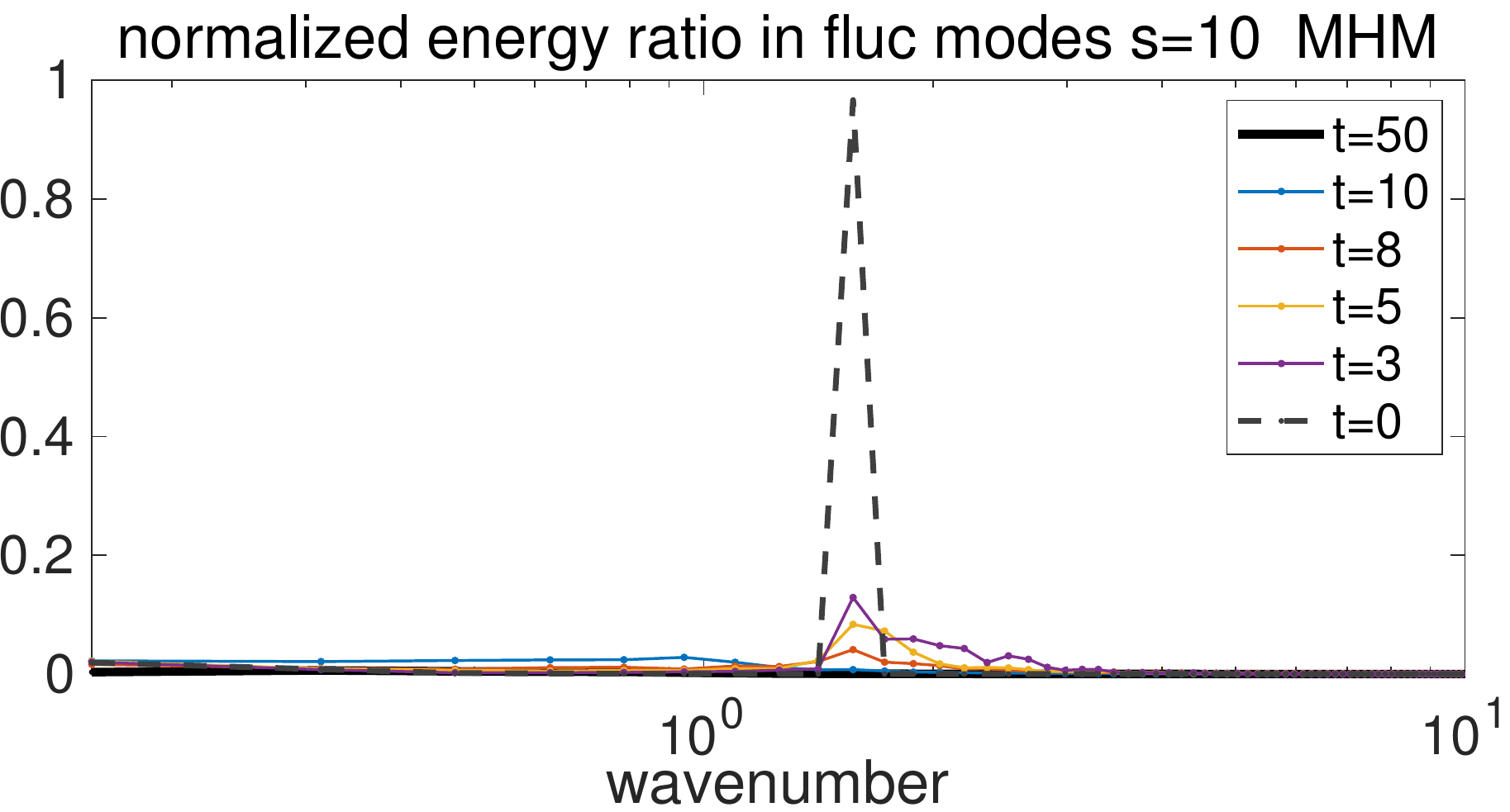}}

\caption{Selective decay of the fluctuation modes to pure zonal modes in the
MHM model from the energy ratios captured at several representative
time instant during the model evolution. The total energy is normalized
to one to emphasize the portion of energy in each mode.\label{fig:Selective-decay}}

\end{figure}

Finally, to offer a complete picture about the creation of pure
zonal jet structure through the combined effects of secondary instability
and the selective damping, we plot the normalized energy ratio for
the zonal modes $k^{y}=0$ and the non-zonal fluctuation modes at
several time instant in Figure \ref{fig:Selective-decay}. In the
initial state (shown in dashed black lines), all the energy is contained
in the pure drift wave mode with wavenumber two ($s=2$, left) or
wavenumber ten ($s=10$, right). As the starting transient state (at
around time $t=3$, see also the time-series of energy and enstrophy
in Figure \ref{fig:Time-series}), the energy in the zonal modes $k^{y}=0$
begins to grow due to the secondary instability induced by the interactions
between the drift waves and zonal modes. At later time (starts
from time $t=5$), the energy in the non-zonal drift wave modes begins
to cascade to smaller scales and gets dissipated by the selective
damping. In accordance with the time-series of energy plotted in Figure
\ref{fig:Time-series}, the energy in the zonal modes grows rapidly
between the time window $t\in\left[3,5\right]$. Then the selective
damping effect takes over to drive the state to purely zonal jets.
In addition, it can be observed from the energy ratios in the zonal
modes that there exist several intermediate metastable saddle
points which the solution visited before the convergence to the final stable
single selective decay zonal mode (see \cite{qi2018selective} for
a complete description of the selective decay).

\section{Concluding discussion}

In this paper, we perform secondary instability analysis about a background
base state to explain the zonal jet creation mechanism generally observed
in plasma edge turbulence. The one-state modified Hasegawa-Mima model
without internal drift wave instability is adopted to identify the
central drift wave -- zonal flow nonlinear interactions, and the
results are compared with the Charney-Hasegawa-Mima model. Together
with the selective decay principle developed previously in \cite{qi2018selective},
a complete picture for the generation and persistence of a dominate
zonal jet structure can be drawn. Starting from a drift wave base
state created from the first linear drift wave instability, secondary
instability due to nonlinear coupling with the fluctuation modes gradually
takes over and transfers the energy in the non-zonal drift wave states
to the zonal states. The induced zonal mode as a background state
is further stabilized from the negative secondary damping effect from
interacting with the perturbations. The small scale fluctuations from
the initial state are maintained if no dissipation exists in the system,
otherwise the selective decay effect will strongly dissipate the smaller
scale modes while it does not alter the dominant zonal structure created
from the instability. Direct numerical simulations of the MHM model
display the creation of zonal flows from a pure non-zonal drift wave
state with only small perturbation and without the effect of selective
decay. When dissipation is also added, secondary instability
is effective before the selective decay to generate the same number
of zonal jets, and the selective decay effect finally drives the state
to a clean single mode zonal jet structure. In contrast, the CHM model
cannot create zonal flows automatically and has no instability along the
zonal model direction.

Here we focus on the main energy mechanism for the creation of zonal structures,
thus a single mode base mode is always used throughout the paper in
illustrating the central role of secondary instability. As an immediate generalization,
the secondary instability with combined effects with multiple background
modes can be investigated. The multiple background base modes should
show stronger dominant exponential growth along the zonal direction
since different base modes enforce the instability in the zonal modes
together and have cancellation effect in the non-zonal directions. As a
further generalization, it is useful to consider the instability in
the two-state Hasegawa-Wakatani models. There we need to consider
the first linear instability in the base mode together with the secondary
instability on top of the stable/unstable base modes. Especially,
it is interesting to investigate the regime with large values of adiabatic
resistivity $\alpha$, where the model is on its way to approach the
MHM model discussed here.

\begin{acknowledgements}
This research of A. J. M. is partially supported by the Office of
Naval Research through MURI N00014-16-1-2161. D. Q. is supported as
a postdoctoral fellow on the grant.
\end{acknowledgements}

\bibliographystyle{spmpsci}      
\bibliography{ref}

\begin{thebibliography}{10}
\providecommand{\url}[1]{{#1}}
\providecommand{\urlprefix}{URL }
\expandafter\ifx\csname urlstyle\endcsname\relax
  \providecommand{\doi}[1]{DOI~\discretionary{}{}{}#1}\else
  \providecommand{\doi}{DOI~\discretionary{}{}{}\begingroup
  \urlstyle{rm}\Url}\fi

\bibitem{dewar2007zonal}
Dewar, R.L., Abdullatif, R.F.: Zonal flow generation by modulational
  instability.
\newblock In: Frontiers in Turbulence and Coherent Structures, pp. 415--430.
  World Scientific (2007)

\bibitem{diamond2005zonal}
Diamond, P.H., Itoh, S., Itoh, K., Hahm, T.: Zonal flows in plasma---a review.
\newblock Plasma Physics and Controlled Fusion \textbf{47}(5), R35 (2005)

\bibitem{fujizawa2009}
Fujisawa, A.: A review of zonal flow experiments.
\newblock Nuclear Fusion \textbf{49}(1), 013001 (2009).
\newblock \urlprefix\url{http://stacks.iop.org/0029-5515/49/i=1/a=013001}

\bibitem{hasegawa1978pseudo}
Hasegawa, A., Mima, K.: Pseudo-three-dimensional turbulence in magnetized
  nonuniform plasma.
\newblock The Physics of Fluids \textbf{21}(1), 87--92 (1978)

\bibitem{hasegawa1983plasma}
Hasegawa, A., Wakatani, M.: Plasma edge turbulence.
\newblock Physical Review Letters \textbf{50}(9), 682 (1983)

\bibitem{hortonreview1999}
Horton, W.: Drift waves and transport.
\newblock Rev. Mod. Phys. \textbf{71}, 735--778 (1999).
\newblock \doi{10.1103/RevModPhys.71.735}.
\newblock \urlprefix\url{https://link.aps.org/doi/10.1103/RevModPhys.71.735}

\bibitem{lee2003stability}
Lee, Y., Smith, L.M.: Stability of rossby waves in the $\beta$-plane
  approximation.
\newblock Physica D: Nonlinear Phenomena \textbf{179}(1-2), 53--91 (2003)

\bibitem{lin1998}
Lin, Z., Hahm, T.S., Lee, W.W., Tang, W.M., White, R.B.: Turbulent transport
  reduction by zonal flows: Massively parallel simulations.
\newblock Science \textbf{281}(5384), 1835--1837 (1998).
\newblock \doi{10.1126/science.281.5384.1835}.
\newblock \urlprefix\url{http://science.sciencemag.org/content/281/5384/1835}

\bibitem{majda2003introduction}
Majda, A.: Introduction to PDEs and Waves for the Atmosphere and Ocean, vol.~9.
\newblock American Mathematical Soc. (2003)

\bibitem{majda2016introduction}
Majda, A.J.: Introduction to turbulent dynamical systems in complex systems.
\newblock Springer (2016)

\bibitem{majda2018strategies}
Majda, A.J., Qi, D.: Strategies for reduced-order models for predicting the
  statistical responses and uncertainty quantification in complex turbulent
  dynamical systems.
\newblock SIAM Review \textbf{60}(3), 491--549 (2018)

\bibitem{majda2018flux}
Majda, A.J., Qi, D., Cerfon, A.J.: A flux-balanced fluid model for collisional
  plasma edge turbulence: Model derivation and basic physical features.
\newblock Physics of Plasmas \textbf{25}(10), 102307 (2018).
\newblock \doi{10.1063/1.5049389}.
\newblock \urlprefix\url{https://doi.org/10.1063/1.5049389}

\bibitem{majda2000selective}
Majda, A.J., Shim, S.Y., Wang, X.: Selective decay for geophysical flows.
\newblock Methods and applications of analysis \textbf{7}(3), 511--554 (2000)

\bibitem{manfredi2001zonal}
Manfredi, G., Roach, C., Dendy, R.: Zonal flow and streamer generation in drift
  turbulence.
\newblock Plasma physics and controlled fusion \textbf{43}(6), 825 (2001)

\bibitem{manz2013}
Manz, P., Ramisch, M., Stroth, U.: Physical mechanism behind zonal-flow
  generation in drift-wave turbulence.
\newblock Phys. Rev. Lett. \textbf{103}, 165004 (2009).
\newblock \doi{10.1103/PhysRevLett.103.165004}.
\newblock
  \urlprefix\url{https://link.aps.org/doi/10.1103/PhysRevLett.103.165004}

\bibitem{meshalkin1962investigation}
Meshalkin, L.: Investigation of the stability of a stationary solution of a
  system of equations for the plane movement of an incompressible viscous
  liquid.
\newblock J. Appl. Math. Mech. \textbf{25}, 1700--1705 (1962)

\bibitem{numata2007bifurcation}
Numata, R., Ball, R., Dewar, R.L.: Bifurcation in electrostatic resistive drift
  wave turbulence.
\newblock Physics of Plasmas \textbf{14}(10), 102312 (2007)

\bibitem{pedlosky2013geophysical}
Pedlosky, J.: Geophysical fluid dynamics.
\newblock Springer Science \& Business Media (2013)

\bibitem{pushkarev2013}
Pushkarev, A.V., Bos, W.J.T., Nazarenko, S.V.: Zonal flow generation and its
  feedback on turbulence production in drift wave turbulence.
\newblock Physics of Plasmas \textbf{20}(4), 042304 (2013).
\newblock \doi{10.1063/1.4802187}.
\newblock \urlprefix\url{https://doi.org/10.1063/1.4802187}

\bibitem{qi2016low}
Qi, D., Majda, A.J.: Low-dimensional reduced-order models for statistical
  response and uncertainty quantification: Two-layer baroclinic turbulence.
\newblock Journal of the Atmospheric Sciences \textbf{73}(12), 4609--4639
  (2016)

\bibitem{qi2018selective}
Qi, D., Majda, A.J.: Transient metastability and selective decay for the
  coherent zonal structures in plasma edge turbulence.
\newblock submitted to Journal of Nonlinear Science  (2018)

\bibitem{2018arXiv181200131Q}
{Qi}, D., {Majda}, A.J., {Cerfon}, A.J.: {A Flux-Balanced Fluid Model for
  Collisional Plasma Edge Turbulence: Numerical Simulations with Different
  Aspect Ratios}.
\newblock submitted to Physics of Plasmas arXiv:1812.00131 (2018)

\bibitem{rhines1975waves}
Rhines, P.B.: Waves and turbulence on a beta-plane.
\newblock Journal of Fluid Mechanics \textbf{69}(3), 417--443 (1975)

\bibitem{smolyakov2000coherent}
Smolyakov, A., Diamond, P., Malkov, M.: Coherent structure phenomena in drift
  wave--zonal flow turbulence.
\newblock Physical review letters \textbf{84}(3), 491 (2000)

\bibitem{xanthopoulos2011}
Xanthopoulos, P., Mischchenko, A., Helander, P., Sugama, H., Watanabe, T.H.:
  Zonal flow dynamics and control of turbulent transport in stellarators.
\newblock Phys. Rev. Lett. \textbf{107}, 245002 (2011).
\newblock \doi{10.1103/PhysRevLett.107.245002}.
\newblock
  \urlprefix\url{https://link.aps.org/doi/10.1103/PhysRevLett.107.245002}

\end{thebibliography}

\end{document}